\documentclass[sn-mathphys,Numbered]{sn-jnl}

\usepackage{graphicx}%
\usepackage{multirow}%
\usepackage{amsmath,amssymb,amsfonts}%
\usepackage{amsthm}%
\usepackage[referable]{threeparttablex}
\usepackage{threeparttable}
\usepackage{rotating}

\usepackage{mathrsfs}%
\usepackage[title]{appendix}%
\usepackage{xcolor}%
\usepackage{textcomp}%
\usepackage{manyfoot}%
\usepackage{booktabs}%
\usepackage{algorithm}%
\usepackage{algorithmicx}%
\usepackage{algpseudocode}%
\usepackage{listings}%
\usepackage{caption}
\usepackage{subcaption}
\usepackage{courier}
\usepackage{tabularx}
\usepackage{adjustbox}

\theoremstyle{thmstyleone}%
\theoremstyle{thmstyletwo}%

\theoremstyle{thmstylethree}%

\begin{document}

\title[Article Title]{Performance analysis of Deep Learning-based Lossy Point Cloud Geometry Compression Coding Solutions}


\author{\fnm{Joao} \sur{Prazeres}}\email{joao.prazeres@ubi.pt}
\equalcont{These authors contributed equally to this work.}

\author{\fnm{Rafael} \sur{Rodrigues}}\email{rafael.rodrigues@ubi.pt}
\equalcont{These authors contributed equally to this work.}

\author{\fnm{Manuela} \sur{Pereira}}\email{mpereira@di.ubi.pt}
\equalcont{These authors contributed equally to this work.}

\author{\fnm{António M.G} \sur{Pinheiro}}\email{pinheiro@ubi.pt}
\equalcont{These authors contributed equally to this work.}

\affil{\orgdiv{Instituto de Telecomunicacoes}, \orgname{Universidade da Beira Interior}, \orgaddress{\street{Rua Marques D'Ávila e Bolama}, \city{Covilha}, \postcode{6201-001}, \state{Castelo Branco}, \country{Portugal}}}


\abstract{The quality evaluation of three deep learning-based coding solutions for point cloud geometry, notably ADLPCC, PCC GEO CNNv2, and PCGCv2, is presented.
The MPEG G-PCC was used as an anchor.
Furthermore, LUT SR, which uses multi-resolution Look-Up tables, was also considered.
A set of six point clouds representing landscapes and objects were used.
As point cloud texture has a great influence on the perceived quality, two different subjective studies that differ in the texture addition model are reported and statistically compared.
In the first experiment, the dataset was first encoded with the identified codecs. Then, the texture of the original point cloud was mapped to the decoded point cloud using the \textit{Meshlab} software, resulting in a point cloud with both geometry and texture information. Finally, the resulting point cloud was encoded with G-PCC using the \texttt{lossless-geometry-lossy-atts} mode, while in the second experiment the texture was mapped directly onto the distorted geometry.
Moreover, both subjective evaluations were used to benchmark a set of objective point cloud quality metrics. The two experiments were shown to be statistically different, and the tested metrics revealed quite different behaviors for the two sets of data. The results reveal that the preferred method of evaluation is the encoding of texture information with G-PCC after mapping the texture of the original point cloud to the distorted point cloud.
The results suggest that current objective metrics are not suitable to evaluate distortions created by machine learning-based codecs.
Finally, this paper presents a study on the compression performance stability of the tested machine learning-based codecs using different training sessions.
The obtained results show that the tested codecs revealed a high level of stability across all training sessions for most of the content, although some undesirable exceptions may be found.}




\maketitle

\section{Introduction}\label{sec1}

Point cloud technology has emerged as a popular method for 3D data representation. A wide variety of applications may benefit from point cloud technology, such as virtual, augmented, and mixed reality applications; 3D printing; automation and robotics; computer graphics and gaming; and medical applications, among others.

In point clouds, objects or scenes are represented by a set of Cartesian coordinates $(x, y, z)$, with each containing a list of attributes, such as RGB color components, reflectance values, normal vectors, or physical sensor information. These allow an accurate representation of objects or scenes from any viewing position or distance, thus making them a very powerful representation model. However, an accurate representation of a building or an artifact may require several millions of points and their associated attributes, leading to huge amounts of information. Hence, there is a need for efficient point cloud coding solutions.
The main contributions of this paper are as follows:
\begin{itemize}
    \item A comparison between two different methods of adding texture to point clouds encoded by learning based solutions that only encode geometry, in order to benchmark them. Two subjective evaluations were conducted with each method, and the result of both was compared.
    \item  An analysis on the stability of machine learning-based coding solutions across different training sessions. This methodology follows the model of a previous study~\cite{euvip22}, aiming to study the reproducibility of learning based point cloud coding solutions.
\end{itemize}

For the comparison between evaluations, the MPEG G-PCC~\cite{G-PCC} was used as a benchmarking anchor for all studies. LUT SR, a fractional super-resolution method for G-PCC reconstructed point clouds \cite{QueirozTechRxiv2021} was also included. It should be emphasized, the main goal is to assert the best way of evaluating point cloud coding solutions that encode only geometry.
The selected learning based solutions for this study were three DL-based codecs that only compress the geometry information. For a subjective study, texture information is required since it is highly important for the subjective quality definition. However, the method used for adding texture might be controversial, as it may introduce texture artifacts not created by the codecs under evaluation~\cite{Silva2019a,MMSP2021}. For that reason, two different models were tested.

In the first model, the texture is compressed with G-PCC, and a balance between the compression rate of the texture and geometry is established.
In the second model, the texture information was mapped from the original point clouds without further encoding onto the coded geometry.
These two models are compared in order to understand which was the best model to add the texture information vital to an appropriate subjective study.
While most previous studies only considered point cloud representations of small objects, point cloud representations of landscapes were included as well.

It is important to emphasize that the quality analysis could be done with point clouds without any texture~\cite{Alexiou2018a}. However, during JPEG Pleno development~\cite{astola2020jpeg}, it was observed that the texture has a strong influence on subjective evaluations, and point cloud visualization without textures leads to results that are difficult to analyze.
Moreover, the emerging deep learning codecs produce distortions that are extremely different from the ones created by the typical codecs (projection-based and octree-based). The typical subjective evaluation models were developed before the emergence of such technology, and their performance should be analyzed to verify which model is more adequate to evaluate this new generation of codecs.

Finally, this paper also provides a study on the stability of machine learning-based coding solutions across different training sessions, following the method of a previous study~\cite{euvip22}.
The remainder of this paper provides a short description of the state of the art in point cloud technology, notably in subjective quality evaluation, objective quality computation, and coding solutions, including the most popular DL-based solutions.
The two subjective quality evaluation experiments are then described and analyzed, followed by a study on the performance of the selected objective metrics.
In Section \ref{sec:stabExp}, the stability of the three DL-based solutions is analyzed. A conclusion section finalizes the paper.

\section{State of the Art}

\subsection{Point Cloud Coding}\label{sec:pcc}
The most traditional coding model for point clouds is based on the octree pruning method~\cite{rusu2011a}.
Recently, MPEG defined the Geometry-based Point Cloud Compression (G-PCC)~\cite{G-PCC} based on the octree point cloud representation. 
G-PCC also defines the trisoup method based on surface reconstruction for geometry compression. The point cloud attributes are compressed either with RAHT, the lifting transform or the predictive transform.
For this study, only the octree method is considered for geometry encoding and the lifting transform for texture encoding. It has been previously shown that subjects tend to prefer the lifting transform~\cite{alexiou2019a} over the RAHT algorithm. This method was selected in another evaluation~\cite{icip2020}, presenting good results.

Another trend for point cloud coding is the encoding of point cloud projections, which can be coded by any image coding codec.
MPEG also explored that approach, resulting in the Video-based Point Cloud Compression (V-PCC)~\cite{vpcc} defined for dynamic point clouds.
V-PCC relies on HEVC (and more recently on VVC) to encode 2D projections of a given point cloud. Despite being developed for dynamic point clouds, its intra-coding has been revealed to be the most efficient for static point cloud coding~\cite{icip2020,EI2022}.

Following the good performance in image coding, several machine learning-based coding solutions for point clouds have been proposed recently~\cite{8954537,9287060,Jianqiang-PCGC,Jianqiang-PCGCv2,quach2019-geocnn,quach2020improved,GuardaDL,pang2022grasp,WangPAMI2022}.
These solutions usually cause distortions that are quite different from those caused by common codecs, which typically create holes in point cloud surfaces.
Hence, there is a need to analyze the reliability of the quality models for learning-based codecs.

The DL-based codecs selected for this study are in concordance with an MPEG document~\cite{MPEGDOC}: Multiscale Point Cloud Geometry Compression (PCGC)~\cite{Jianqiang-PCGCv2}, Deep Point Cloud Geometry Compression~\cite{quach2020improved}, and Adaptive Deep Learning Point Cloud Coding (ADLPCC)~\cite{GuardaDL}. The codecs were compared against the MPEG anchor G-PCC~\cite{G-PCC} (V.14), using the octree mode, and the LUT SR~\cite{QueirozTechRxiv2021}.

PCGCv2~\cite{Jianqiang-PCGCv2} performs block-wise multi-resolution encoding. The point cloud is downsampled three times, and the encoding is done recurring to the Inception Residual Network~\cite{SzegedyInceptionv4}. At the bottleneck, the geometry coordinates are encoded with G-PCC, and entropy coding is used for the attributes. The decoding branch architecture mirrors the encoder.
The implementation available at\footnote{available at https://github.com/NJUVISION/PCGCv2} was used.

Deep Point Cloud Geometry Compression~\cite{quach2020improved} learns an encoding function from three sequential convolution layers. The first two use ReLU activation. The latent representation of the third label is quantized through element-wise integer rounding and then compressed through a combination of algorithms. The decoding architecture mirrors the encoding. The output of the last layer is converted to the distorted point cloud using element-wise minimum, maximum, and rounding functions.
The implementation available at\footnote{available at https://github.com/mauriceqch/pcc\_geo\_cnn\_v2} was used.

ADLPCC~\cite{GuardaDL} partitions the point cloud into regular-sized 3D blocks. Several models separately code those blocks. The codec contains an autoencoder (AE) and a variatonal autoencoder (VAE) with three convolutional layers of both encoding and reconstruction, with sigmoid and ReLU activations, respectively.
The implementation available at\footnote{https://github.com/aguarda/ADLPCC} was used.

Moreover, the LUT SR Look-Up Tables~\cite{QueirozTechRxiv2021} solution is also considered. Based on G-PCC, it creates a hierarchical tree-like dictionary, mapping the occupancy relationships between downsampled geometry and the reference. A second downsampling is performed, storing the occupancy in the dictionary as well as the neighborhood configuration. The point cloud is then upsampled by applying nearest-neighbor interpolation to find all the possible child nodes of the input point cloud. The resulting geometry is obtained by following the respective dictionary entries.
The implementation available at\footnote{https://github.com/digitalivp/PCC\_LUT SR} was used.

\subsection{Point Cloud Subjective Quality Evaluation}
Several point cloud quality evaluation studies have been proposed, considering different coding methodologies and setups. Several studies established quality models for geometry-only encoding methods, such as octree-based~\cite{Javaheri2017b, Alexiou2018a, alexious2018point, Silva2019a}, graph-based~\cite{Javaheri2017b}, and projection-based encoding~\cite{Silva2019a}.
Honglei \textit{et al.}~\cite{Su2019a} carried out a subjective evaluation of MPEG test models V-PCC and also S-PCC and L-PCC, which were earlier proposals that led to the final version of G-PCC. In the same year, Alexiou \textit{et al.} also reported on an early subjective evaluation of G-PCC and V-PCC~\cite{alexiou2019a}, before these were standardized. The previously mentioned study~\cite{icip2020} reports a subjective quality evaluation of MPEG Point Cloud codecs using a 2D visualization setup. An initial quality study of DL-based point cloud coding quality was presented, targeting machine learning codecs~\cite{acm22}.

Subjective evaluation using augmented or virtual reality (AR/VR) environments has been previously researched~\cite{AlexiouMMSP2017, ViolaVr, MekuriaVR, SubramanyamQoMEX2022}. Alexiou \textit{et al.} proposed PointXR~\cite{Alexiou:277378}, a toolbox for visualization and subjective evaluation of point clouds in VR environments. Recently, a subjective quality evaluation conducted in a 3D environment was presented in~\cite{ICIP2022}. The obtained results were compared to a previous study using 2D displays~\cite{EI2022}, and showed no statistical differences.
Crowdsourcing methodologies have also been studied~\cite{ak2023basics} as a method of subjective evaluation.
A subjective evaluation was conducted using a light field display and compared to an evaluation conducted with a 2D display~\cite{LazzarottoQoMEX2022}. The evaluations were highly correlated but presented statistical differences, and the authors concluded that no benefits were gained from using a light field display.

Large scale subjective evaluation provides databases with annotated mean opinion scores (MOS), which are then used for subjective evaluation. Such databases include the Waterloo~\cite{Liu2022TVCG}, SJTU-PCQA~\cite{projectionYANG}, LS-PCQA~\cite{LiuYipengACM2023} and BASICS~\cite{ak2023basics}. These databases are important for developing new point cloud quality objective metrics.

\subsection{Point Cloud Objective Quality Evaluation}\label{sec:metrics}

Objective quality metrics aim at accurately predicting the visual quality of content representations and may be used to set up codecs for an improved quality of experience without the need for subjective studies.
These may be classified as image-based or model-based metrics specifically developed for point cloud quality evaluation~\cite{7272102}. Image-based metrics, such as PSNR and SSIM, operate directly on representative 2D views.

The aforementioned study~\cite{icip2020} tested a group of point-based metrics and concluded that point-to-point and point-to-plane metrics~\cite{Dtian} using the Mean Squared Error (MSE) were the best performing ones and provided a good representation of the subjective evaluation. Later, a benchmarking study using a 2D experimental setup for the subjective assessment, which included a broader selection of objective quality metrics, was reported~\cite{LazzarottoMMSP2021}, with PCQM and PointSSIM showing the best performances in terms of correlation with the Mean Opinion Score (MOS). Moreover, the image metric Multiscale Structural Similarity Index (MS-SSIM) computed over the video generated for the 2-D visualization of the point cloud also revealed a good representation of the subjective evaluation.
Recently, no reference metrics are also being proposed~\cite{Yang_2022_CVPR,ZhangCSVT2022,Liu2022TVCG,LiuYipengACM2023}. These metrics use features extracted from the distorted point cloud and deep learning technology to evaluate the compression quality.

For this paper, a set of objective metrics was considered, namely the MSE PSNR D1 and MSE PSNR D2~\cite{DongICIP2017}, the Point Cloud Structural Similarity (PointSSIM) Metric~\cite{AlexiouPSSIM}, the Point Cloud Quality Metric (PCQM)~\cite{Meynet2020PCQM}, the Point to Distribution Metric~\cite{javaheri2021pointtodistribution}, the Reduced Reference Point Cloud Metric~\cite{Viola2020PCMRR} and the GraphSIM~\cite{QiYangGraphSIM2022} metric. These metrics are widely used in subjective evaluation, and they usually provide a good representation of subjective results~\cite{LazzarottoMMSP2021}. The MSE PSNR D1 and D2 allow evaluations of the performance of the coding solutions regarding geometry, and the remaining metrics evaluate the performance based on both geometry and texture information.
In the following, a short description of these metrics is presented.

\subsubsection{Point-to-Point - MSE PSNR D1~\cite{Dtian}}\label{D1}
The MSE PSNR D1 metric measures geometric distortions by computing the Euclidean distance between every point $b_k$ in the distorted input and the nearest corresponding point $a_i$ in the reference point cloud, $E(a_i , b_k) = |(\overrightarrow{v}^{a_i}_{b_k} )|_2$, and the MSE considering $E(a_i, b_k)$ is taken. The final output is given by the PSNR as follows: $D1 = 10 \cdot \log_{10} \left( \frac{max^2}{MSE} \right)$.
The available MPEG implementation was used\footnote[1]{http://mpegx.int-evry.fr/software/MPEG/PCC/mpeg-pcc-dmetric/tree/master}.

\subsubsection{Point-to-Plane - MSE PSNR D2~\cite{Dtian}}\label{D2} 
The D2 metric considers the projection of the error vector $\overrightarrow{v}^{a_i}_{b_k}$ along the surface normal of the nearest neighbor $a_i$ ($N_{a_i}$). The MSE considering the projected errors $E(a_i,b_k)=\left|{\overrightarrow{v}^{a_i}_{b_k} \cdot N_{a_i}}\right|$ is taken, followed by PSNR. The available MPEG implementation was used\footnotemark[1].

\subsubsection{Point Cloud Structural Similarity (PointSSIM)~\cite{AlexiouPSSIM}}
This metric measures the statistical dispersion of attributes (either present or estimated) such as geometry information, color, normal vectors, or curvature.
For each point $p$, a similarity index $S_{Y}(p)$ between the reference ($X$) and the distorted point cloud ($Y$), considering the $K$ nearest neighbors, is given by:
\begin{equation}
    S_Y(p)=\frac{|F_X(q) - F_Y(p|}{\max\{|F_X(q),F_Y(p)|\}} 
\end{equation}

Where $F_X$ and $F_Y$ are the feature values of the reference and distorted point clouds, respectively. The final quality score ${S_Y}$ is obtained by pooling all points.
\begin{equation}
    S_Y = \frac{1}{N_p}\sum_{p=1}^{N_p}S_y(p)^K.
\end{equation}

The implementation available online\footnote{https://github.com/mmspg/pointssim} was used, with the covariance $(COV)$ as an estimator and $K=12$~\cite{AlexiouPSSIM}. Also, color attributes were used, as they led to the best results in~\cite{AlexiouPSSIM}.

\subsubsection{Point Cloud Quality Metric~\cite{Meynet2020PCQM}}\label{sec:PointSSIM}
The Point Cloud Quality Metric (PCQM) considers both geometry features based on the local mean curvature and color features computed on the LAB2000HL perceptual color space. Before feature computation, all points of the reference point cloud $p^X \in X$ are projected onto the 3D quadric surface subtended by the distorted point cloud $Y$, computed on a neighborhood of the closest point in $Y$. Each individual feature $f_i$ takes into account the analogous values of $p$ and its corresponding projected point $\hat{p}$. The final quality index is given by a weighted linear combination, $\mbox{PCQM} = \sum_{i \in S} {w_i f_i}$, where $S$ is the set of indices of features and $w_i$ are their associated weights. These are obtained by optimizing the linear model via logistic regression.
The implementation made available by the authors\footnote{https://github.com/MEPP-team/PCQM} was used with the recommended weights.

\subsubsection{Point-to-Distribution~~\cite{javaheri2021pointtodistribution}}
This metric provides a joint geometry and color quality score ($P2D$) by computing the \textit{Mahalanobis} distance between a point in point cloud $X$ and the distribution of points in its $K$ nearest neighborhood in point cloud $Y$. For geometry ($P2D-G$) and YUV color components ($P2D-C_m$), the distances are computed in two directions, i.e., reference to distorted and vice versa, with both results averaged across all points. The maximum between the two computations is then considered. The joint distance ($P2D-JGC_m$) is also obtained by averaging $P2D-G$ and $P2D-C_m$, and a final $P2D$ quality score is given by $P2D = \log_{10} \left( 1 + \frac{1}{P2D-JGC_m} \right)$.
In this work, the implementation was made available by the authors\footnote{https://github.com/AlirezaJav/Point\_to\_distribution\_metric} was used, which considers only the luminance (Y) channel.

\subsubsection{Reduced-Reference Point Cloud Metric ($\mbox{PCM}_{\mbox{\small RR}})$~\cite{Viola2020PCMRR}}
PCM\textsubscript{RR} is based on features derived from geometry, luminance, and normal attributes. Notably, the mean, standard deviation, median, mode, entropy, energy, and sparsity are computed from the attributes and the occurrence histogram of both geometry and luminance. For normal attributes, the angular similarity $\theta$ between the normal of a given point and its \textit{K}-nearest neighborhood is considered, along with the probability histogram of $\theta$ averaged across $k \in K$. The normal-based feature vector includes the mean of means, mean of standard deviations, mean of medians, standard deviation of means, entropy, energy, and sparsity. The final quality score is given by $\mbox{PCM}_{\mbox{RR}}. = \sum_i w_id_i$, where $d_i$ is the distance between the reference and distorted features with weights $w_i$, which are obtained by training on a point cloud set, maximizing the Pearson Linear Correlation Coefficient.
The metric made available online\footnote{https://github.com/cwi-dis/PCM\_RR} was used.

\subsubsection{GraphSIM~\cite{QiYangGraphSIM2022}}

This metric first establishes a 3D keypoint skeleton $\Vec{P}_S$, by re-sampling the reference point cloud through high-pass graph filtering. For each keypoint, a graph is constructed for both the reference and the distorted point clouds, connecting it to neighboring points below a certain threshold of the Euclidean distance. The gradient mass ($m_g$), the gradient mean ($\mu_g$) and the gradient variance ($\sigma^2_g$) and the co-variance ($c_g$) are computed, with Euclidean distance-based weights being used to compute $m_g$ and $\sigma^2_g$. Similarity measures are obtained for the color attributes and pooled channel-wise.
This is done by aggregating it across all color components using the following equation: 

\begin{equation}\label{eq:GraphPooling}
    S_{\overrightarrow{s}_k}=\frac{1}{\gamma} \sum_C\gamma C \cdot |S_{\overrightarrow{s}_k,C}|
\end{equation}
where $\gamma C$ is the pooling factor of the color channel.
The metric is computed by averaging the different keypoint similarity contributions $S_{\overrightarrow{s}_k}$.
The implementation available online\footnote{https://github.com/NJUVISION/GraphSIM} was used.

\section{Quality Assessment of Deep Learning-based Codecs}
In this section, details concerning both quality assessment models are provided.
Section \ref{sec:subjExp} explains the experimental setup and procedure for both evaluations and the conclusions drawn from the analysis of the results.
Section \ref{sec:objEval} discusses the performance of the selected objective metrics in predicting the scores of the subjective evaluation.

\subsection{Subjective Quality Evaluation}\label{sec:subjExp}

\subsubsection{Experimental Setup}

\begin{figure}[t!]
    \centering
    \begin{subfigure}[t]{0.16\textwidth}
        \centering
        \includegraphics[width=\textwidth]{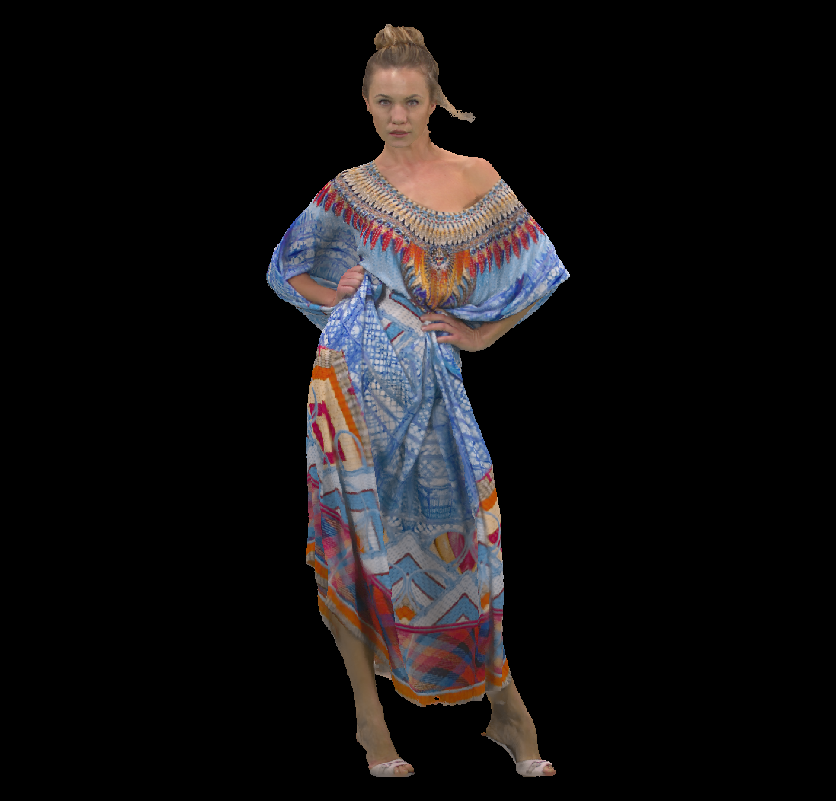}
        \caption{\textit{Longdress}}
        \label{longdresspc}
    \end{subfigure}
    \begin{subfigure}[t]{0.16\textwidth}
        \centering
        \includegraphics[width=\textwidth]{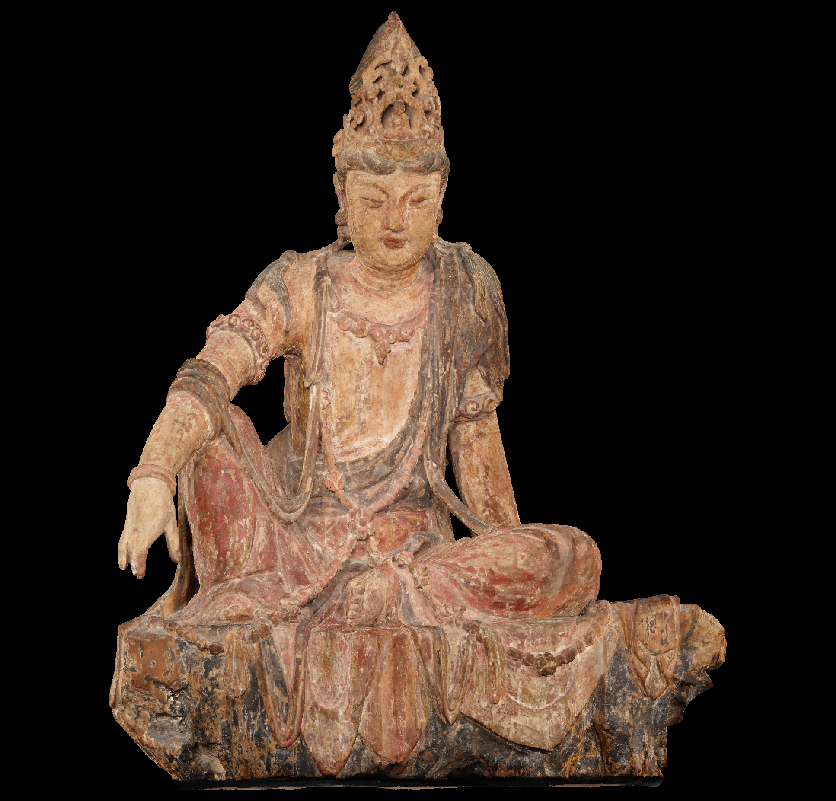}
        \caption{\textit{Guanyin}}
        \label{guanyinpc}
    \end{subfigure}
   \begin{subfigure}[t]{0.16\textwidth}
        \centering
        \includegraphics[width=\textwidth]{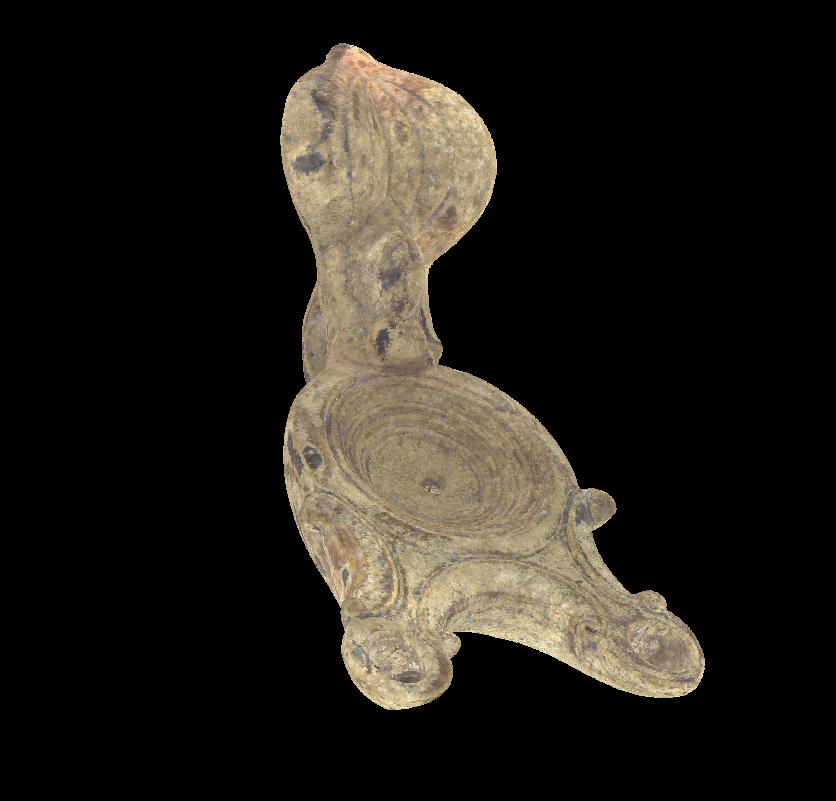}
        \caption{\textit{Romanoillamp}}
        \label{romanoillamppc}
    \end{subfigure}
    \begin{subfigure}[t]{0.16\textwidth}
        \centering
        \includegraphics[width=\textwidth]{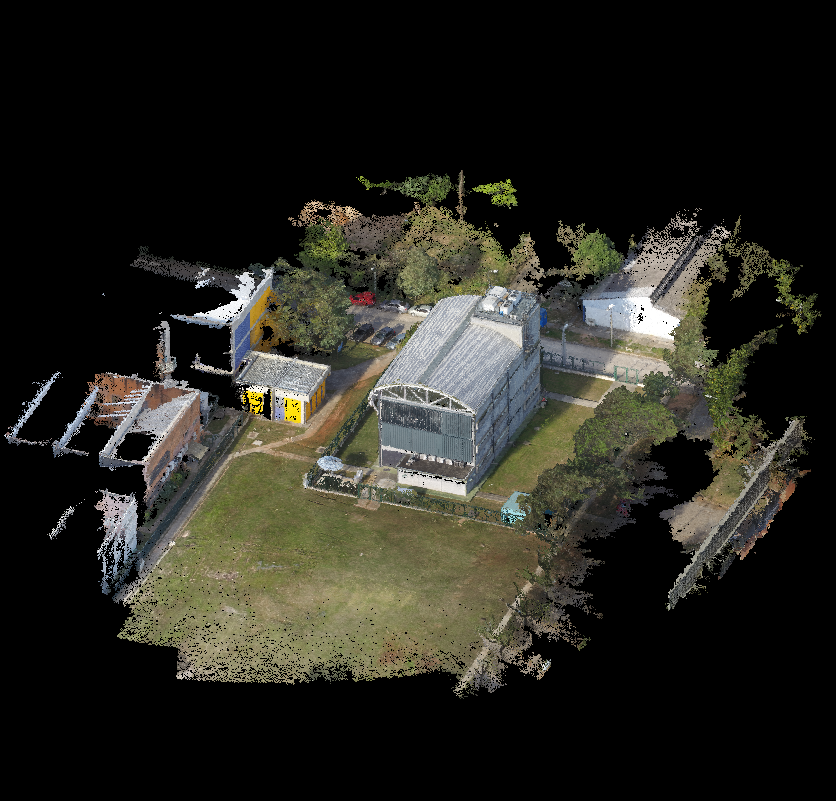}
        \caption{\textit{Citiusp}}
        \label{citiusppc}
    \end{subfigure}
    \begin{subfigure}[t]{0.16\textwidth}
        \centering
        \includegraphics[width=\textwidth]{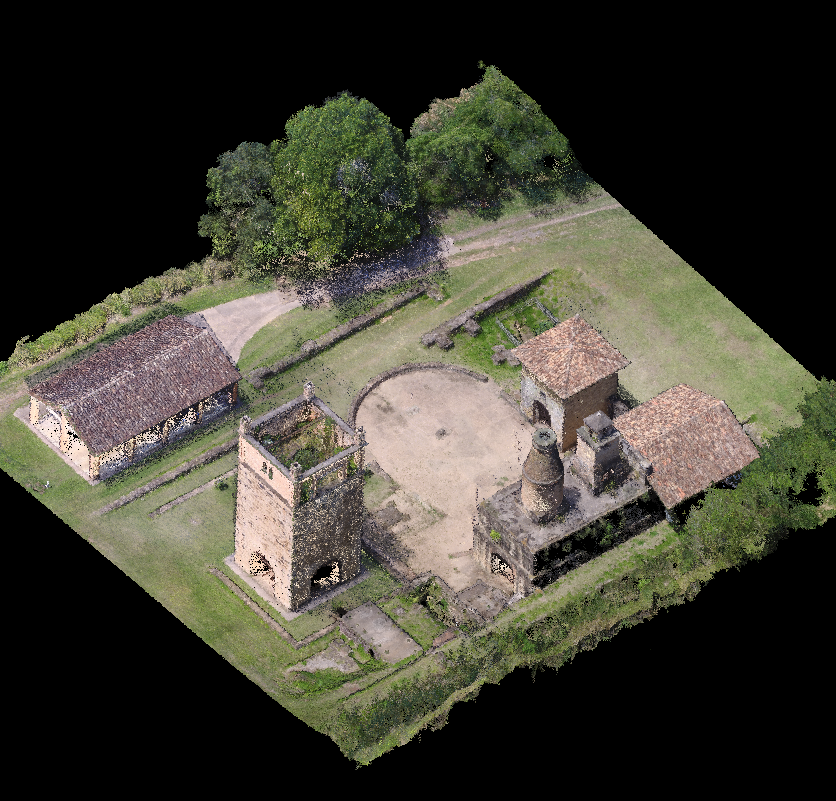}
        \caption{\textit{IpanemaCut}}
        \label{ipanemacutpc}
    \end{subfigure}
    \begin{subfigure}[t]{0.16\textwidth}
        \centering
        \includegraphics[width=\textwidth]{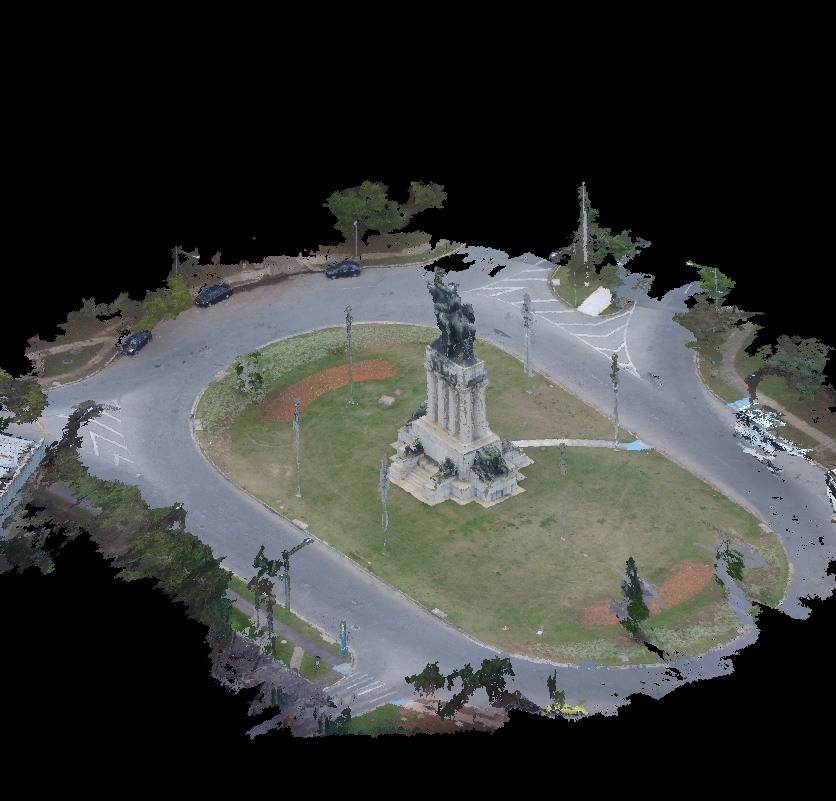}
        \caption{\textit{Ramos}}
        \label{ramospc}
    \end{subfigure}
    \caption{Point Cloud test set.}
    \label{PCS}
\end{figure}

Two subjective quality evaluations were carried out at the test laboratory of the Image and Video Technology Group of the Universidade da Beira Interior.
Both subjective quality evaluations used the same set of six point clouds (Fig. \ref{PCS}), which includes three objects: frame 1300 of the \textit{Longdress} dynamic point cloud available at JPEG Pleno database\footnote{http://plenodb.jpeg.org/pc/8ilabs}, \textit{Guanyin} from the EPFL Dataset, and \textit{Romanoillamp}, and three landscapes: \textit{Citiusp}, \textit{IpanemaCut}, and \textit{Ramos} from the Univ. São Paulo Database\footnote{http://uspaulopc.di.ubi.pt}.
Table \ref{table:charact} shows the selected point cloud sparsity, color gamut volume, and standard deviations of the YCbCr color channels.
The sparsity is defined as the average distance between each point and its 20 nearest neighbors, averaged over the entire point cloud.
The color gamut volume is defined as the volume of the convex hull of the distribution of color points in the YCbCr color space.
The characteristics of Table \ref{table:charact} reflect a suitable degree of diversity for evaluating the coding solutions.
\begin{table}[t!]
    \caption{Point cloud characteristics}

    \centering
   \small
    \begin{tabular}{|c|c|c|c|c|c|}
    \hline
    \textbf{Point Cloud} & \multicolumn{1}{|c|}{\parbox{0.1\textwidth}{\begin{center} \textbf{Sparsity}\\ \textbf{(K=20)} \end{center}}} & \multicolumn{1}{|c|}{\parbox{0.1\textwidth}{\begin{center} \textbf{Colour Gamut}\\ \textbf{Volume} \end{center}}} & \textbf{Y Deviation} & \textbf{Cb Deviation}& \textbf{Cr Deviation} \\ \hline
    \textit{Longdress} & 1.730 & 1.5\% & 0.114 & 0.060 & 0.065\\ \hline
    \textit{Guanyin}  & 1.748 & 1.3\% & 0.144 & 0.025 & 0.027 \\ \hline
   \textit{Romanoillamp}  & 2.204 & 1.5\% & 0.094 & 0.020 & 0.012\\ \hline
    \textit{Citiusp}  & 1.788 & 4.03\% & 0.150 & 0.041 & 0.022 \\ \hline
    \textit{IpanemaCut}  & 1.685 & 2.45\% & 0.161 & 0.032 & 0.022\\ \hline
    \textit{Ramos} &   1.588 & 2.35\% & 0.125 & 0.029 & 0.014\\ \hline
    \end{tabular}
    \label{table:charact}

\end{table}
For the first subjective quality evaluation (\textit{Evaluation 1}), the point cloud data was encoded using the codecs described in Section \ref{sec:pcc}, targeting five different encoding rates, ranging from poor quality (R01) to high quality (R05). As these codecs only encode geometry, there is a need to add texture to the distorted geometry, as it plays a very important role in quality perception. Hence, the texture information of the reference point cloud was mapped onto the distorted geometry. The resulting point clouds were then encoded with G-PCC using the \texttt{lossless-geometry-lossy-atts} mode. To achieve this, the QP parameter (which controls texture encoding) was set to $QP=\{0.25, 0.5, 0.75, 0.875, 0.9375\}$, for R01 to R05 (using the lifting transform), and the positionQuantizationScale (pQs), which controls geometry encoding, is set to 1. This ensures that no further artifacts are introduced by G-PCC in the distorted geometry. A total of 17 subjects (12 males and 5 females, ages 18–58 (24.7$\pm$8.3)) participated in \textit{Evaluation 1}.

In the second subjective quality evaluation (\textit{Evaluation 2}), data preparation was similar, but instead of encoding texture using G-PCC, the texture of the reference point clouds was mapped directly onto the decoded geometry. A total of 17 subjects (11 males and 6 females, ages 21-32 (24.8$\pm$2.8)) participated in \textit{Evaluation 2}.

For both experiments, the texture was mapped using \textit{Meshlab}\footnote{https://www.meshlab.net}. The software maps the color using the nearest neighbor algorithm. For the point cloud without texture information, the nearest neighbor of the point cloud with texture information is identified. The color of the nearest neighbor is then attributed to that point.

The point size was adjusted heuristically, as shown in Table \ref{table:pointSize}. This is important to create continuous surfaces, thus avoiding perceptual effects caused by transparency~\cite{Alexiou2018a,Silva2019a}. However, it should also be carefully adjusted so that the point size does not mask the distortion artifacts created by the codecs. For content representing landscapes, only G-PCC required an increase in the point size for lower bitrates. All other codecs maintained good surface integrity for the visualization using the standard point size without a relevant influence on the perceived quality. In the case of content representing objects, most codecs require adjustment, especially at lower bitrates. This was not verified for LUT SR.

\begin{sidewaystable}
    \centering
    \caption{Point size of each point cloud for visualization in the subjective test.}\label{table:pointSize}
    \begin{tabular}{|l |ccccc|ccccc|ccccc|ccccc|} 
    \hline
    &\multicolumn{5}{c|}{\textbf{ADLPCC}}&\multicolumn{5}{c|}{\textbf{PCC\_GEO\_CNN}} & \multicolumn{5}{c|}{\textbf{PCGC}} \\
    \hline
    \textbf{Content} & R01 & R02 & R03 & R04 & R05 & R01 & R02 & R03 & R04 & R05& R01 & R02 & R03 & R04 & R05\\ \hline 
    \hline
    \textit{Longdress} & 7 & 4 & \multicolumn{3}{|c|}{3} & 8 & 6 & 5 & \multicolumn{2}{|c|}{3} & 12 & 10 & 9 & \multicolumn{2}{|c|}{3} \\ \hline
    \textit{Guanyin} & 7 & 4 & \multicolumn{3}{|c|}{3} & 8 & 6 & \multicolumn{2}{|c|}{5} & 3 & 12 & 10 & 9 & \multicolumn{2}{|c|}{3} \\ \hline 
    \textit{Romanoillamp} & 15 & \multicolumn{2}{|c|}{6} & \multicolumn{2}{c|}{5} & 10 & \multicolumn{2}{|c|}{8} & 7 & 6 & 15 &  14 & 13 & 8 & 7 \\ \hline
    \textit{Citiusp} & \multicolumn{5}{c|}{3} & \multicolumn{5}{|c|}{3}& \multicolumn{5}{|c|}{3}  \\ \hline
    \textit{IpanemaCut} & \multicolumn{5}{c|}{3} & \multicolumn{5}{|c|}{3} & \multicolumn{5}{|c|}{3} \\ \hline
    \textit{Ramos} & \multicolumn{5}{c|}{3} & \multicolumn{5}{|c|}{3} & \multicolumn{5}{|c|}{3} \\ \hline
    & \multicolumn{5}{c|}{\textbf{G-PCC}} & \multicolumn{5}{c|}{\textbf{LUT SR}}\\ \cmidrule{1-11}
     & R01 & R02 & R03 & R04 & R05 & R01 & R02 & R03 & R04 & R05 \\ \cmidrule{1-11}
    \textit{Longdress} & 8 & 4 &\multicolumn{3}{|c|}{3}  & \multicolumn{5}{c|}{3}\\ \cmidrule{1-11}
    \textit{Guanyin} &  12 & 10 & 9 & \multicolumn{2}{|c|}{3}  &\multicolumn{5}{c|}{3} \\ \cmidrule{1-11}
    \textit{Romanoillamp} &  10 & 6 & \multicolumn{3}{|c|}{5}  &\multicolumn{5}{c|}{5} \\ \cmidrule{1-11}
    \textit{Citiusp} & 12 & 6 & 5 &\multicolumn{2}{|c}{3}  & \multicolumn{5}{|c|}{3} \\ \cmidrule{1-11}
    \textit{IpanemaCut} & 12 & 6 & 5 &\multicolumn{2}{|c}{3}  &\multicolumn{5}{|c|}{3} \\ \cmidrule{1-11}
    \textit{Ramos} & 12 & 6 & 4 &\multicolumn{2}{|c}{3} &\multicolumn{5}{|c|}{3} \\ \cmidrule{1-11}
    \end{tabular}
\end{sidewaystable}

\begin{figure}[t!]
\begin{center}    
    \subfloat[\textit{Longdress}\label{longdressMosText}]{%
        \includegraphics[width=0.17\linewidth]{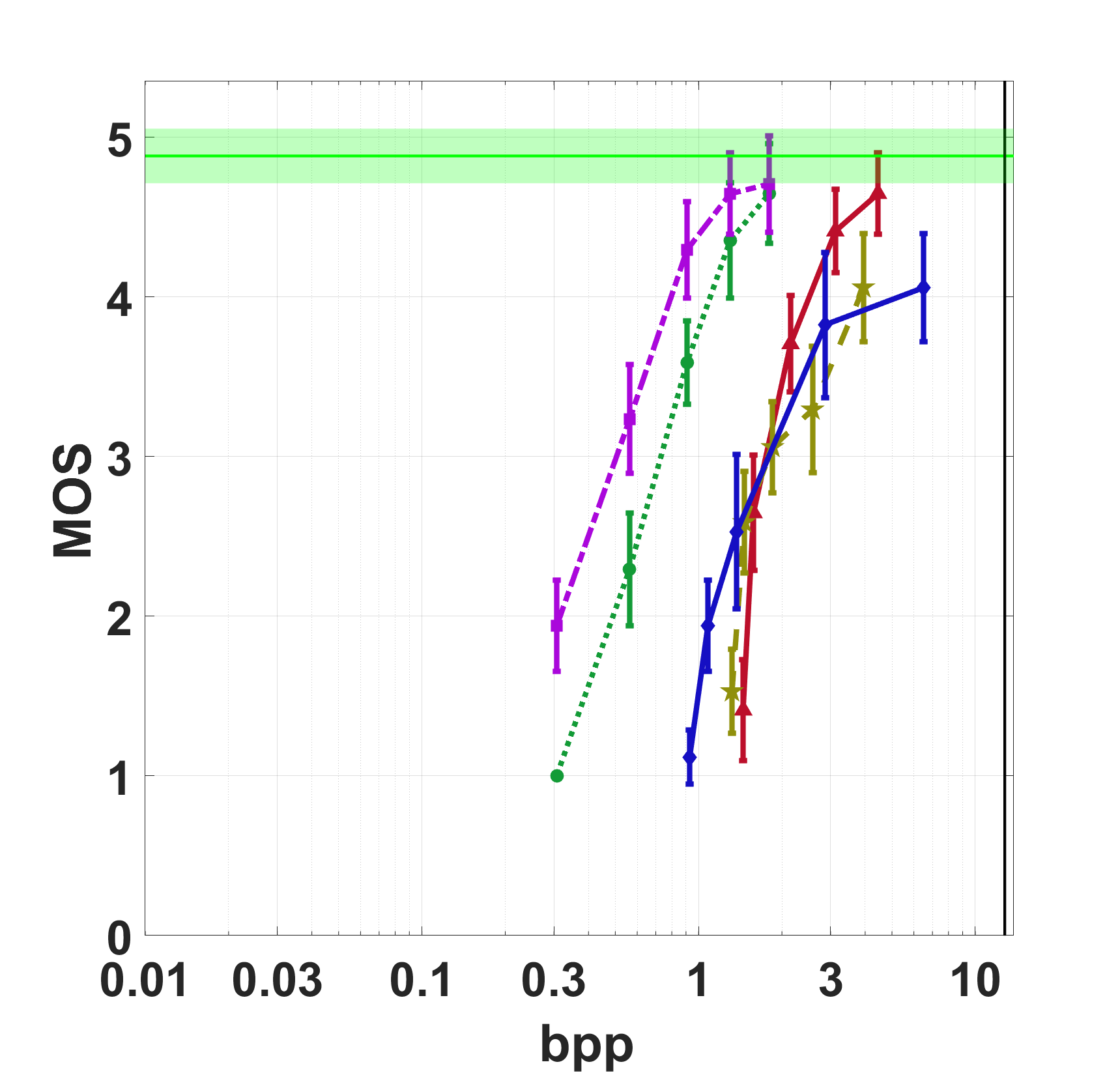}}    
    \subfloat[\textit{Guanyin}\label{guanMosText}]{%
        \includegraphics[width=0.17\linewidth]{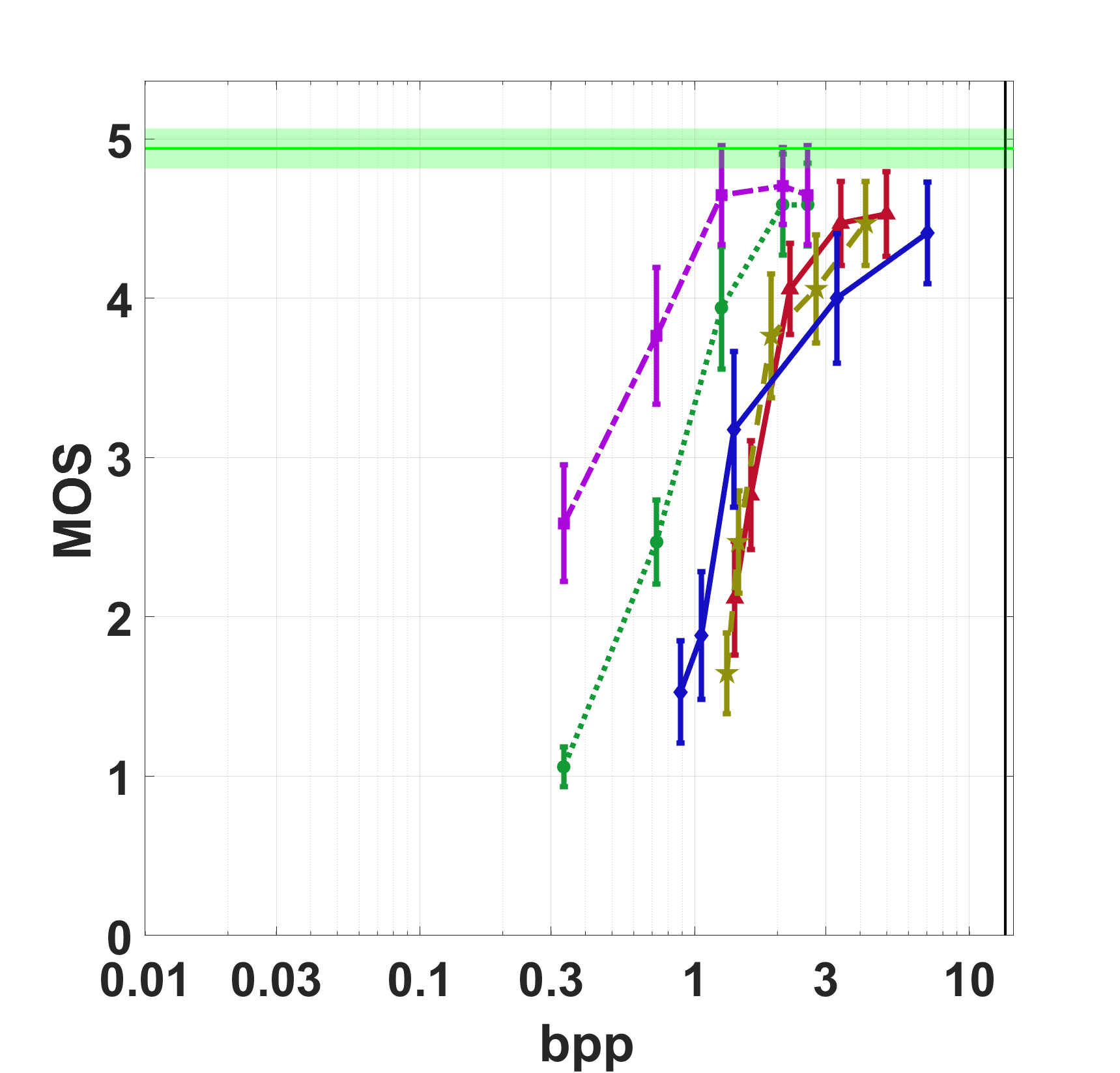}}     
    \subfloat[\textit{Romanoillamp}\label{romanMosText}]{%
        \includegraphics[width=0.17\linewidth]{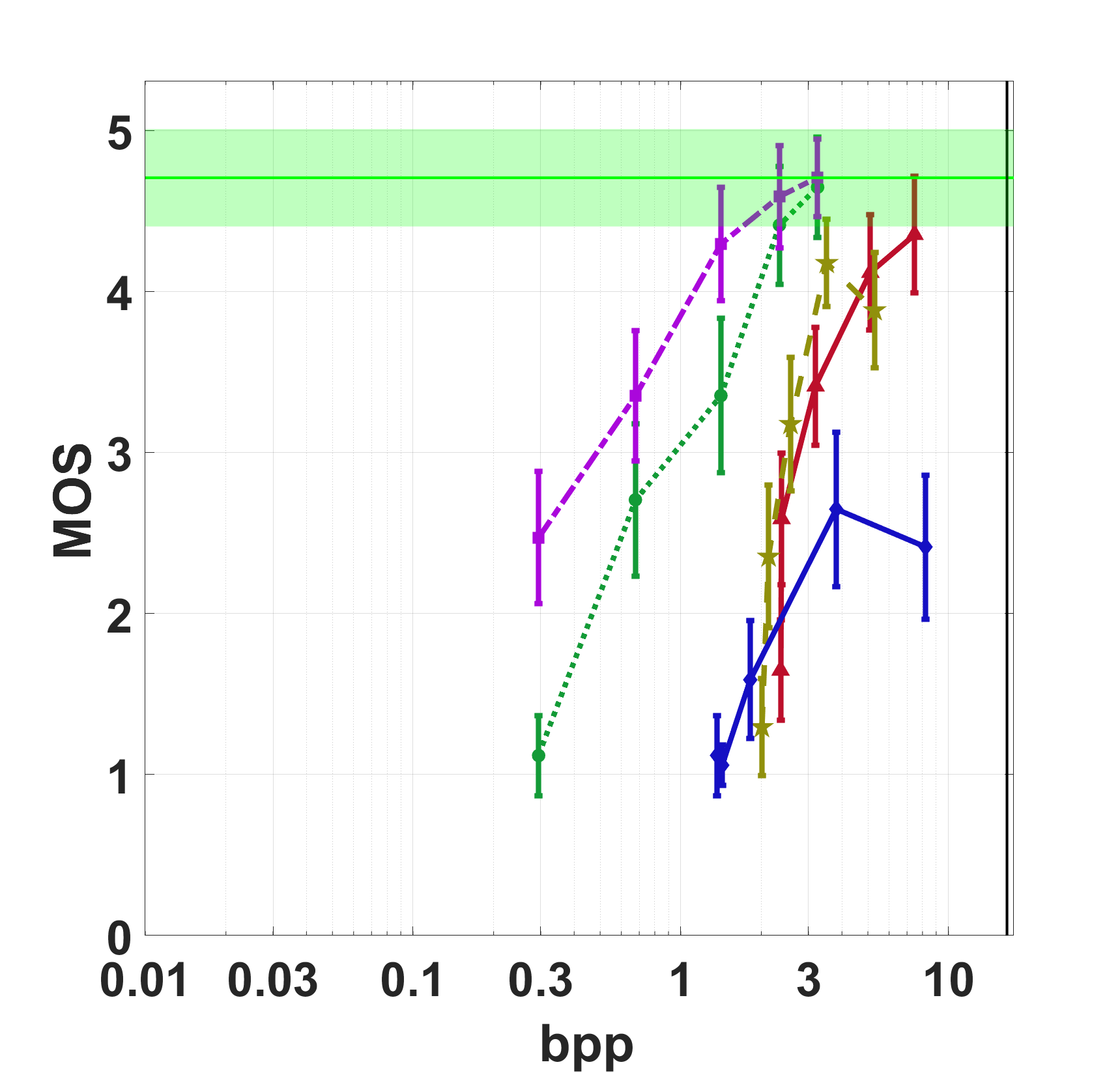}}      
    \subfloat[\textit{Citiusp}\label{citiMosText}]{%
        \includegraphics[width=0.17\linewidth]{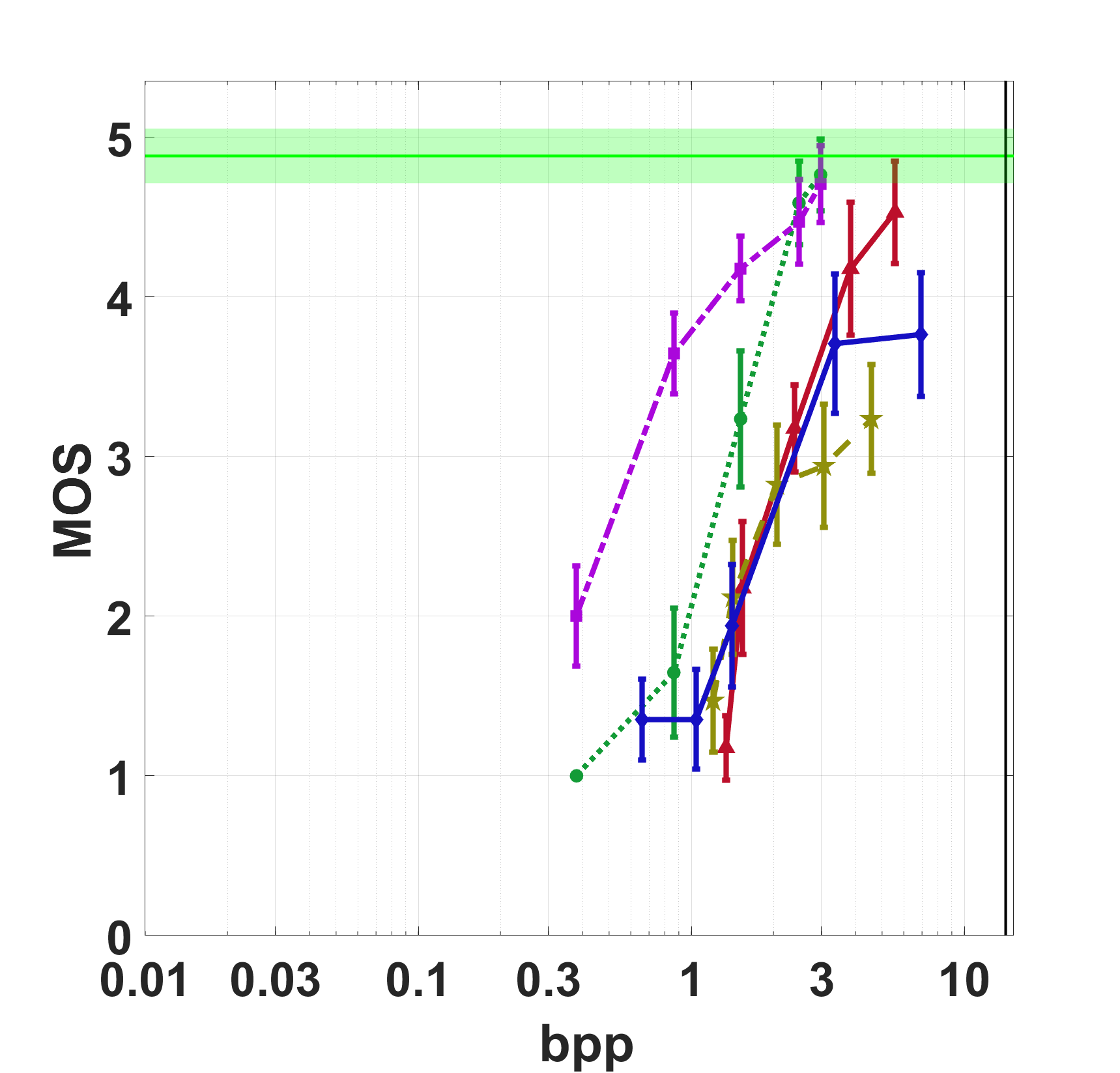}}        
    \subfloat[\textit{IpanemaCut}\label{ipanMosText}]{%
        \includegraphics[width=0.17\linewidth]{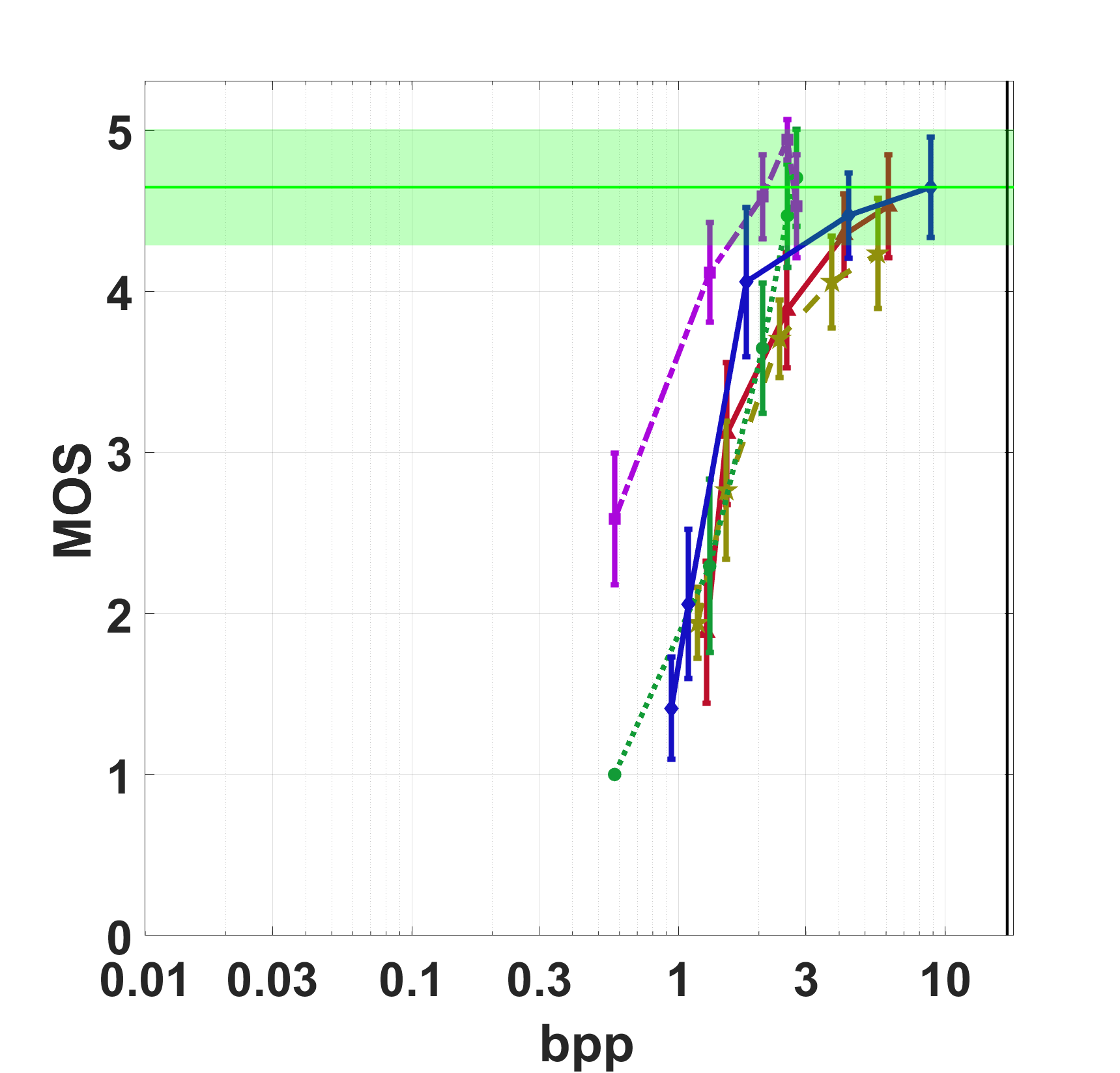}}       
    \subfloat[\textit{Ramos}\label{ramosMosText}]{%
        \includegraphics[width=0.17\linewidth]{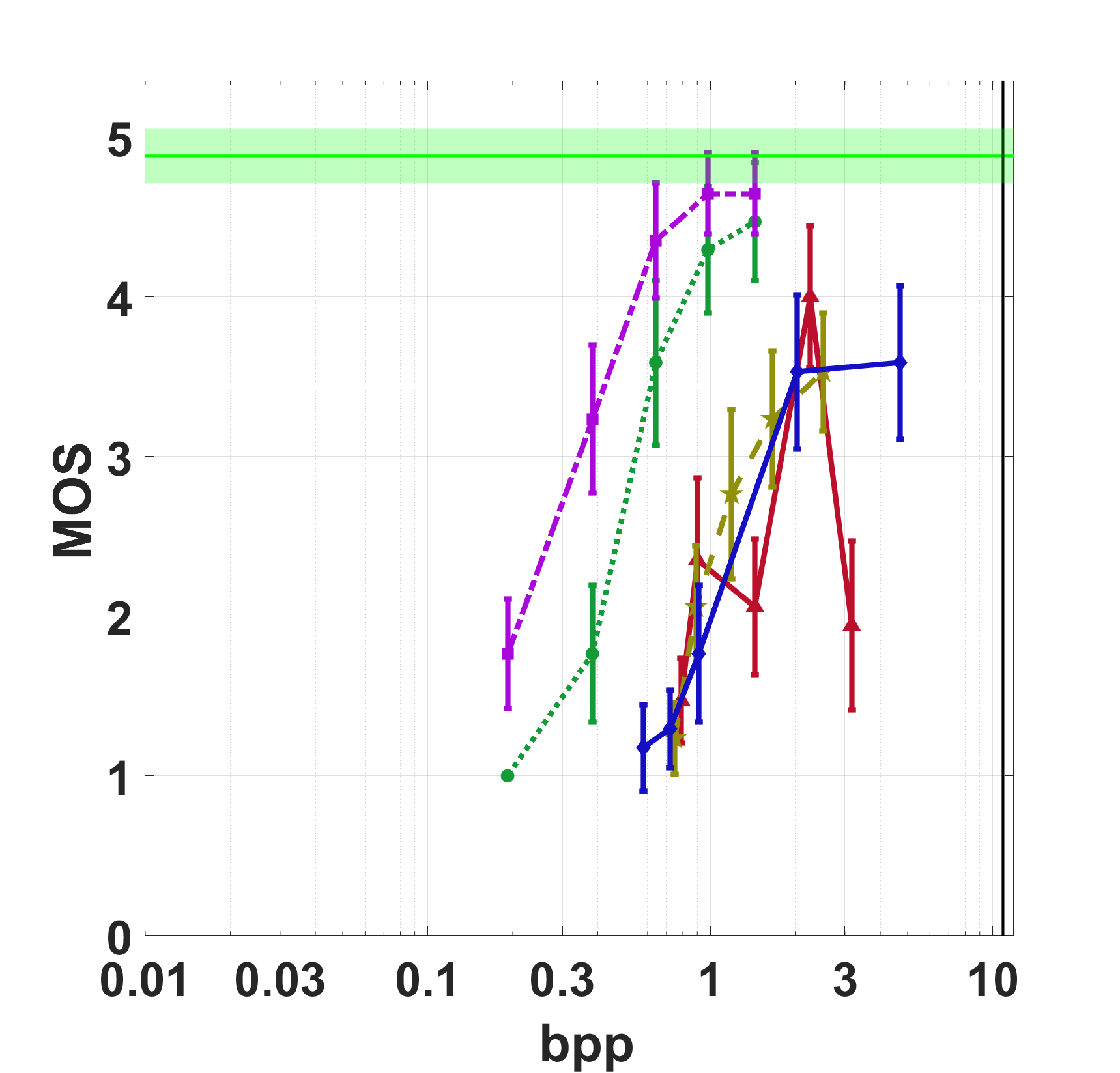}}
    \caption{MOS vs bpp with 95\% CI for \textit{Evaluation 1} (texture encoded with G-PCC). The bitrate results from geometry and texture.}
    \label{MOSBPPtextgeo}
    \end{center}
\end{figure}

\begin{figure}[t!]
   \begin{center}
   \subfloat[\textit{Longdress}]{%
        \includegraphics[width=0.17\linewidth]{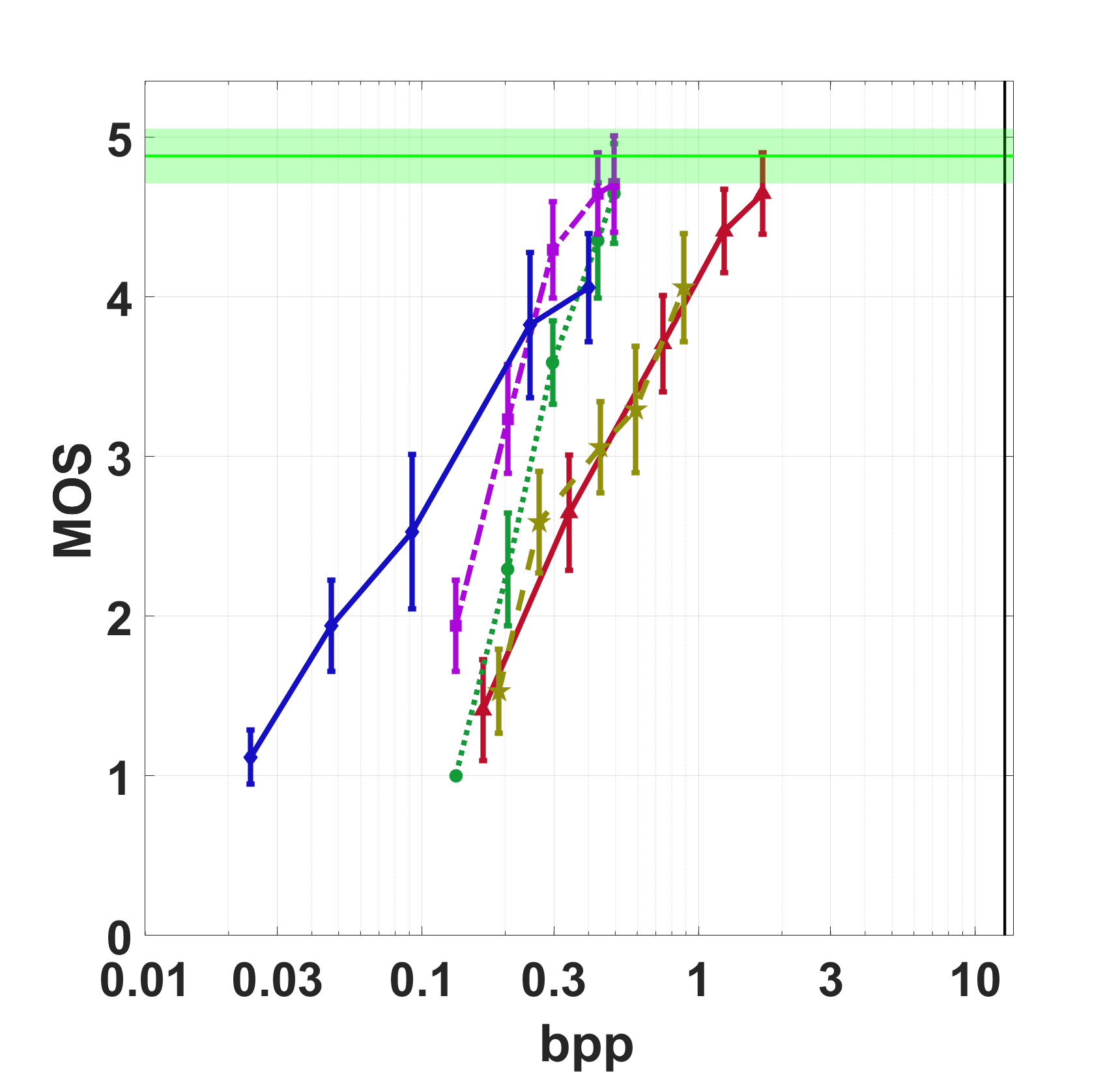}}
    \subfloat[\textit{Guanyin}]{%
        \includegraphics[width=0.17\linewidth]{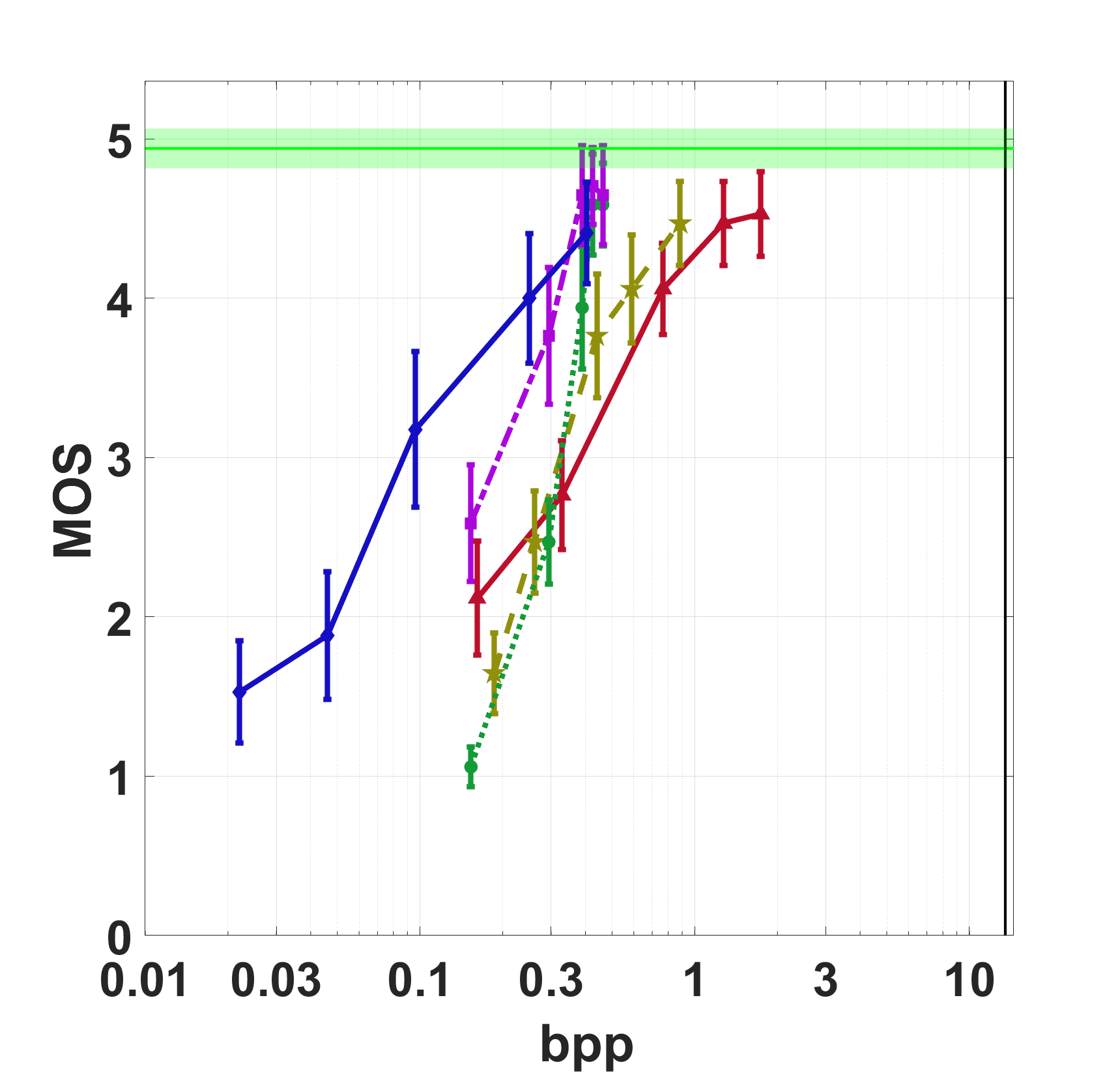}}
    \subfloat[\textit{Romanoillamp}]{%
        \includegraphics[width=0.17\linewidth]{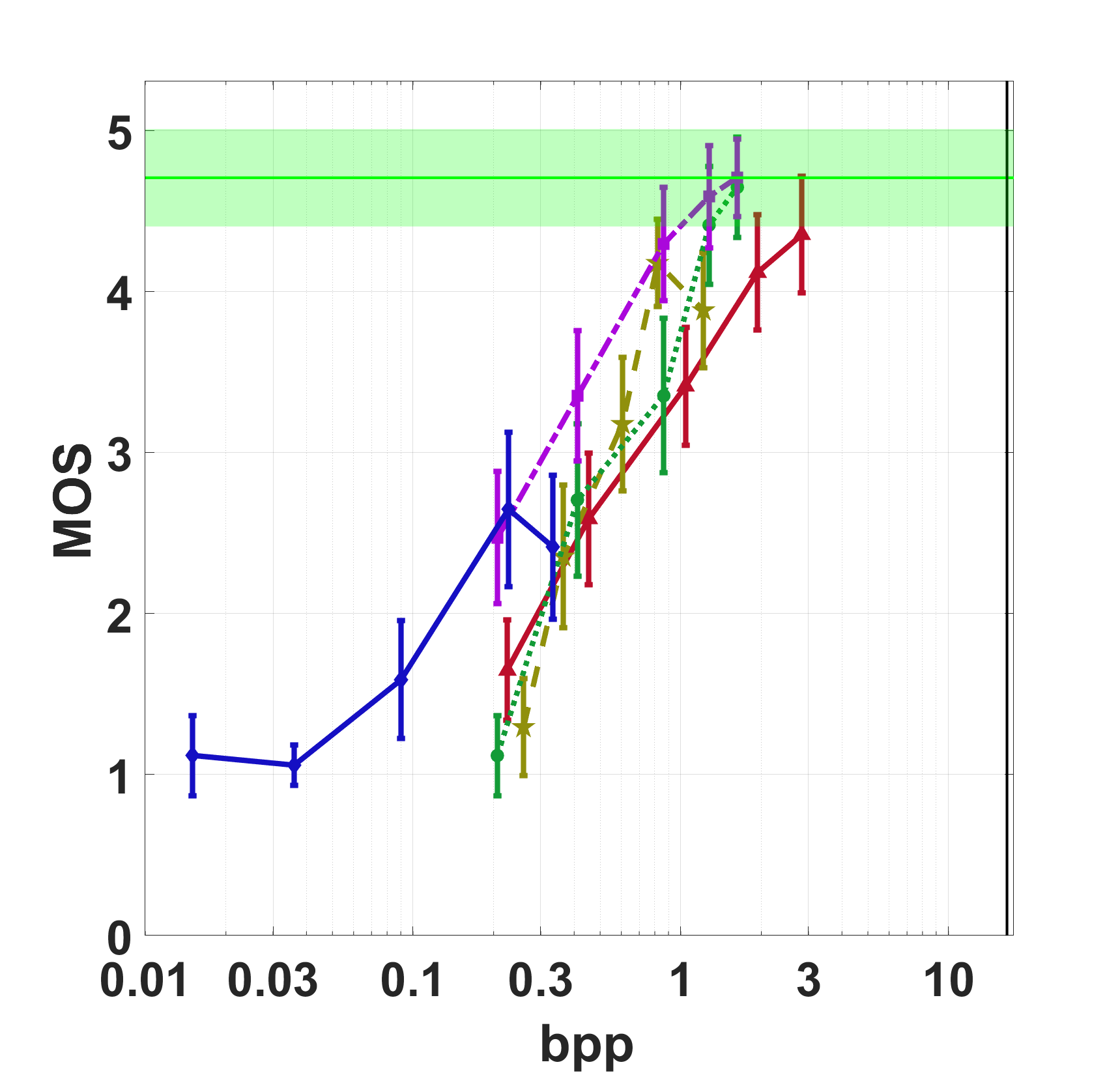}}
    \subfloat[\textit{Citiusp}]{%
        \includegraphics[width=0.17\linewidth]{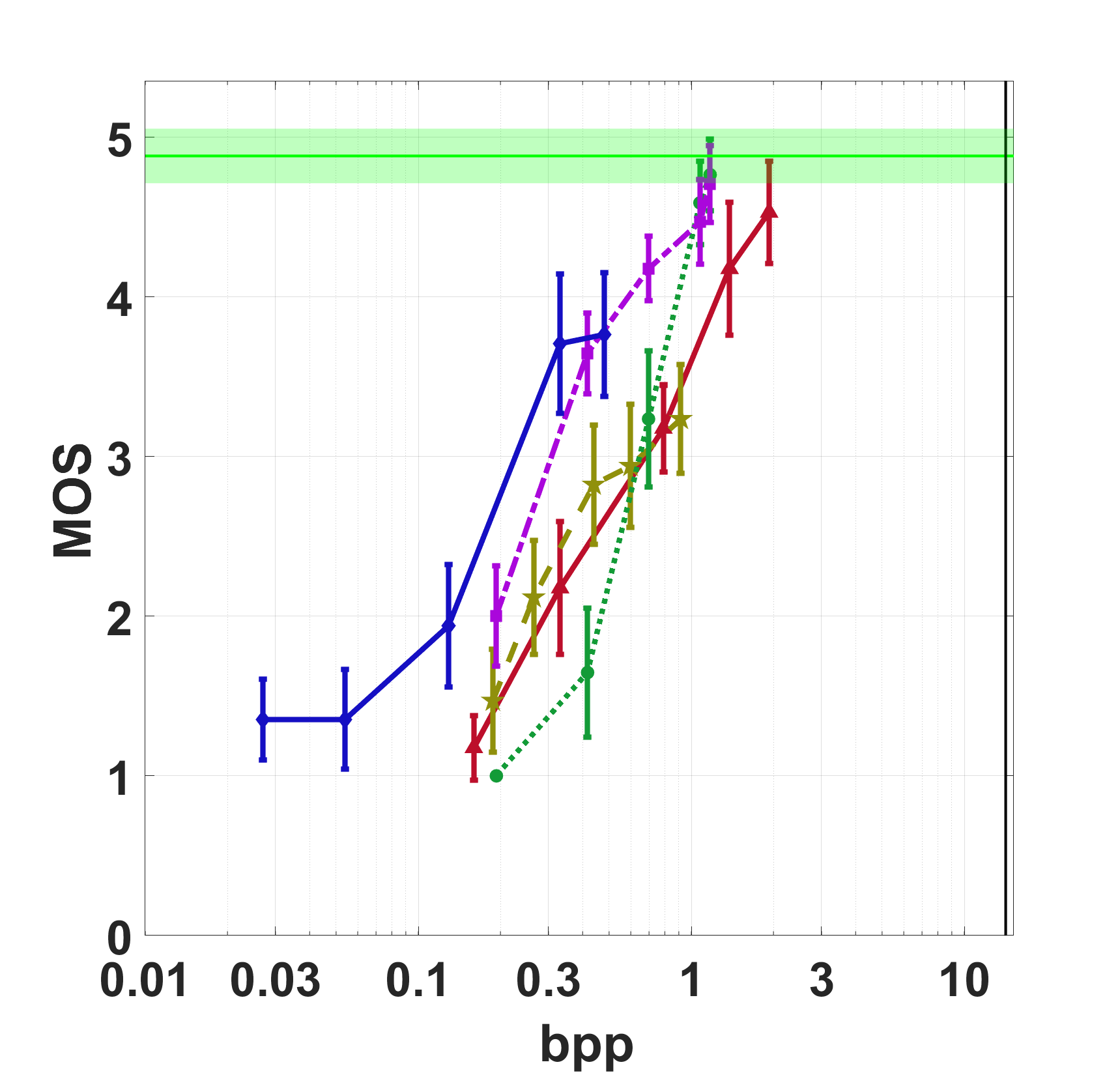}}
    \subfloat[\textit{IpanemaCut}]{%
        \includegraphics[width=0.17\linewidth]{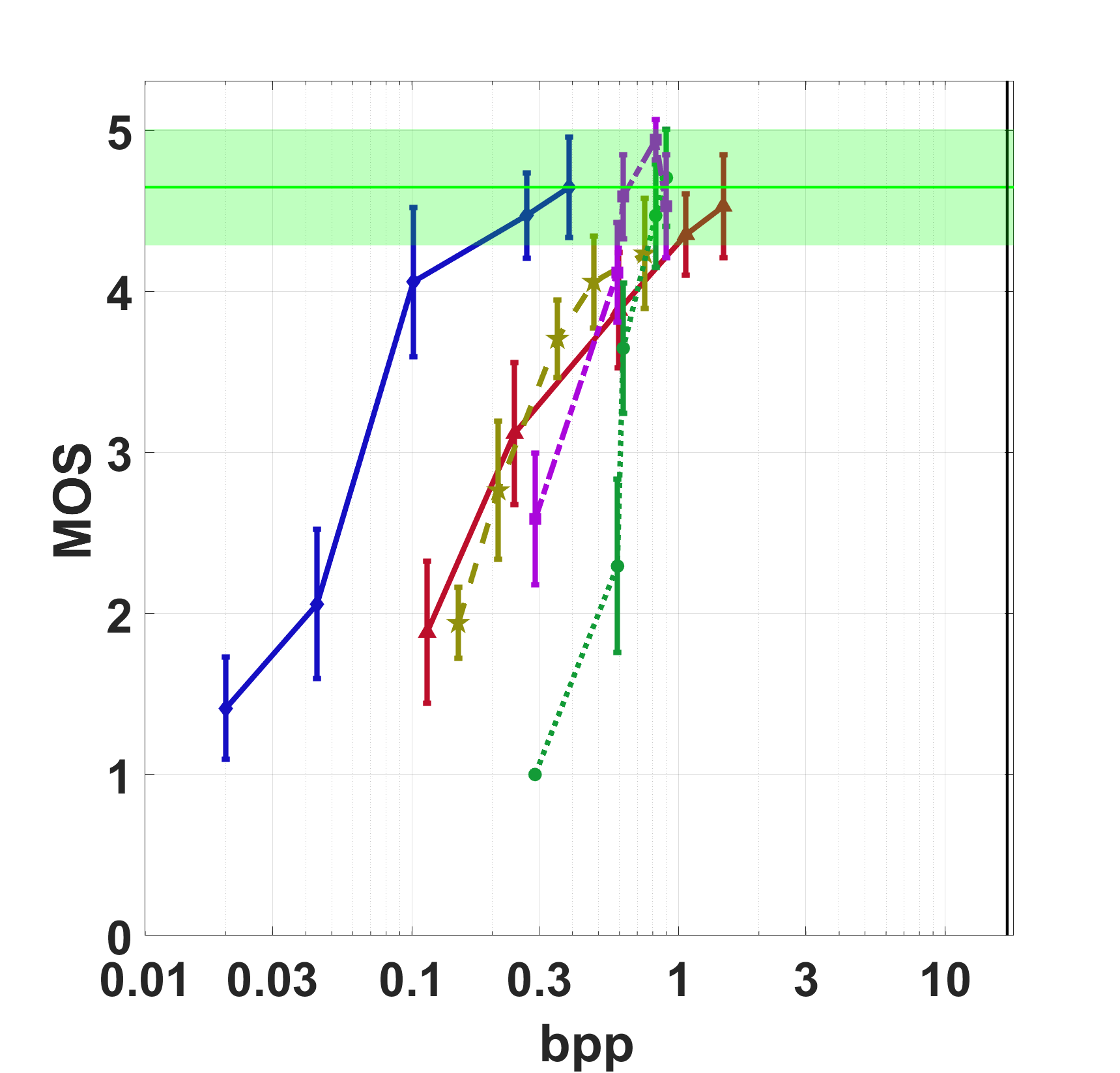}}
    \subfloat[\textit{Ramos}]{%
        \includegraphics[width=0.17\linewidth]{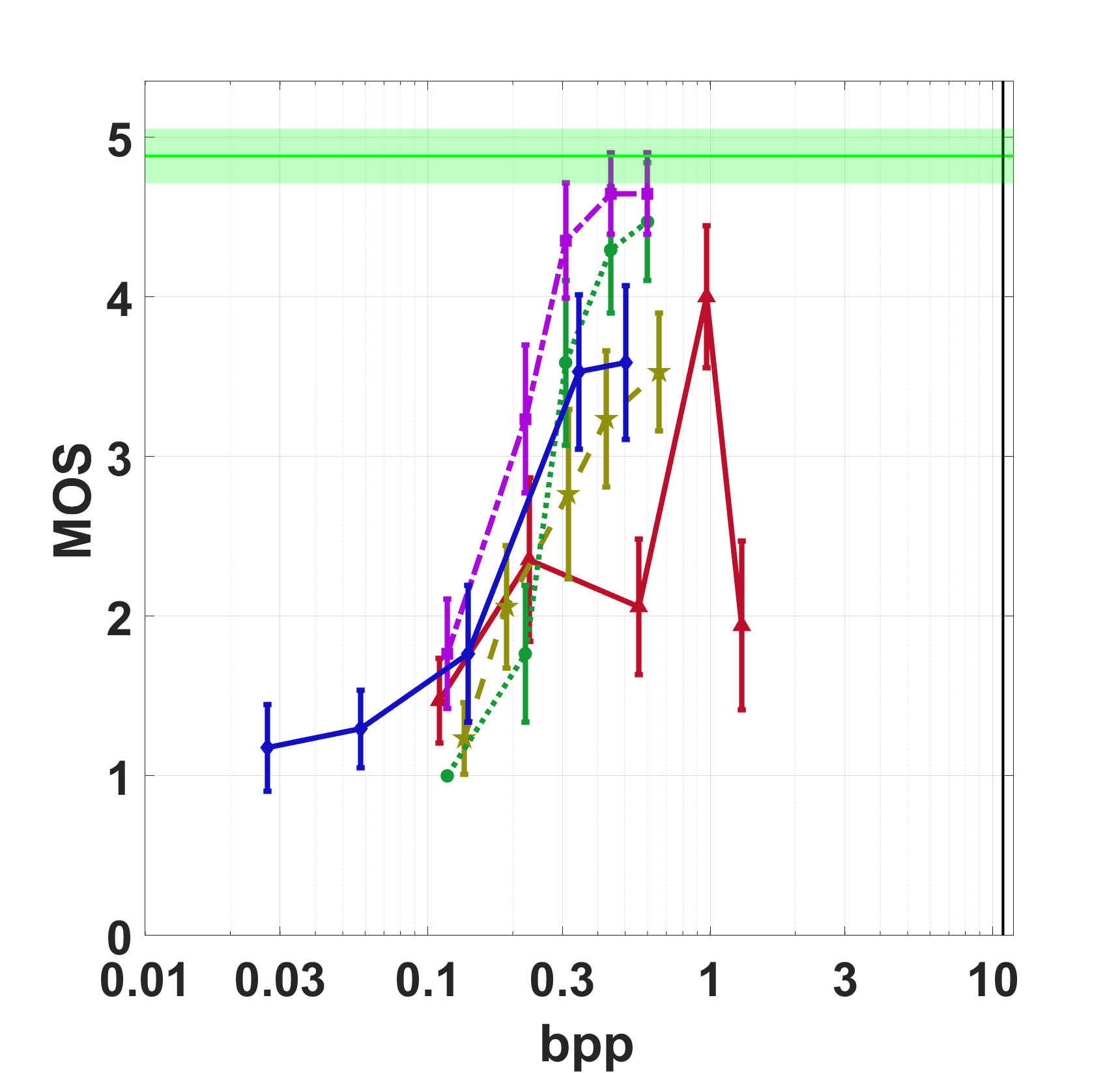}}
    \caption{MOS vs bpp with 95\% CI for \textit{Evaluation 1} (texture encoded with G-PCC). Here, bpp refers to the geometry bitrate only.}
    \label{MOSBPPtextgeoBPP}
    \end{center}
\end{figure}

\begin{figure}[t!]
    \begin{center}
    \subfloat[\textit{Longdress}\label{longdressMosGeo}]{%
        \includegraphics[width=0.17\linewidth]{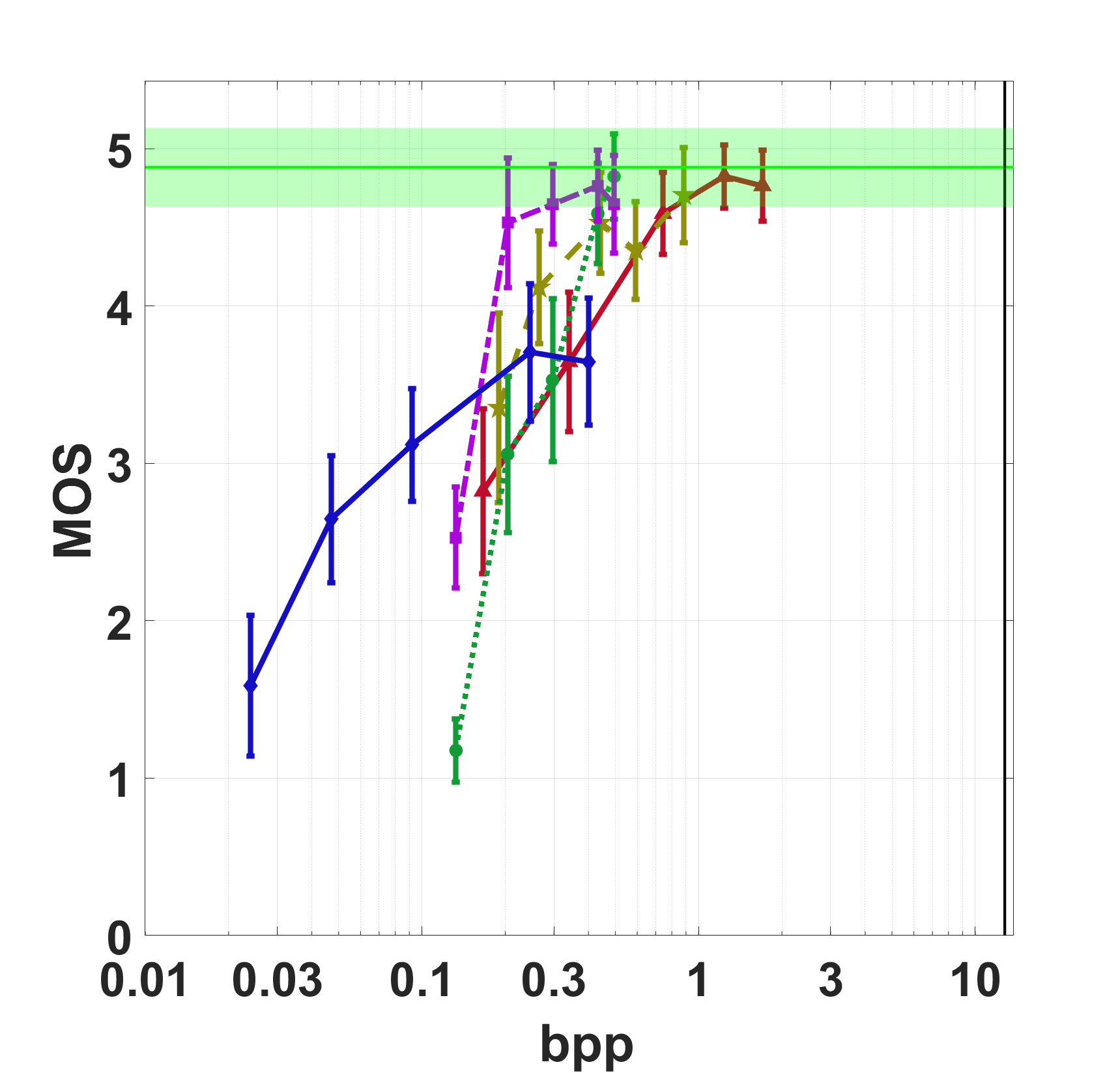}}    \subfloat[\textit{Guanyin}\label{guanMosGeo}]{%
        \includegraphics[width=0.17\linewidth]{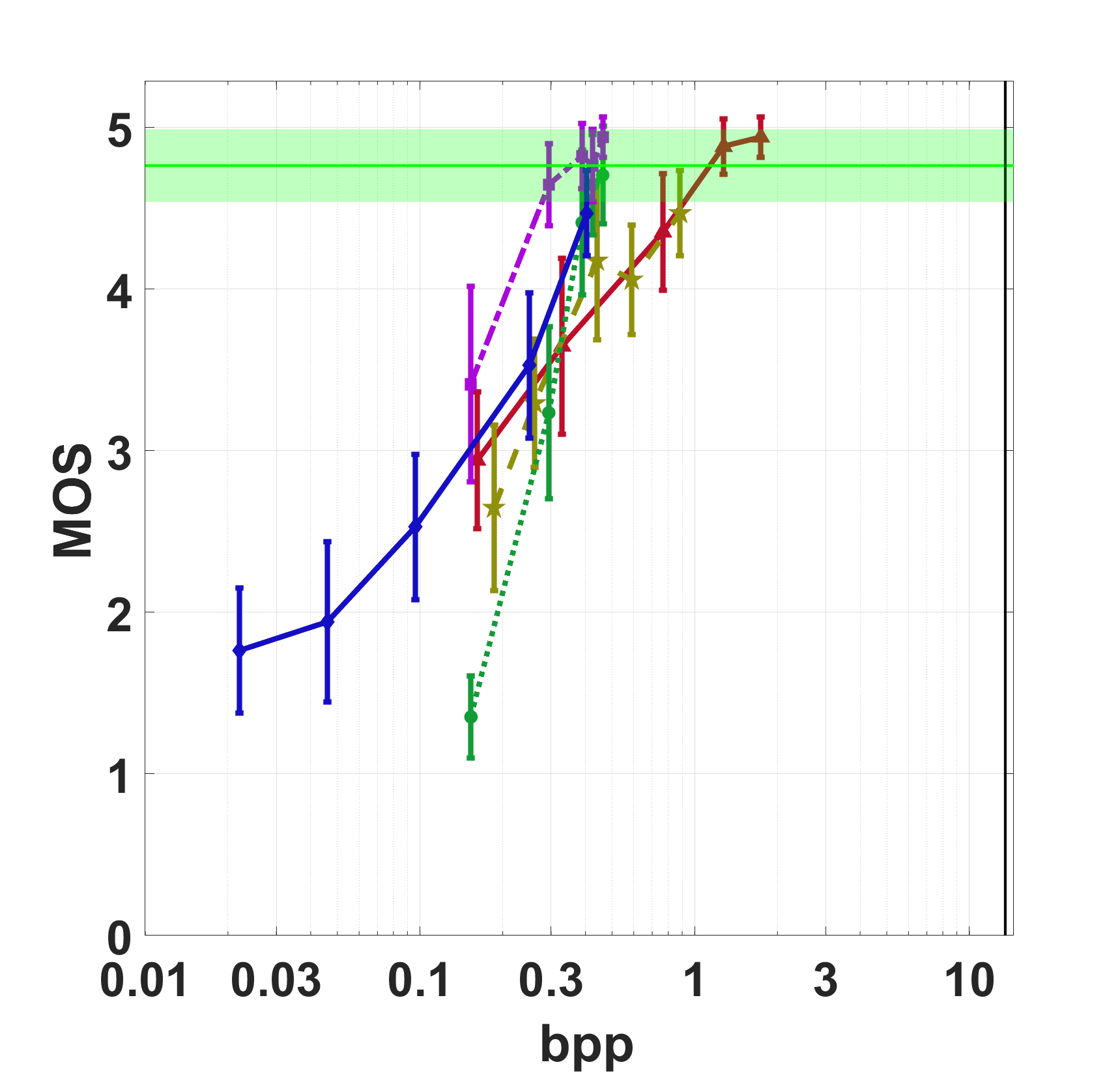}}    \subfloat[\textit{Romanoillamp}\label{romanMosGeo}]{%
        \includegraphics[width=0.17\linewidth]{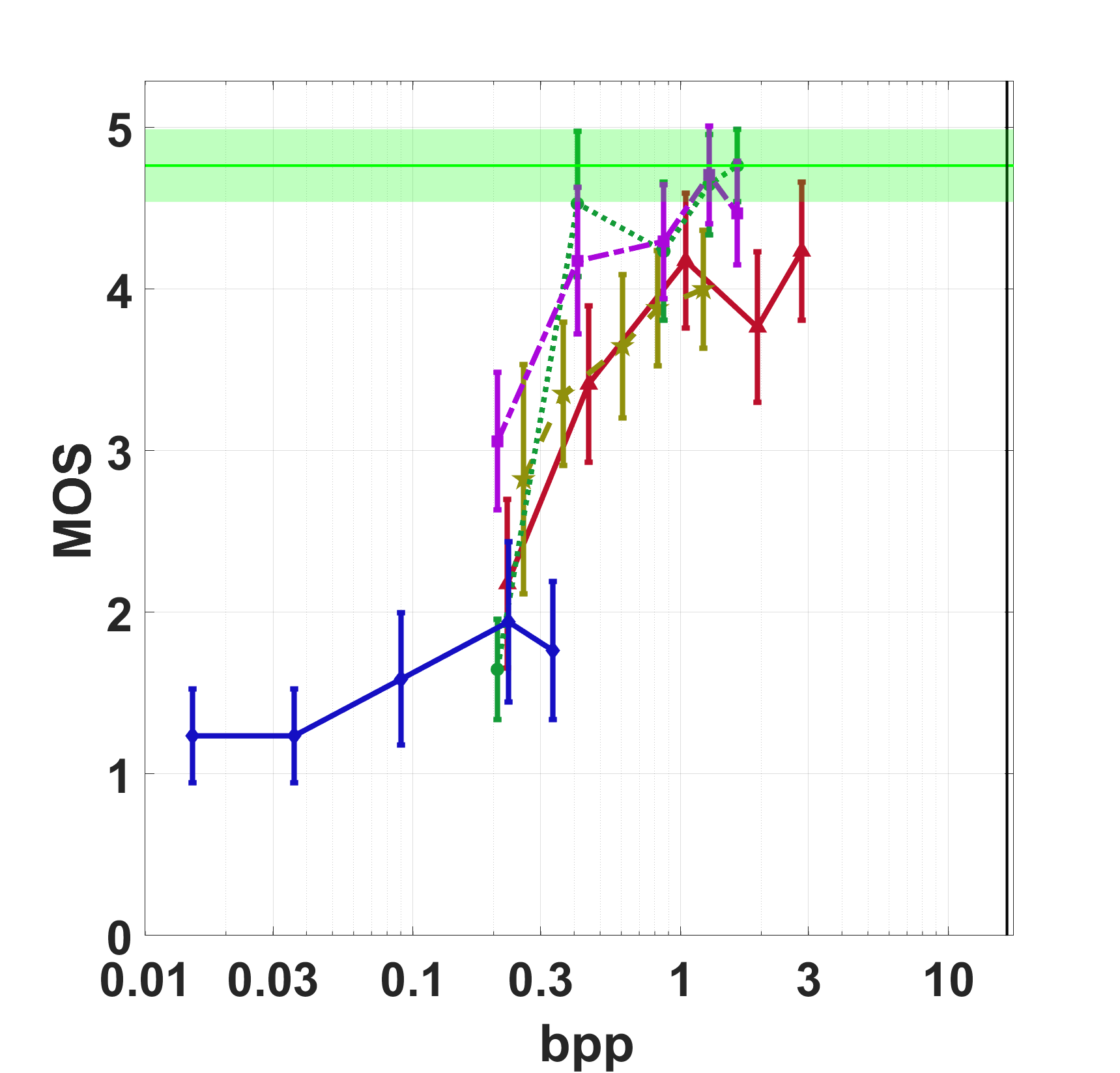}}    \subfloat[\textit{Citiusp}\label{citiMosGeo}]{%
        \includegraphics[width=0.17\linewidth]{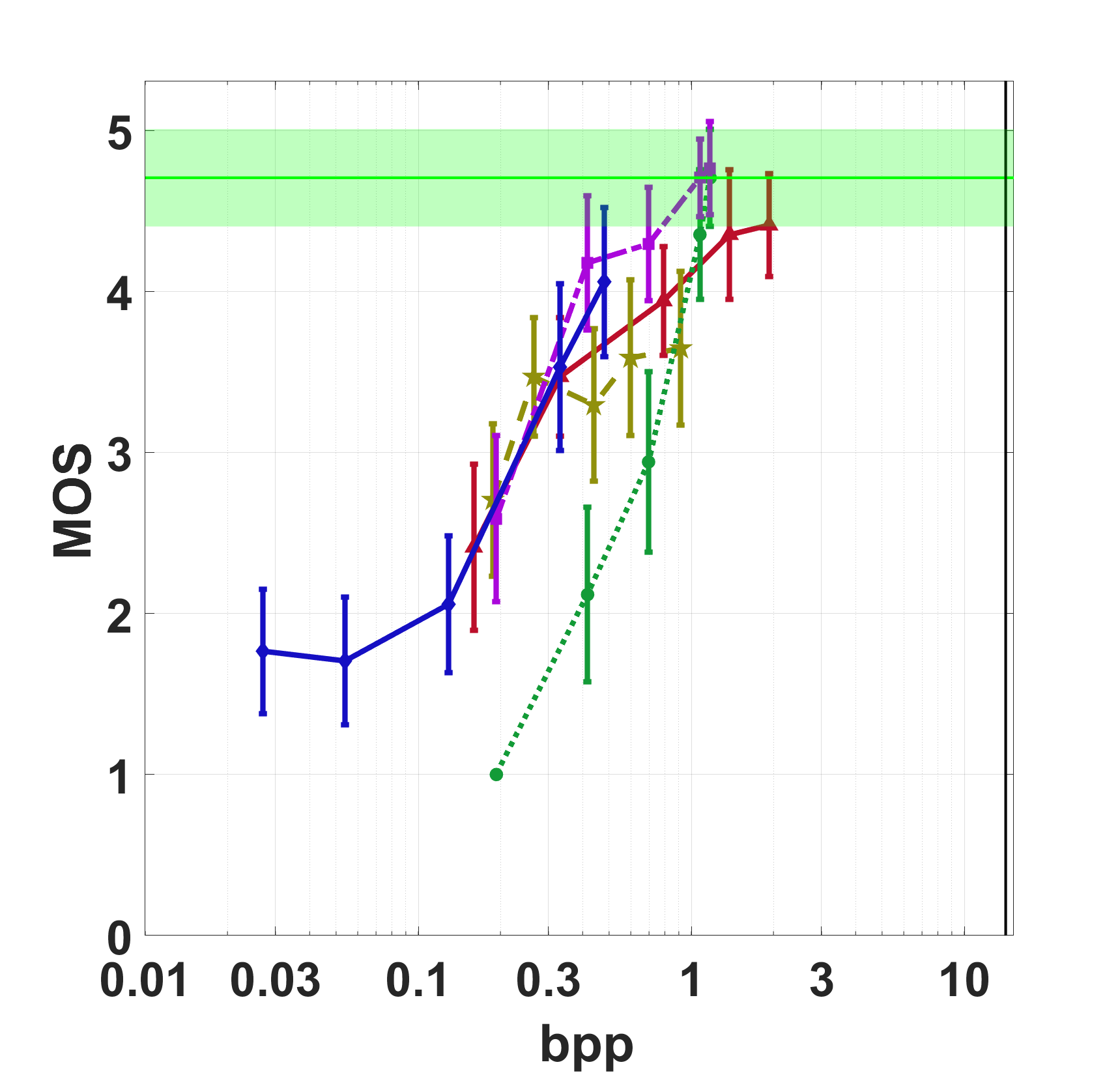}}    \subfloat[\textit{IpanemaCut}\label{ipanMosGeo}]{%
        \includegraphics[width=0.17\linewidth]{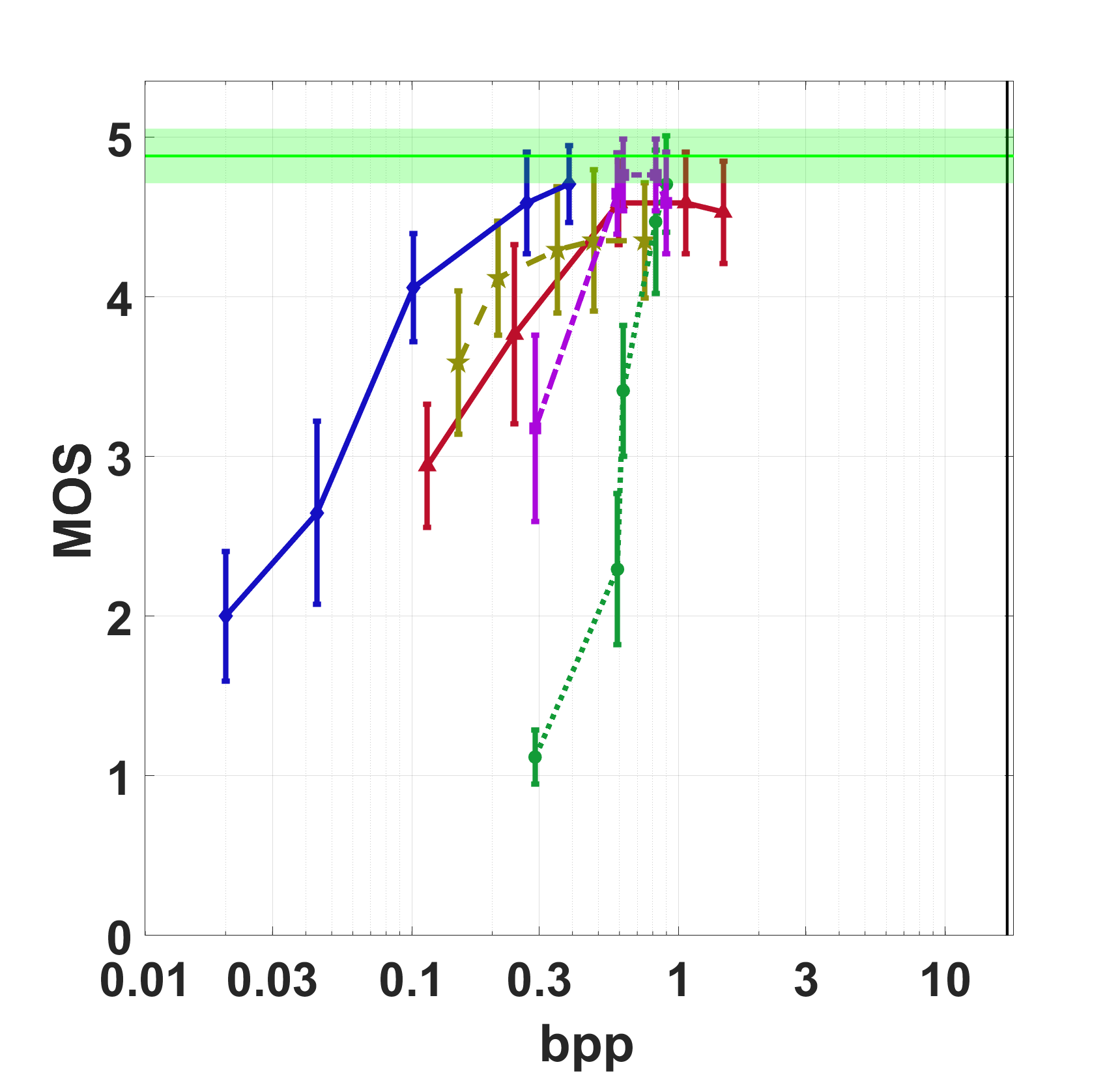}}    \subfloat[\textit{Ramos}\label{ramosMosGeo}]{%
        \includegraphics[width=0.16\linewidth]{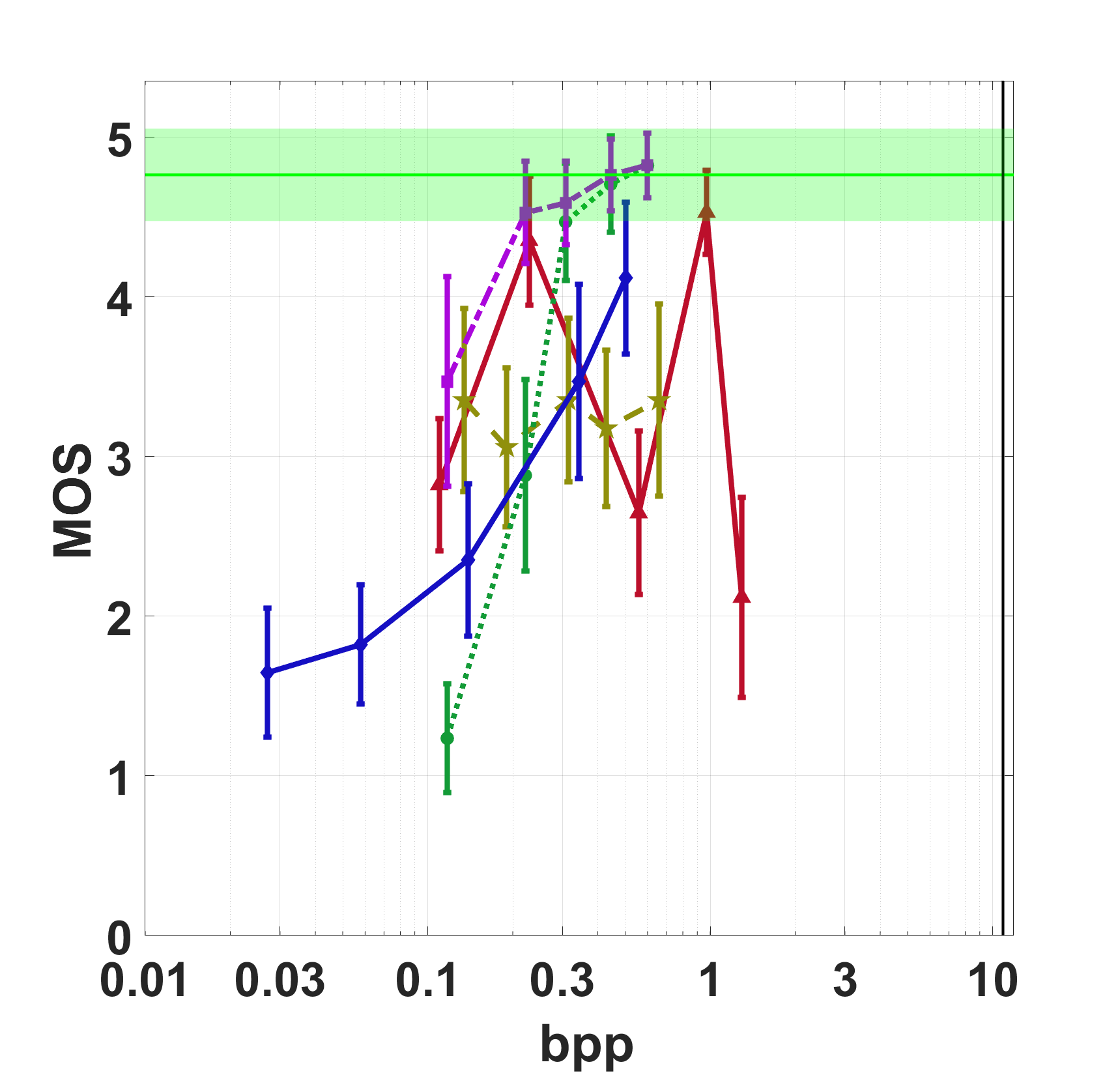}}\\
    \includegraphics[width=0.7\textwidth]{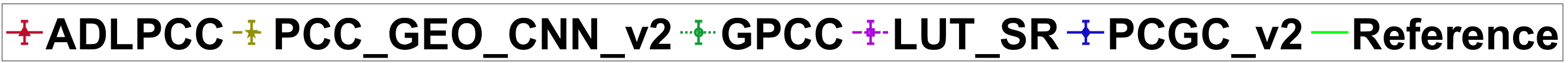}
   \caption{MOS vs bpp with 95\% CI for \textit{Evaluation 2} (texture directly mapped onto the distorted geometry). Here, bpp refers to the geometry bitrate only.}
    \label{MOSBPPgeo}

    \end{center}
\end{figure}

For both evaluations, videos depicting the reference and distorted point clouds were prepared.
A point cloud view was captured for each 1$^{\circ}$ degree rotation using PCL Visualizer~\cite{rusu2011a}, completing a full rotation around the vertical axis. For point clouds depicting objects, the frontal view was chosen as the initial view. Furthermore, in the case of landscapes, the viewing point was rotated to an angle of 43$^{\circ}$ degrees with the xy plane, allowing a top view visualization, as their frontal view is not suitable for subjective evaluation.

The full sequences of frames were then rendered with FFMPEG\footnote{https://ffmpeg.org/}, using the H.264 codec~\cite{H264Text} at 30 fps, resulting in 12-second videos.
To ensure that no compression was applied, the CRF (Constant Rate Factor) and $q$ were set to 0. The \texttt{libx264rgb} option was used to prevent any RGB to YUV conversion. The subjective test setup used a 31.1-inch Eizo ColorEdge CG318-4K with a full resolution of 4096x2160 and followed the specifications in~\cite{conditions}.

Before starting any quality evaluation, all participants were shown eight training videos with encoded versions of two point clouds not included in the test set, i.e., \textit{Airplane} from the PointNet Database and \textit{Villalobospark}, from the University of Sao Paulo Database. This allowed an adaptation to typical encoding distortions, to the evaluation scale, and also to the user interface. During the evaluation, each participant was shown a unique, randomized sequence of videos of the distorted point clouds and the respective reference, side by side. Reference/reference pairs were also included for hidden reference evaluation.
Distortions of the same point cloud were never shown one after the other. Moreover, half of the subjects performed the subjective quality evaluation with the reference on the right, whereas the other half had the reference on the left.

A Double Stimulus Impairment Scale method was adopted, with the subjects being prompted to evaluate the quality of the distorted point cloud in comparison to the provided reference according to a five-level rating scale (1: very annoying, 2: annoying, 3: slightly annoying, 4: perceptible but not annoying, 5: imperceptible). After the subjective test, the MOS for all stimuli was computed.

\subsubsection{Results and Discussion}\label{sec:resultsSubjQuality}

\begin{figure}[t!]
    \begin{center}
        \subfloat[R01]{\includegraphics[width=0.15\textwidth]{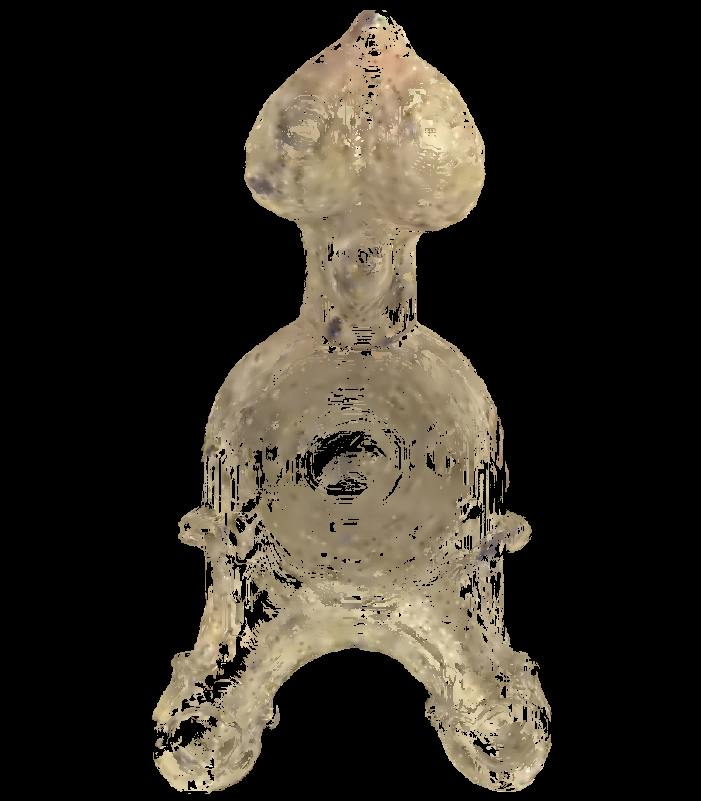}}
        \subfloat[R03]{ \includegraphics[width=0.15\textwidth]{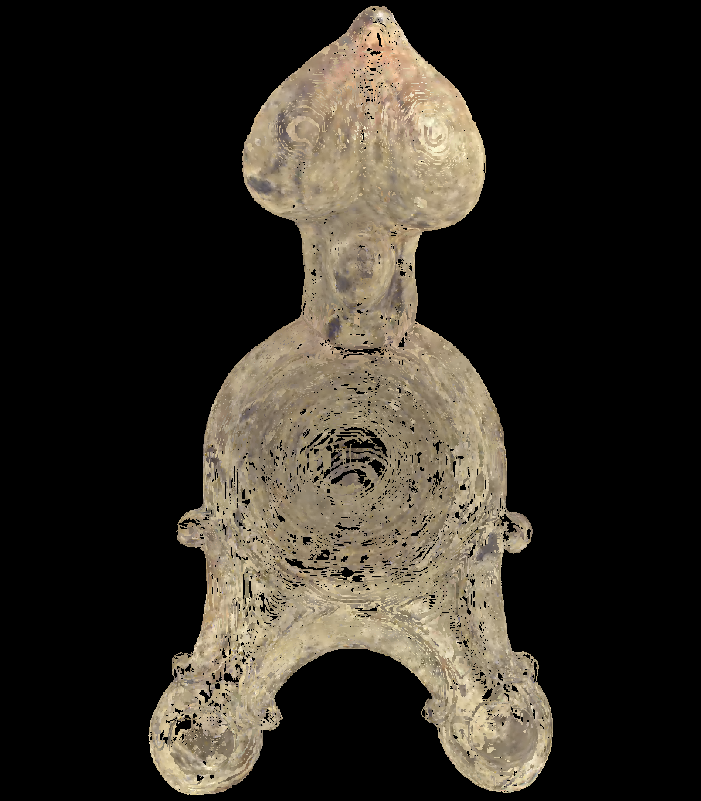}}
        \subfloat[R05]{ \includegraphics[width=0.15\textwidth]{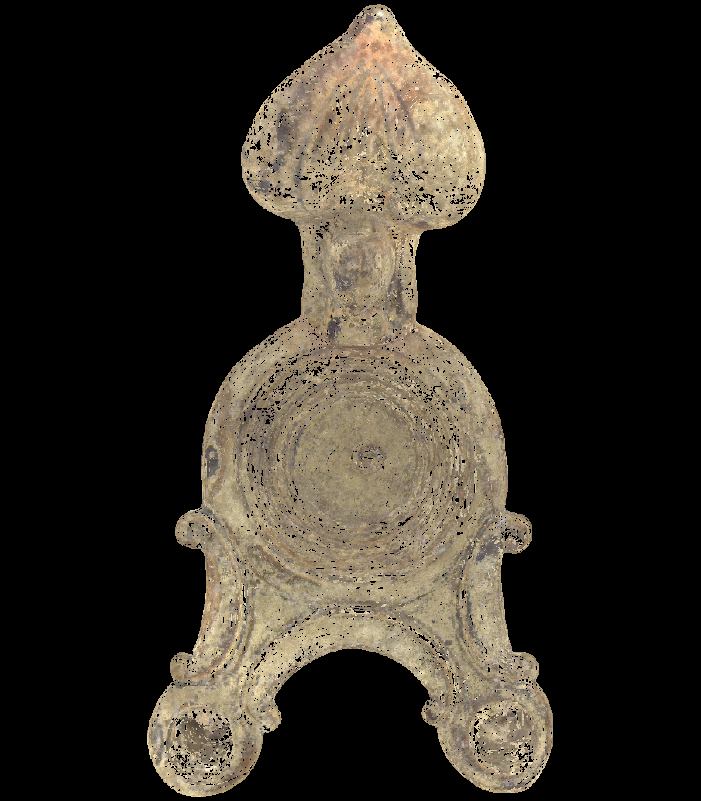}}
        \caption{\textit{Romanoillamp} encoded with PCGcv2}
        \label{fig:romanPCGCv2}
    \end{center}
    \begin{center}
        \subfloat[R01]{\includegraphics[width=0.15\textwidth]{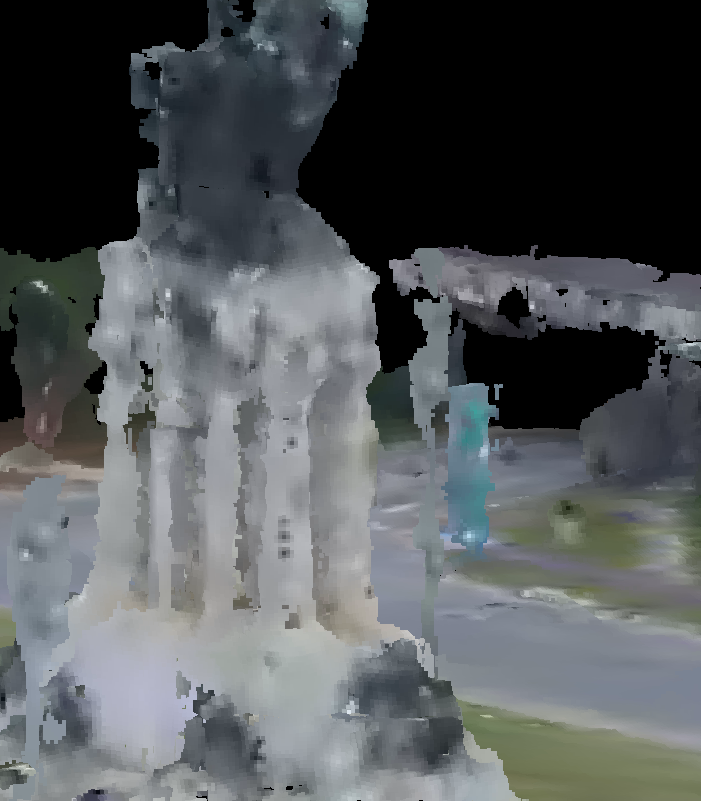}}
        \subfloat[R03]{ \includegraphics[width=0.15\textwidth]{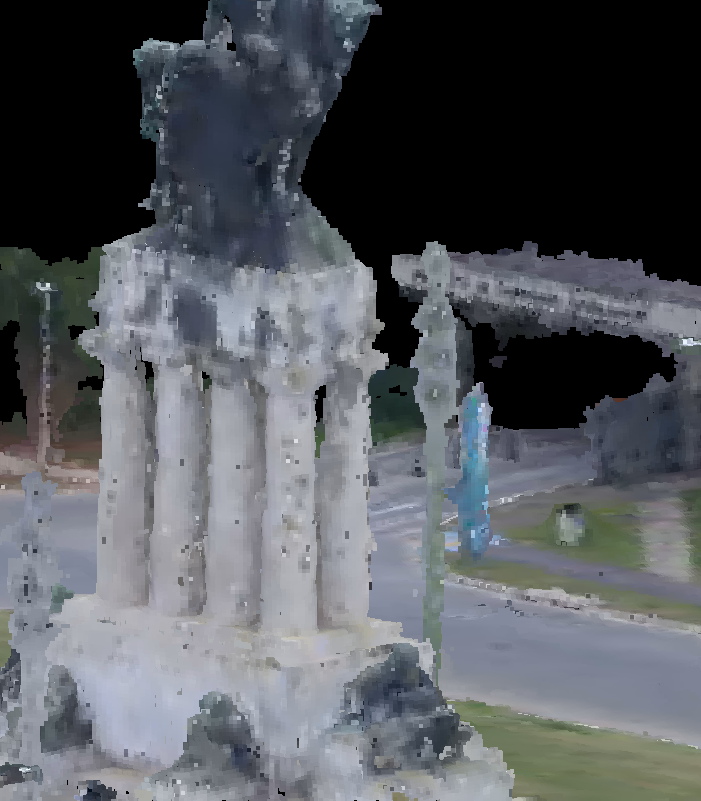}}
        \subfloat[R05]{ \includegraphics[width=0.15\textwidth]{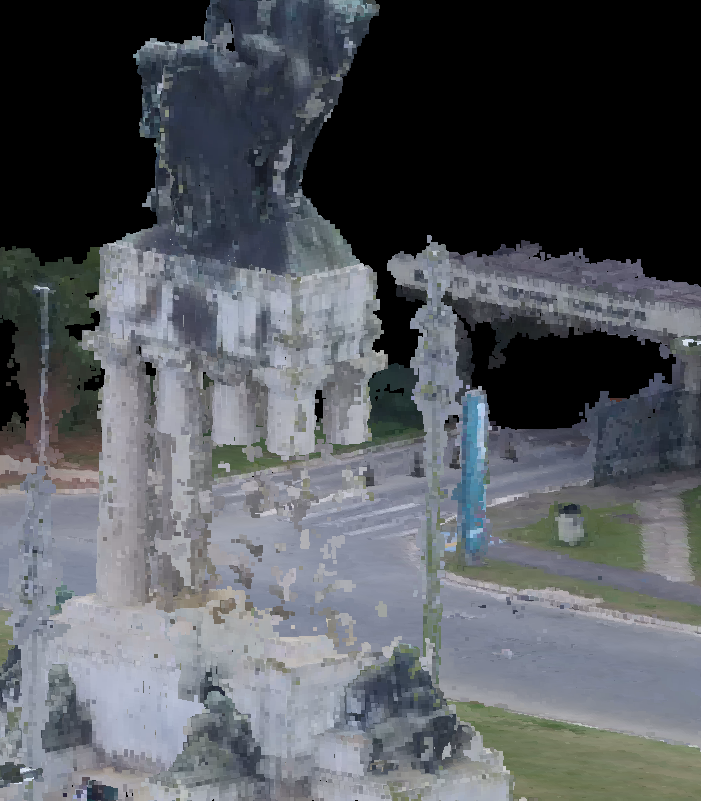}}
        \caption{\textit{Ramos} encoded with ADLPCC (crop)}
        \label{fig:ramosADLPCC}
    \end{center}
    \begin{center}
        \subfloat[R01]{\includegraphics[width=0.15\textwidth]{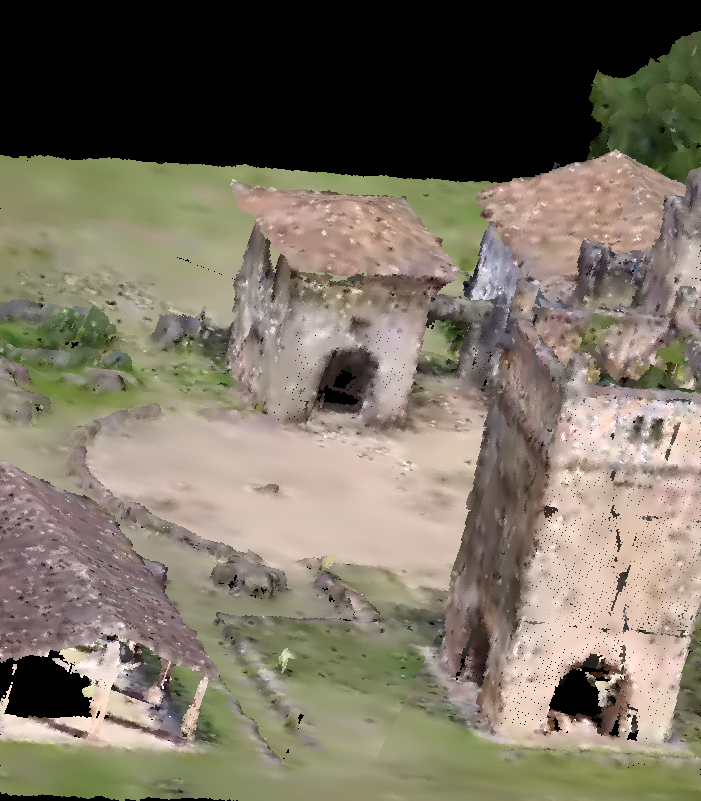}}
        \subfloat[R03]{ \includegraphics[width=0.15\textwidth]{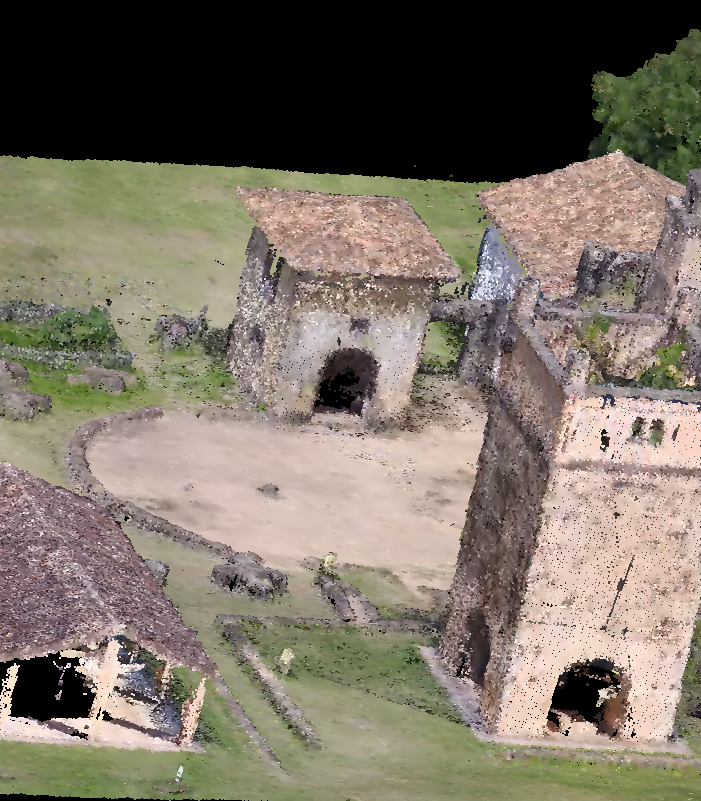}}
        \subfloat[R05]{ \includegraphics[width=0.15\textwidth]{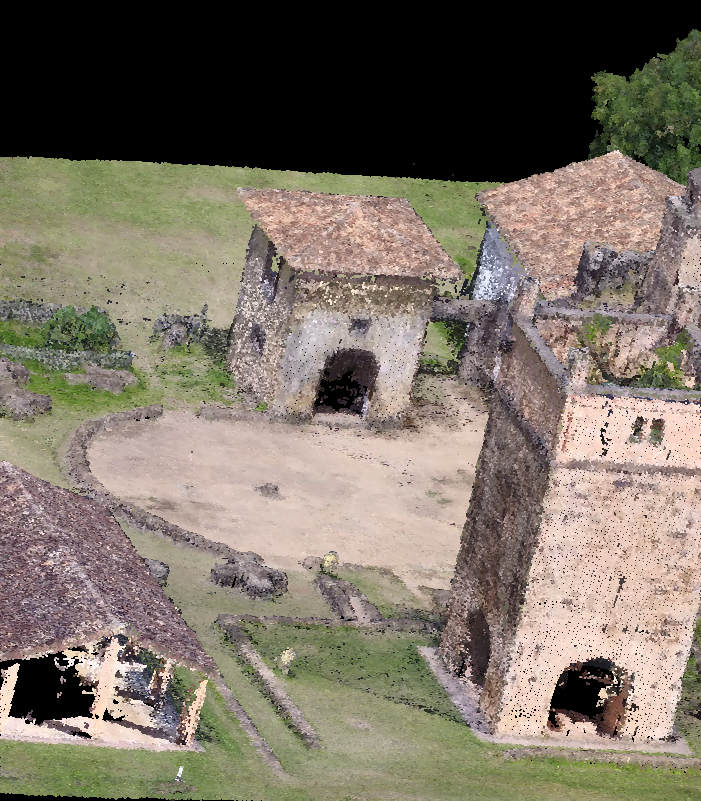}}
        \caption{\textit{IpanemaCut} encoded with PCC GEO CNNv2 (crop)}        
        \label{fig:ipanemaGEOCNN}
    \end{center}
\end{figure}
In \textit{Evaluation 1}, since the texture of the original point cloud was mapped to the distorted one and then encoded using G-PCC, the texture bitrate is added to the geometry bitrate. Fig. \ref{MOSBPPtextgeoBPP} also refers to the results of \textit{Evaluation 1}, but only the geometry bitrate is considered. In \textit{Evaluation 2}, represented in Fig. \ref{MOSBPPgeo}, the texture information was directly mapped onto the distorted geometry and no texture coding was applied, hence the bitrate is relative to geometry only.
Although the scores do not follow a Gaussian distribution, their 95\% Confidence Interval (CI) was computed, assuming a Student's t-distribution. The horizontal green line at the top of each plot refers to the MOS for the hidden references, whereas the green bar around it represents its 95\% CI.
The vertical black line on the right side of each plot represents the lossless encoding with G-PCC. This was computed to assure that the tested bitrates were not larger than the lossless bitrate of G-PCC. It should be noted here that G-PCC and LUT SR were tested for similar bitrates, allowing a more direct comparison.
For the deep learning-based codecs, i.e., PCC GEO CNNv2, PCGCv2 and ADLPCC, the resulting bitrates are highly dependent on their training, so it is not possible to select the bitrates freely. Nevertheless, these are directly comparable within their selected range of bitrates and are simultaneously comparable to the higher bitrates of G-PCC and LUT SR.

It can be concluded that none of the machine-learning codecs can reach the performance of G-PCC when the encoded texture is added. The bitrate achieved by them is always superior to G-PCC, except for the \textit{IpaemaCut} point cloud.
PCGCv2 only reaches a MOS similar to the reference for the \textit{IpanemaCut} (R04 and R05) and the \textit{Guanyin} (R05) point clouds. However, it seems to have slightly better performance at low bitrates than ADLPCC and PCC GEO CNNv2 for some of the tested content, notably \textit{Longdress}, \textit{Guanyin}, and \textit{IpanemaCut}.
The \textit{Romanoillamp} point cloud is an outlier to this general behavior, as PCGCv2 performed quite poorly with it.
The MOS did not reach 3 at any bitrate, which may be related to a lack of suitable data in the training set.
In the case of ADLPCC, the \textit{Ramos} point cloud exhibits strange behavior, as the higher bitrate results in very low performance. This might also be caused by a lack of suitable training data or overfitting.
Another strange behavior is observed for the higher bitrate of the \textit{Romanoillamp} point cloud encoded with PCC GEO CNNv2. The plot shows a slight decrease in MOS in that case. Since it is a very small decrease, it can be concluded that it is due to the randomness of the test sequence, as it can also influence the scores given by the subjects.

%

\begin{minipage}{\textwidth}
  \begin{minipage}[t]{0.49\textwidth}
     \centering
     \includegraphics[width=0.9\linewidth]{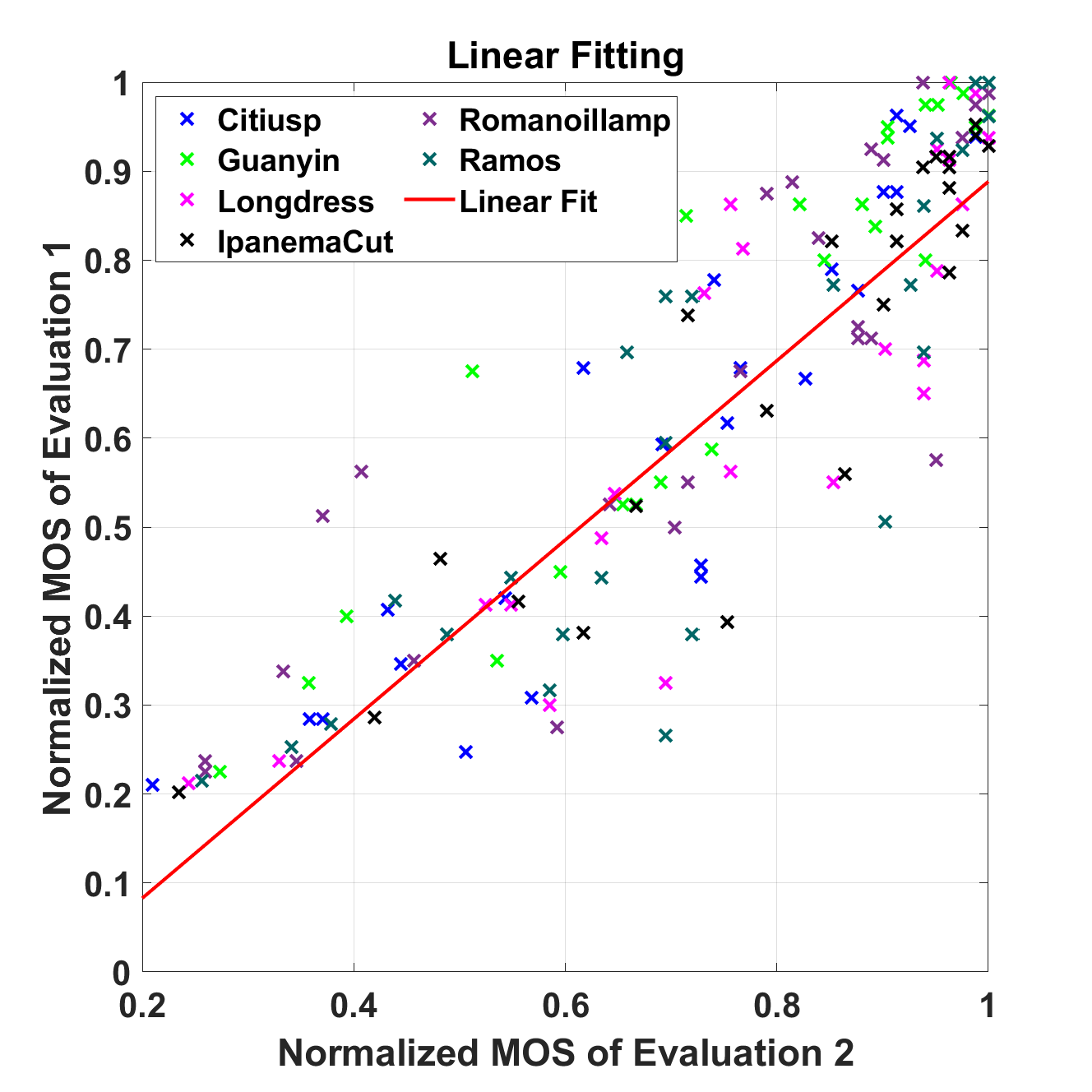}
     \captionof{figure}{MOS scatter plot (geometry-only vs. geometry and texture) with the respective linear fitting. The MOS scores were normalized between 0 and 1.}\label{fig:LinCorrel}
  \end{minipage}
  \hfill
  \begin{minipage}[b]{0.49\textwidth}
    \centering
    \begin{tabular}{|c|c|c|c|}
     \hline
        \multicolumn{4}{|c|}{\textbf{Linear Fitting}}\\
        \hline
        \textbf{PCC} & \textbf{SROCC} & \textbf{RMSE} & \textbf{OR}\\
        \hline
        0.894 & 0.908 & 0.740 & 0.139\\
        \hline
    \end{tabular}%
      \captionof{table}{Statistical analysis between both conducted subjective quality evaluations.}
      \label{table:StatCompar}
    \end{minipage}
  \end{minipage}
Fig. \ref{MOSBPPtextgeoBPP} shows the influence of the geometry bit rate alone, without the influence of the texture that is coded with G-PCC in the case of the deep-learning codecs, shown in Fig. \ref{MOSBPPtextgeo}.
It allows observing how effective the deep learning solution was in compressing the subjective quality.
PCGCv2 exhibits the best performing DL-based solution overall, except for \textit{Romanoillamp}, where PCC GEO CNNv2 achieved the best performance.

Fig. \ref{MOSBPPgeo} shows that most codecs achieve very competitive results, outperforming G-PCC for almost every point cloud. The only exception is observed for the \textit{Romanoillamp} point cloud. The best performing codec is PCGCv2.
As expected, the same drop in the MOS at the highest rate is observed for the \textit{Ramos} point cloud encoded with ADLPCC, as well as the poor overall evaluation of the \textit{Romanoillamp} point cloud encoded with PCGCv2.
It should also be noted that for the \textit{Guanyin} point cloud, PCGCv2 outperformed LUT SR, with similar scores at lower bitrates.
From the plots, it can also be concluded that \textit{Evaluation 1} usually presents lower confidence intervals than \textit{Evaluation 2}, revealing a more stable grading from the subjects.

Figs. \ref{fig:romanPCGCv2} to \ref{fig:ipanemaGEOCNN} show examples of three point clouds, each encoded with a different codec and at three quality levels, from lower (left) to higher (right).

Fig. \ref{fig:romanPCGCv2} shows several artifacts present across all bitrates for \textit{Romanoillamp} encoded with PCGCv2.
These artifacts are most likely derived from the downsampling and upsampling operations. When downsampling a point cloud at the encoder stage, some information regarding that process can be lost, negatively impacting the upsampling process on the decoder side.
Fig. \ref{fig:ramosADLPCC} shows that a large part of the column is missing in the R05 rate of the\textit{Ramos} point cloud encoded with ADLPCC.
The codec relies on dividing the point cloud into blocks, followed by coding them separately. Some blocks are flagged as empty blocks after the lossy process, even if they exist in the original point cloud.
Because of that, some parts of the point clouds might be missing. This is common in the lower bitrates, but should not happen for the higher quality levels.

Fig. \ref{fig:ipanemaGEOCNN} shows a cropped area with several artifacts in the tower present across bitrates for the \textit{IpanemaCut} encoded with PCC GEO CNNv2.
PCC GEO CNNv2 uses convolutional layers to encode and decode the point clouds. The resulting artifacts are most likely due to the bit-wise and element-wise rounding that occurs in the final layer of the encoding and decoding architecture.
Sometimes the artifacts appear in the higher bitrates, which is caused by the lack of appropriate training data to handle such point clouds.

A very important part of this work is to analyze how effective the two subjective models for quality evaluation are. For that, the statistical similarity of the two subjective evaluations is analyzed. Fig. \ref{fig:LinCorrel} shows the linear fitting between the MOS scores of both evaluations and Table \ref{table:StatCompar} shows the Pearson Correlation Coefficient (PCC), Spearman Rank Order Correlation Coefficient (SROCC), the Root Mean Square Error (RMSE), and the Outlier Ratio (OR) values between them using linear fitting, as recommended in ITU-T P.1401~\cite{ITU-T}. The results reveal some statistical similarity, but the two subjective evaluations do not seem to provide the same results.

To further assess the statistical similarity between the two evaluations, a Kruskal-Wallis one-way analysis followed by a multiple comparison test~\cite{KruskalWallis} was performed. It is concluded that there are statistical differences between the two subjective evaluations ($p$-value = 7,479E-04 $<$ 0.05).

At this stage, it is important to try to understand which of the two subjective evaluation models is better.
It is important to consider that mapping the texture onto the distorted geometry without any encoding (\textit{Evaluation 2}) results in a subsampling process that is likely to cause aliasing. This may eventually influence the perceived quality. From the observation of the samples, it is observed that encoding the texture provides a better balance between geometry and texture quality, resulting in a more reliable subjective evaluation.

\subsection{Performance of objective quality evaluation} \label{sec:objEval}

\begin{figure}[thb]
    \begin{center}
    \subfloat[\textit{Longdress}\label{longdress1 - PCQM}]{%
        \includegraphics[width=0.17\textwidth]{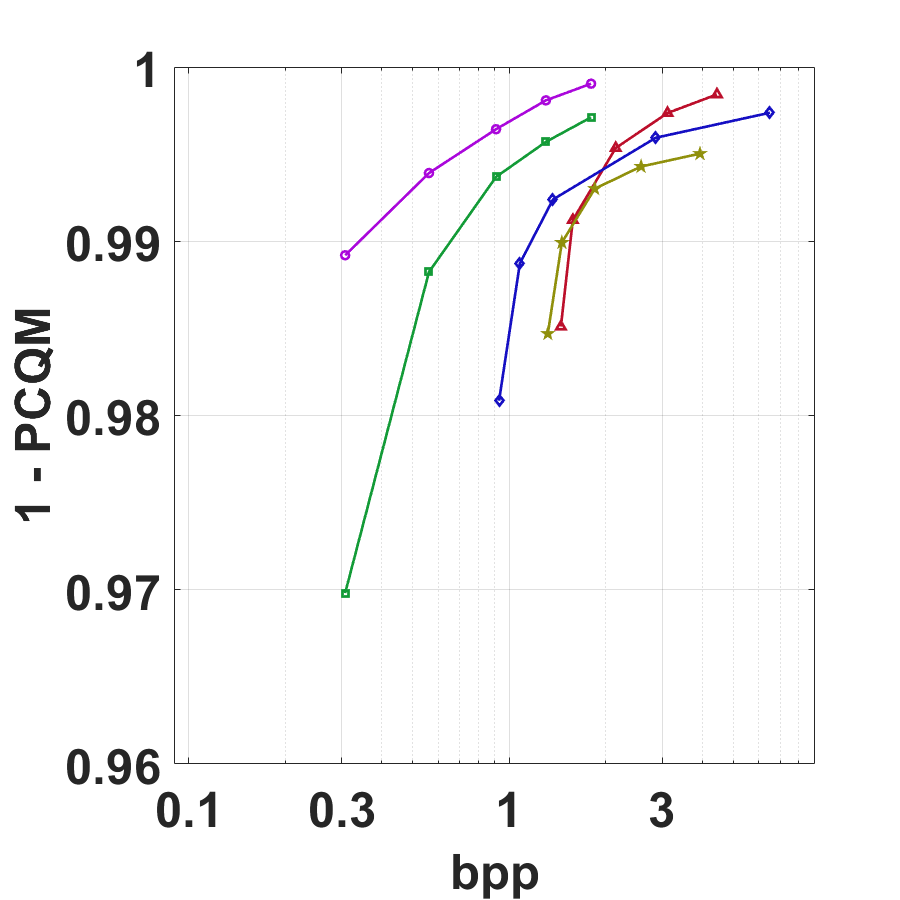}}
    \subfloat[\textit{Guanyin}\label{guan1 - PCQM}]{%
        \includegraphics[width=0.17\textwidth]{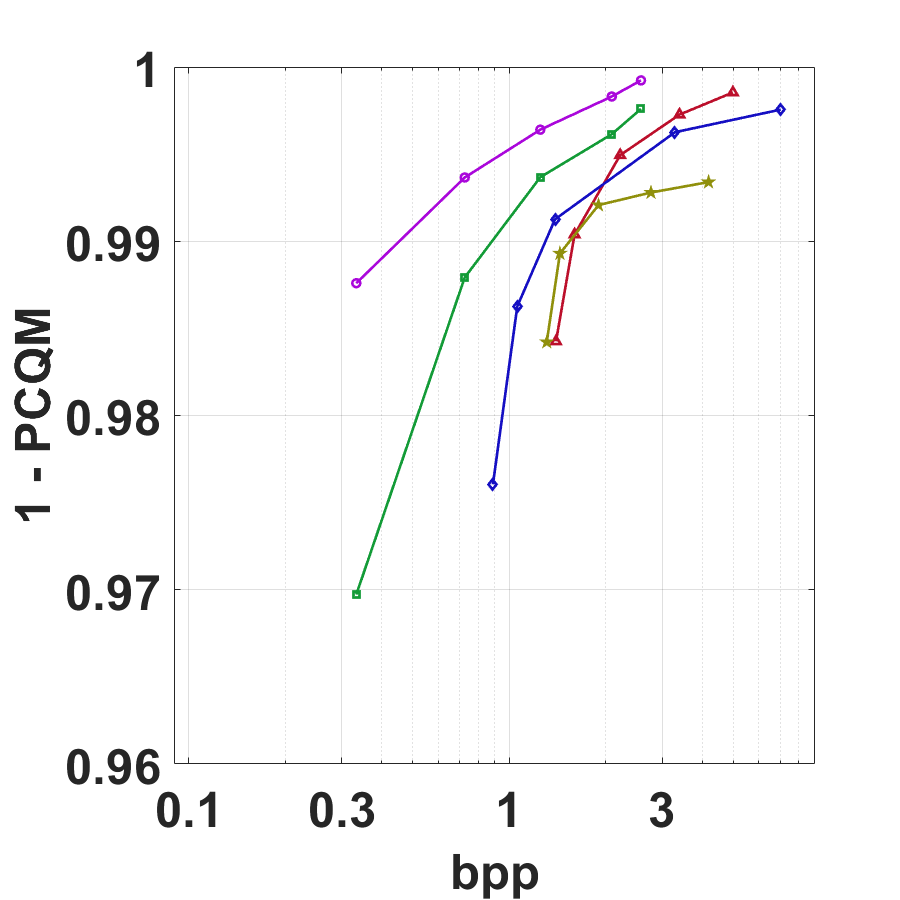}}
    \subfloat[\textit{Romanoillamp}\label{roman1 - PCQM}]{%
        \includegraphics[width=0.17\textwidth]{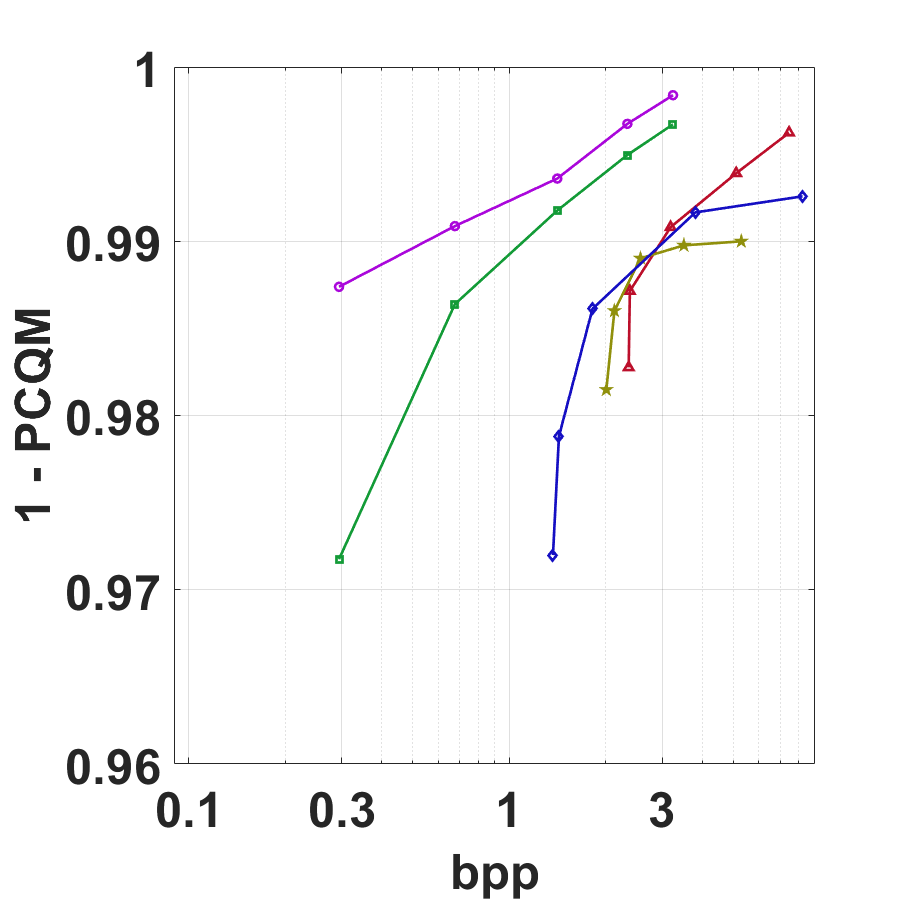}}
    \subfloat[\textit{Citiusp}\label{citi1 - PCQM}]{%
        \includegraphics[width=0.17\textwidth]{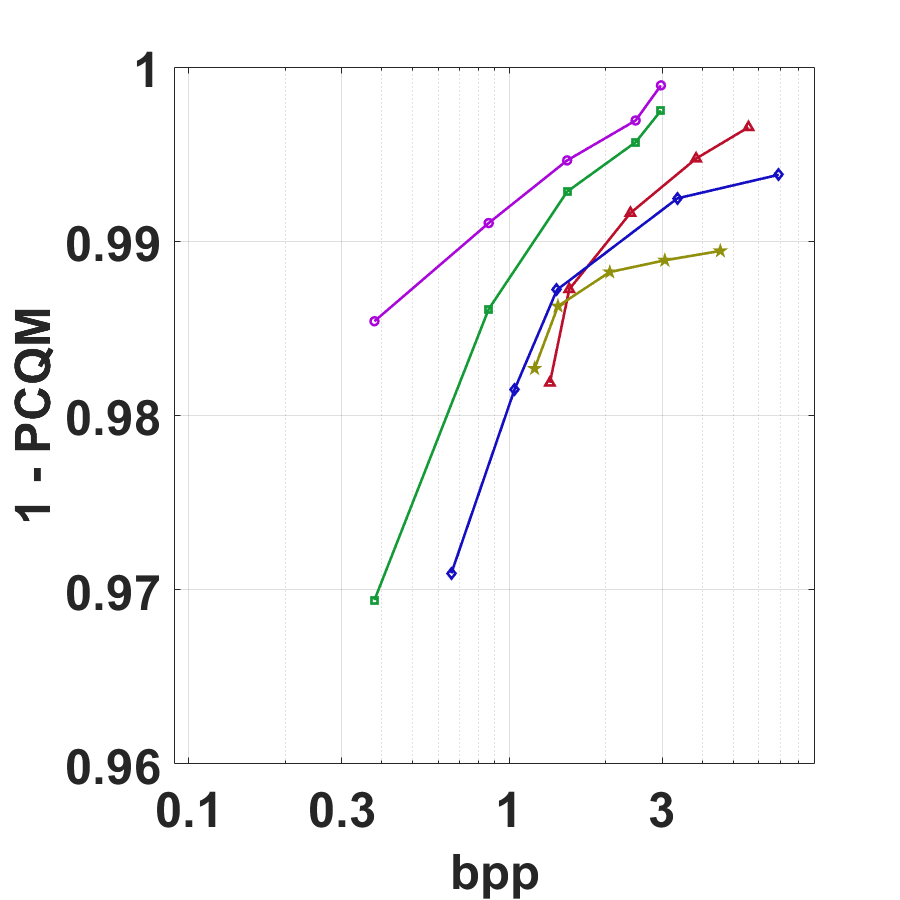}}
    \subfloat[\textit{IpanemaCut}\label{ipan1 - PCQM}]{%
        \includegraphics[width=0.17\textwidth]{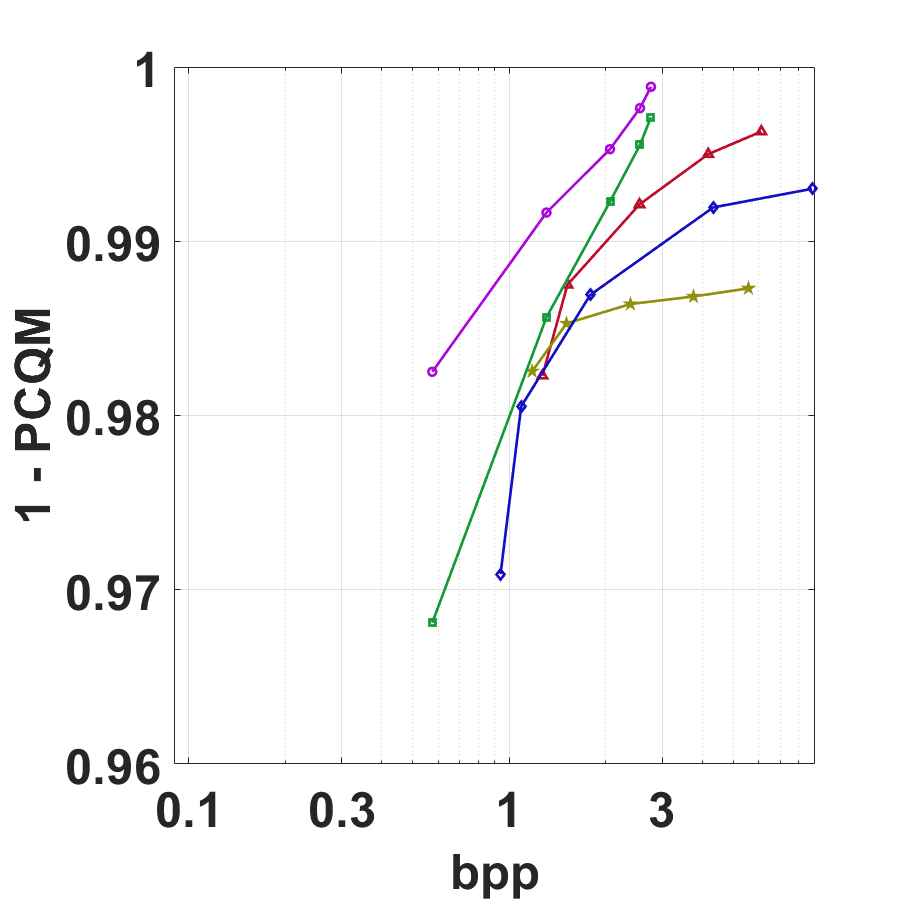}}
    \subfloat[\textit{Ramos}\label{ramos1 - PCQM}]{%
        \includegraphics[width=0.17\textwidth]{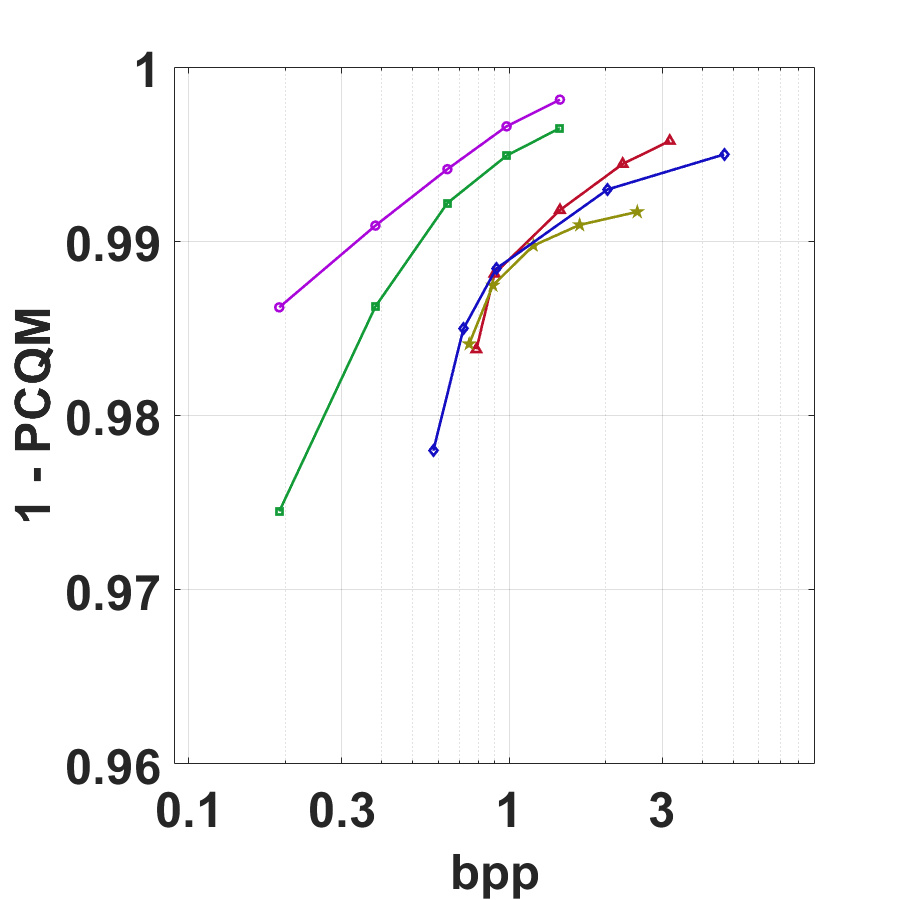}}\\
    \includegraphics[width=0.7\textwidth]{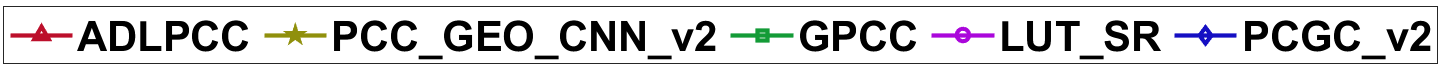}
    \caption{1 - PCQM vs bpp (geometry + texture).}
    \label{1 - PCQMBPP}
    \end{center}
\end{figure}

\begin{figure}
    \begin{center}
    \subfloat[\textit{Longdress}\label{longdressD1GEO}]{%
        \includegraphics[width=0.17\textwidth]{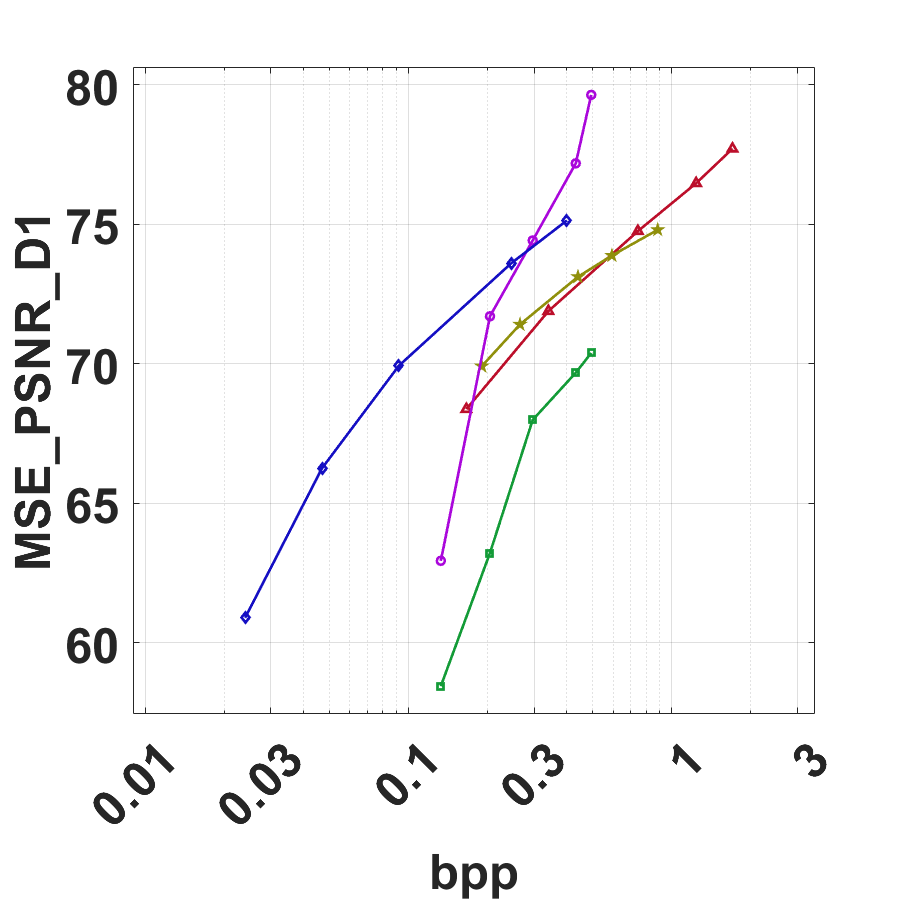}}
    \subfloat[\textit{Guanyin}\label{guanD1GEO}]{%
        \includegraphics[width=0.17\textwidth]{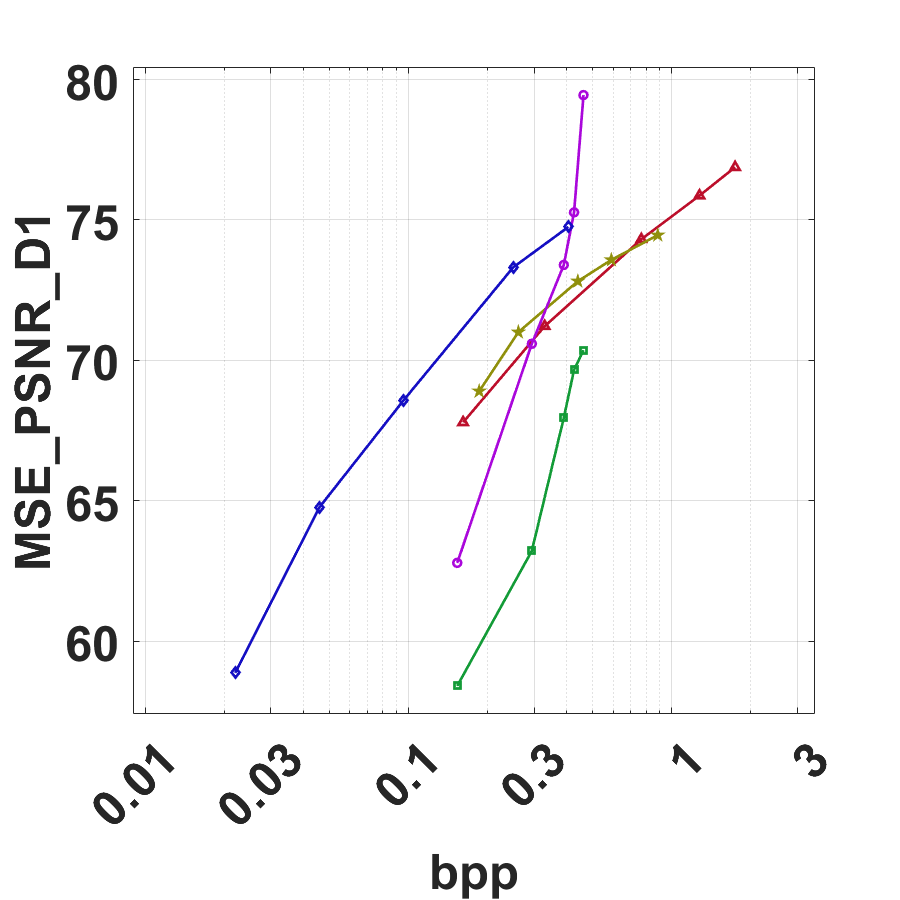}}
    \subfloat[\textit{Romanoillamp}\label{romanD1Geo}]{%
        \includegraphics[width=0.17\textwidth]{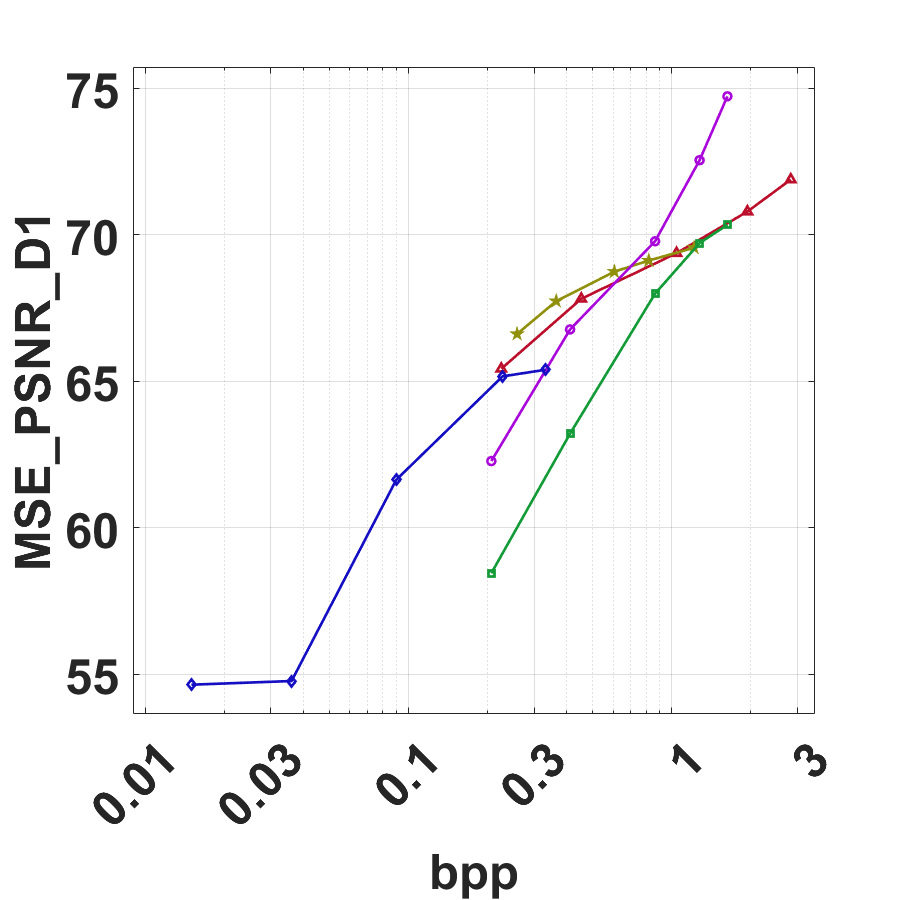}}
    \subfloat[\textit{Citiusp}\label{citiD1Geo}]{%
        \includegraphics[width=0.17\textwidth]{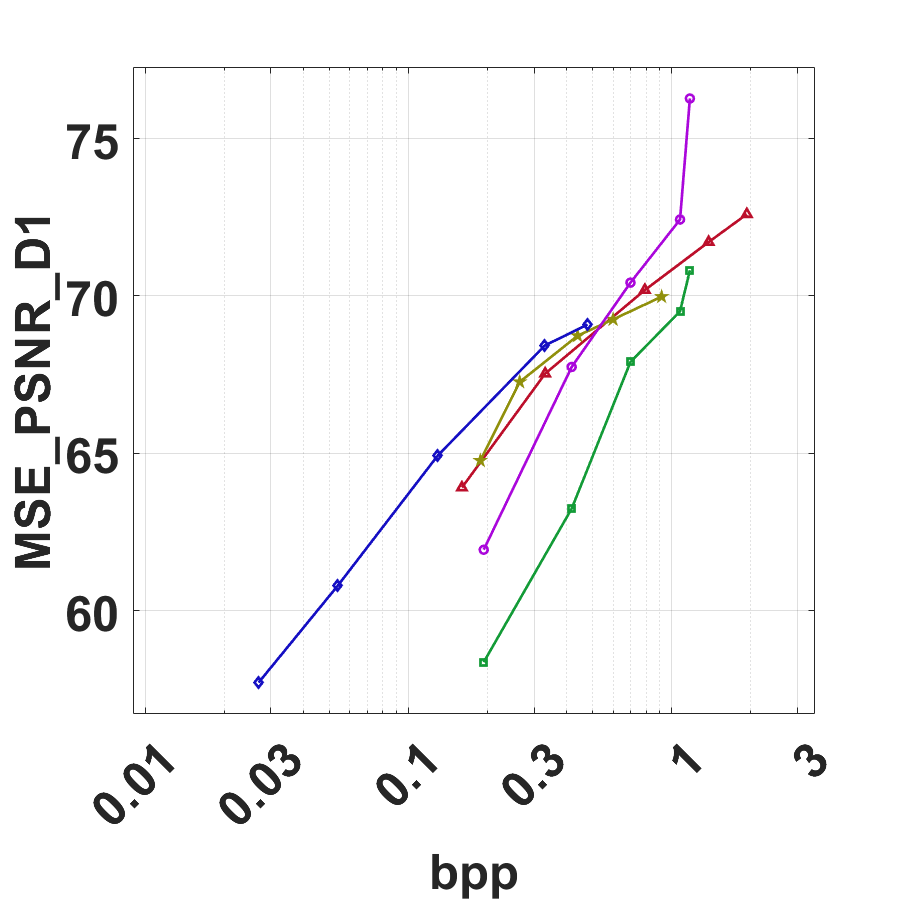}}
    \subfloat[\textit{IpanemaCut}\label{ipanD1Geo}]{%
        \includegraphics[width=0.17\textwidth]{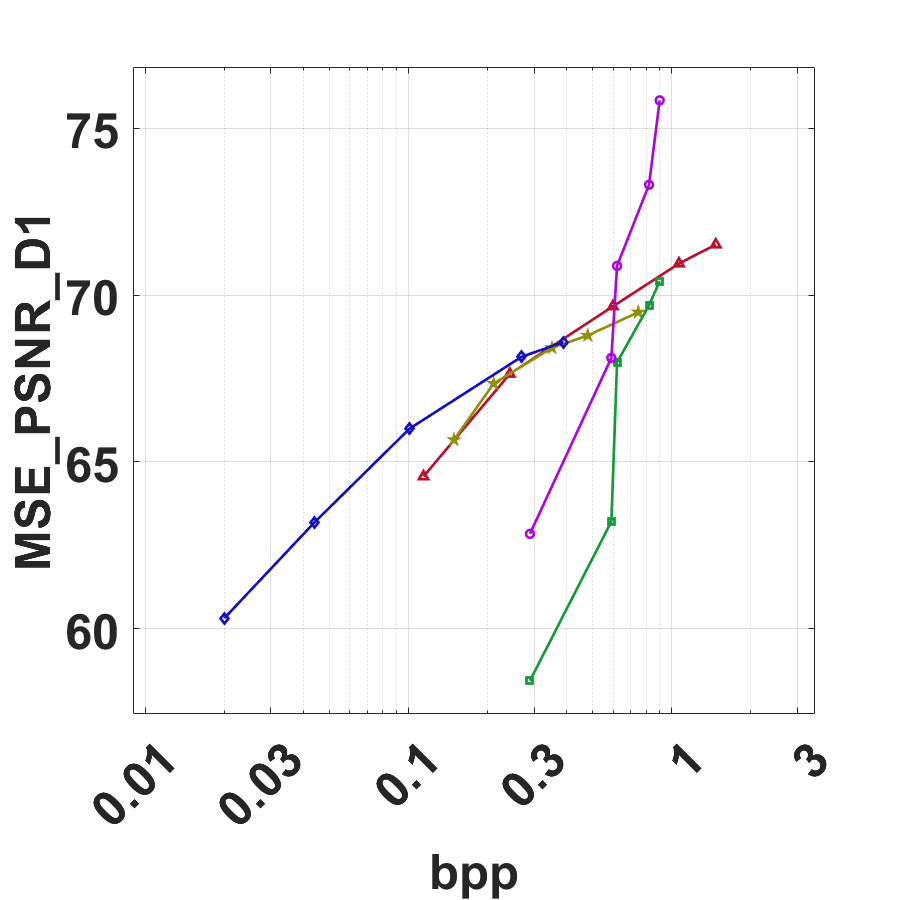}}
    \subfloat[\textit{Ramos}\label{ramosDG1Geo}]{%
        \includegraphics[width=0.17\textwidth]{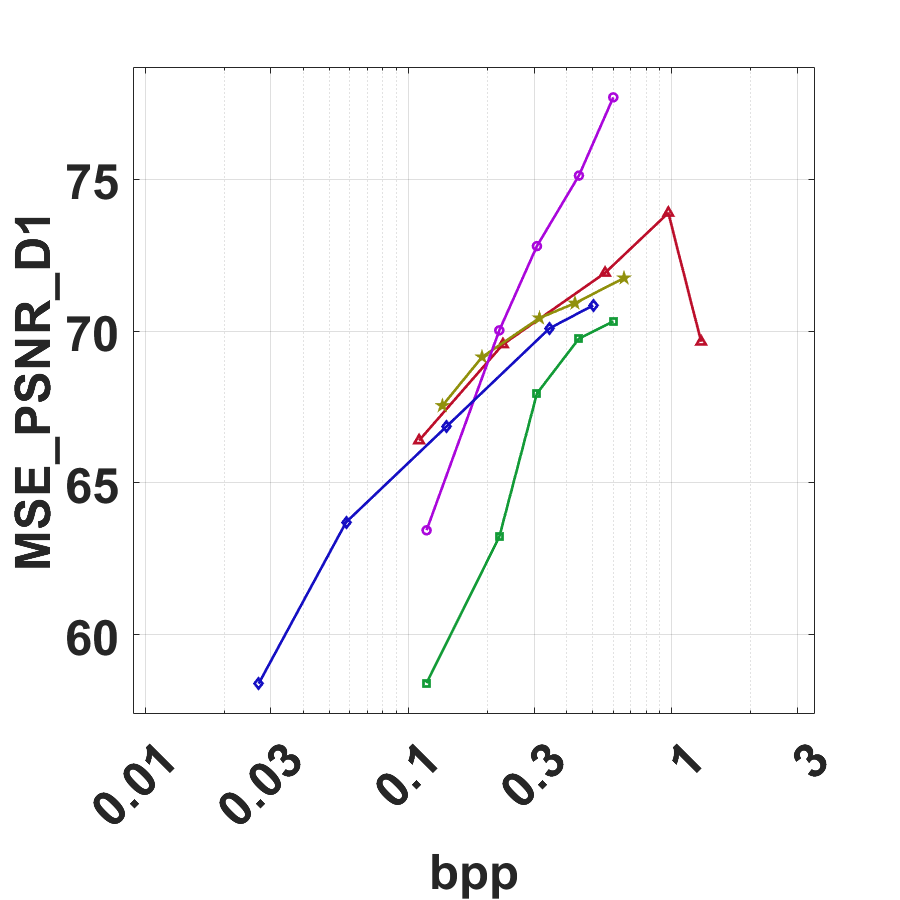}}
    \caption{MSE PSNR D1 vs bpp (geometry only).}

    \label{D1BPP}
    \end{center}
\end{figure}

Subjective quality assessment provides the ground truth for the validation of objective quality metrics, considering the distortions produced by the tested codecs.
Figs. \ref{1 - PCQMBPP} and \ref{D1BPP} show the PCQM and MSE PSNR D1 metrics, respectively, plotted against the coding bitrates. In Fig. \ref{1 - PCQMBPP}, the bitrates consider both geometry and texture, as PCQM is a joint perceptual metric. On the other hand, Fig. \ref{D1BPP} considers only the geometry bitrate, as MSE PSNR D1 only uses the geometry information.

The plots in Fig. \ref{1 - PCQMBPP} indicate that PCQM tends to establish the same performance relations between different codecs as the subjective evaluations, although some exceptions can be identified easily.
The PGCGv2 codec obtained results similar to the other codecs for the \textit{Romanoillamp} point cloud, which is not identified in the subjective evaluation. This metric also failed to predict the unusual behavior observed with the \textit{Ramos} point cloud encoded with ADLPCC at R05.

MSE PSNR D1 is represented in Fig. \ref{D1BPP}. It is observed that it does not follow the performance found in the subjective evaluations represented in Figs. \ref{MOSBPPtextgeoBPP} or \ref{MOSBPPgeo}.
The metric shows that G-PCC is the worst performing codec, which is not the case in any evaluation. The metric predicts that PCGCvs is the best performing codec, which is contrary to what is shown in \ref{MOSBPPtextgeo}.
The point clouds \textit{IpanemaCut} and \textit{Ramos} have some similarity between the metric MSE PSNR D1 and subjective results.
The quality decrease at rate R05 for \textit{Ramos} is predicted, but that does not happen for R03.
The metric also did not predict the quality decrease for the rate R05 for \textit{Romanoillamp} encoded with PCC GEO CNNv2.

\begin{table*}[t]
    \centering
    \caption{Correlation of the objective metrics with the subjective quality evaluation results. The best values are shown in bold, and the second best values are shown in italic.}\label{tab:metricCorrel}
    \begin{tabular}{@{} |l |c|c|c|c|c|  @{}} 
    \hline
    & & \multicolumn{4}{c|}{\textbf{\textit{Evaluation 1}}} \\ \hline
    \textbf{Metric} & Type & PCC & SROCC & RMSE & OR \\ \hline
    MSE PSNR D1 & $FR,\,GEO$ & 0.806 & 0.782 & 0.184 & 0.753  \\ \hline
    MSE PSNR D2 & $FR,\,GEO$ & 0.821 & 0.796 & 0.177 & 0.813  \\ \hline
    PointSSIM & $FR,\,COL$ & \textit{0.859} & \textit{0.857} & \textit{0.159} & \textit{0.720}  \\ \hline
    Point 2 Distribution & $FR,\,GEO+COL$ & 0.851 & 0.828 & 0.164 & 0.640  \\ \hline
    PCM-RR & $FR,\,GEO+COL$ & 0.834 & 0.834 & 0.172 & 0.727  \\ \hline
    GraphSIM & $FR,\,GEO+COL$ & 0.800 & 0.799 & 0.186 & 0.780  \\ \hline
    PCQM & $FR,\,GEO+COL$ & \textbf{0.899} & \textbf{0.903} & \textbf{0.137} & \textbf{0.573}  \\ \hline
    & & \multicolumn{4}{c|}{\textbf{\textit{Evaluation 2}}} \\ \hline
    MSE PSNR D1 & $FR,\,GEO$ & \textbf{ 0.834} & \textbf{0.774} & \textbf{0.152} & \textbf{0.720} \\ \hline
    MSE PSNR D2 & $FR,\,GEO$  & \textit{0.777} &\textit{ 0.740} &\textit{ 0.174} & \textit{0.793} \\ \hline
    PointSSIM & $FR,\,COL$ & 0.188 & 0.143 & 0.271 & 0.920 \\ \hline
    Point 2 Distribution & $FR,\,GEO+COL$ & 0.437 & 0.472 & 0.249 & 0.873 \\ \hline
    PCM-RR & $FR,\,GEO+COL$ & 0.408 & 0.323 & 0.252 & 0.900 \\ \hline
    GraphSIM & $FR,\,GEO+COL$ & 0.560 & 0.573 & 0.229 & 0.907\\ \hline
    PCQM & $FR,\,GEO+COL$ & 0.634 & 0.700 & 0.214 & 0.787 \\ \hline
    \end{tabular}
\end{table*}
\begin{figure}
    \centering
    \subfloat[1 - PCQM\label{fig:PCQM}]{%
        \includegraphics[width=0.14\linewidth]{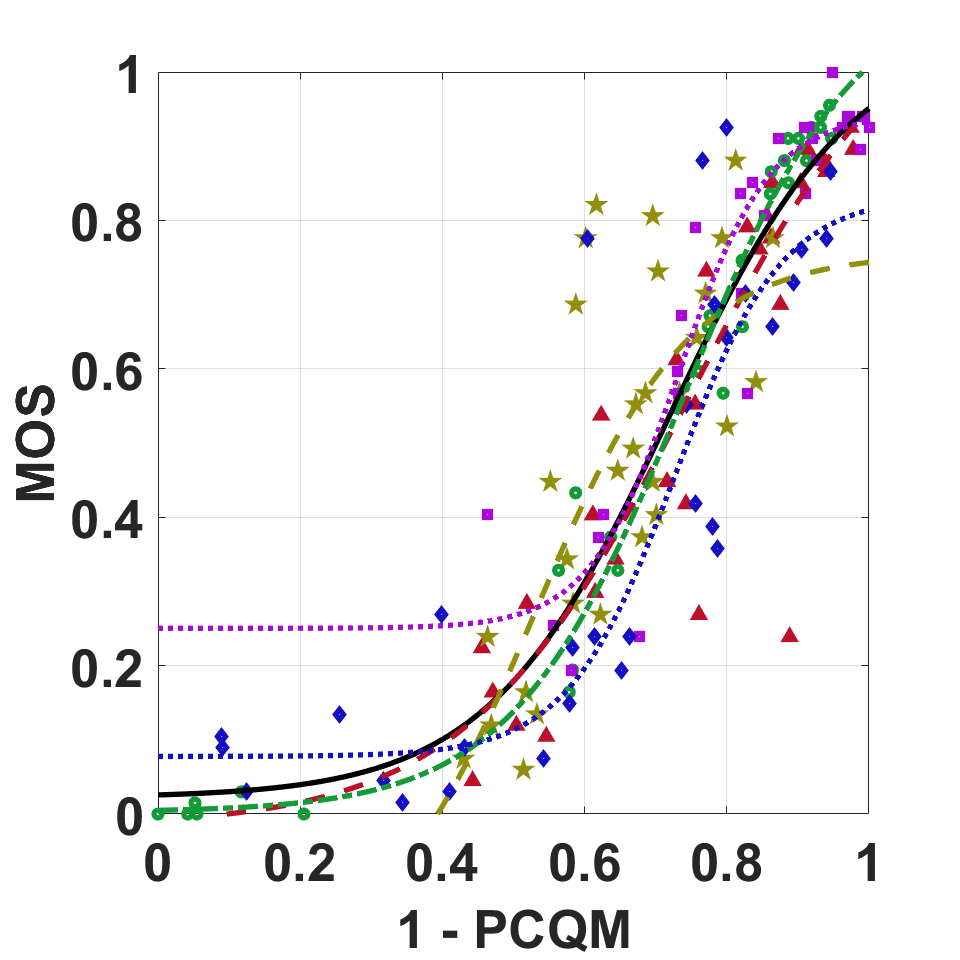}}
    \subfloat[MSE PSNR D1\label{fig:D1MSEPSNR}]{%
        \includegraphics[width=0.14\linewidth]{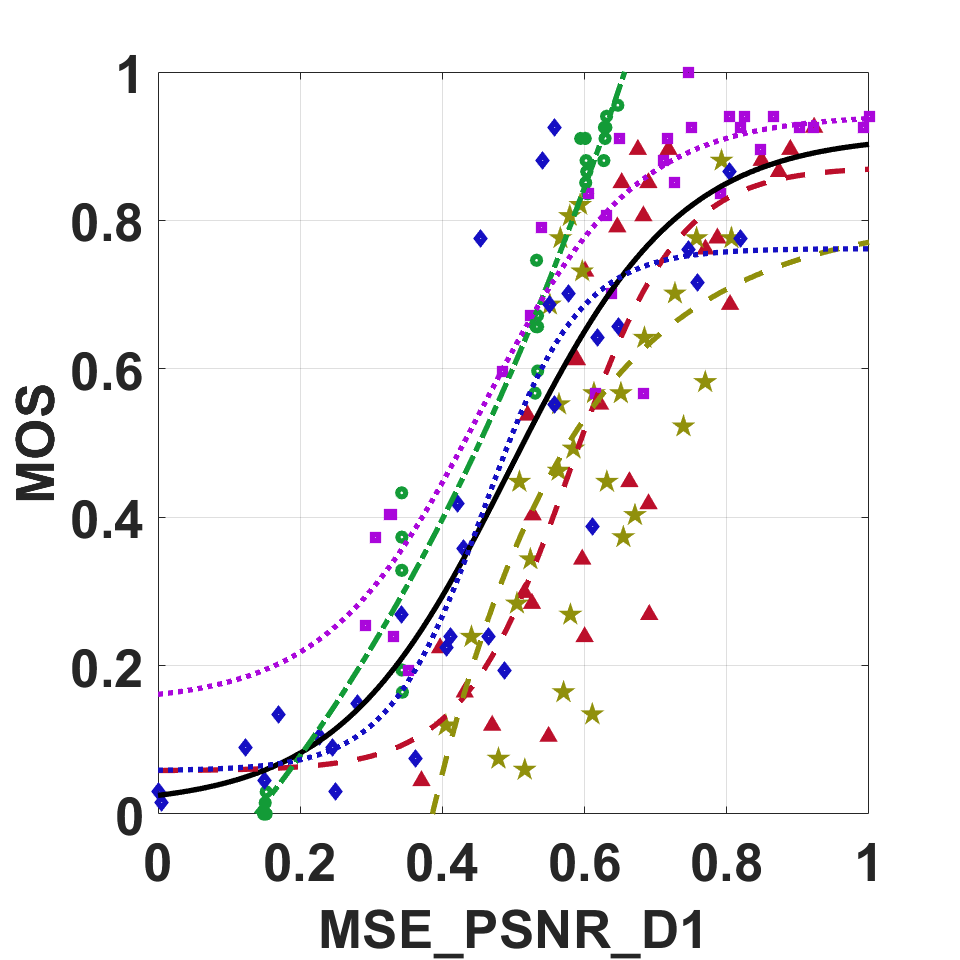}}
    \subfloat[MSE PSNR D2\label{fig:D2MSEPSNR}]{%
        \includegraphics[width=0.14\linewidth]{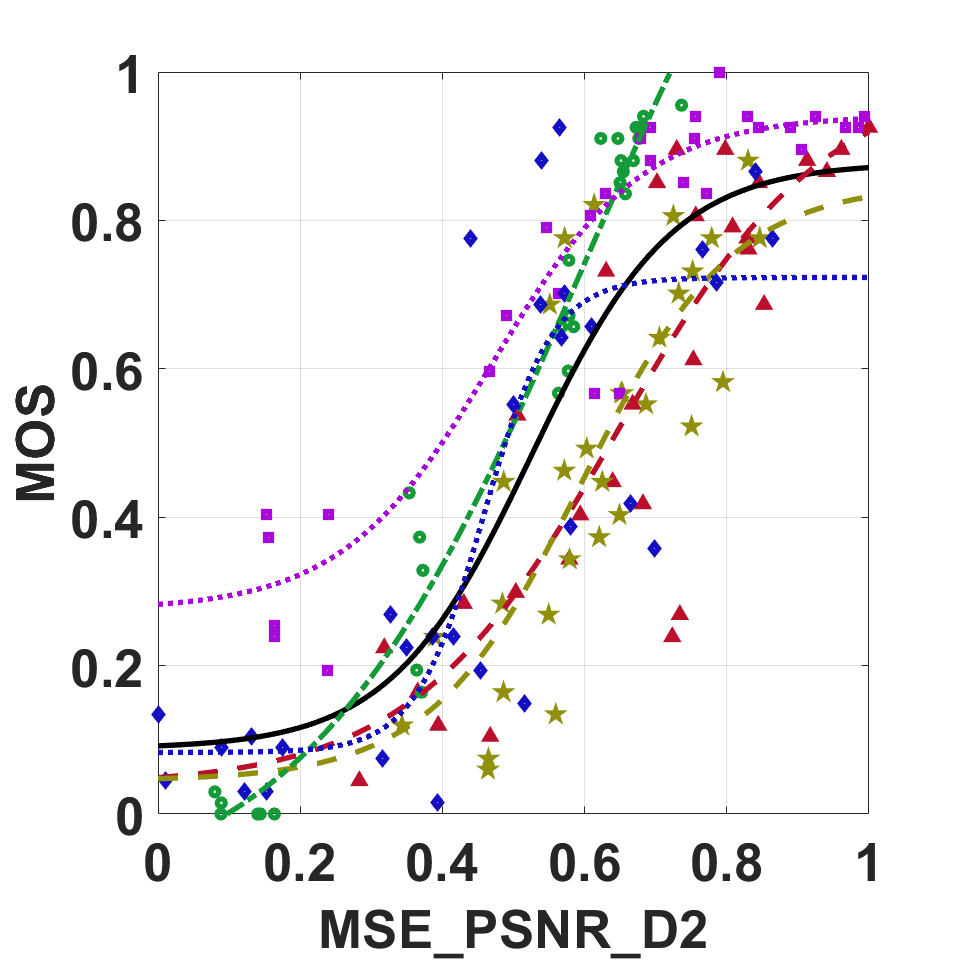}}
    \subfloat[\centering Point 2 \mbox{Distribution}\label{fig:P2Dist}]{%
        \includegraphics[width=0.14\linewidth]{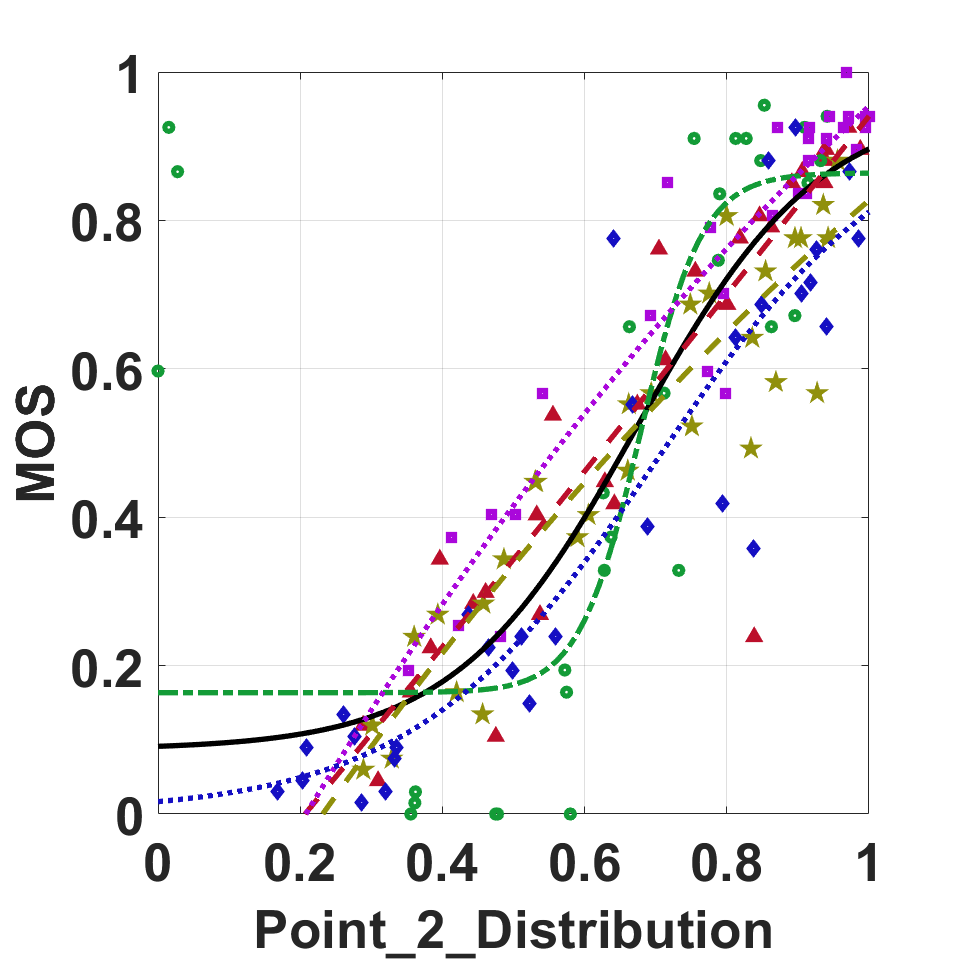}}
    \subfloat[PSSIM\label{fig:PSSIM}]{%
        \includegraphics[width=0.14\linewidth]{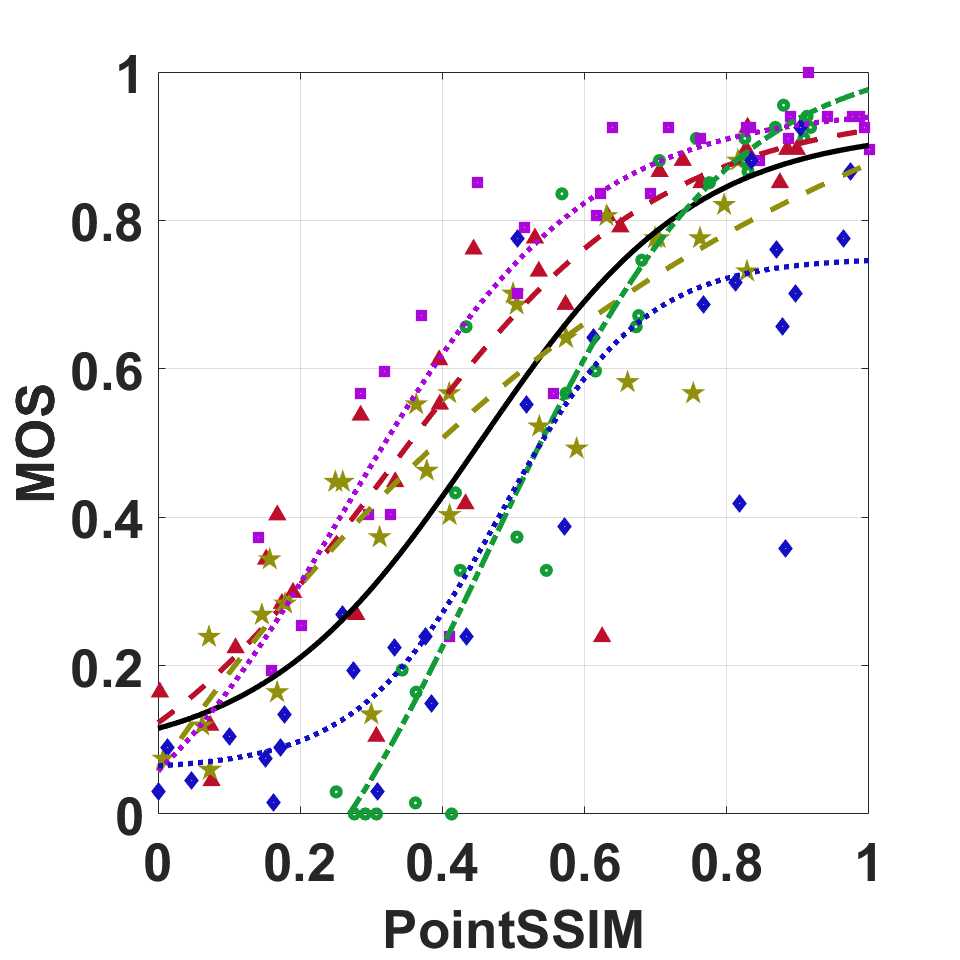}}
    \subfloat[1 - PCM-RR\label{fig:PCMRR}]{%
        \includegraphics[width=0.14\linewidth]{1_-_PCQM_texture.png}}
    \subfloat[GraphSIM\label{fig:GRAPHSIM}]{%
        \includegraphics[width=0.14\linewidth]{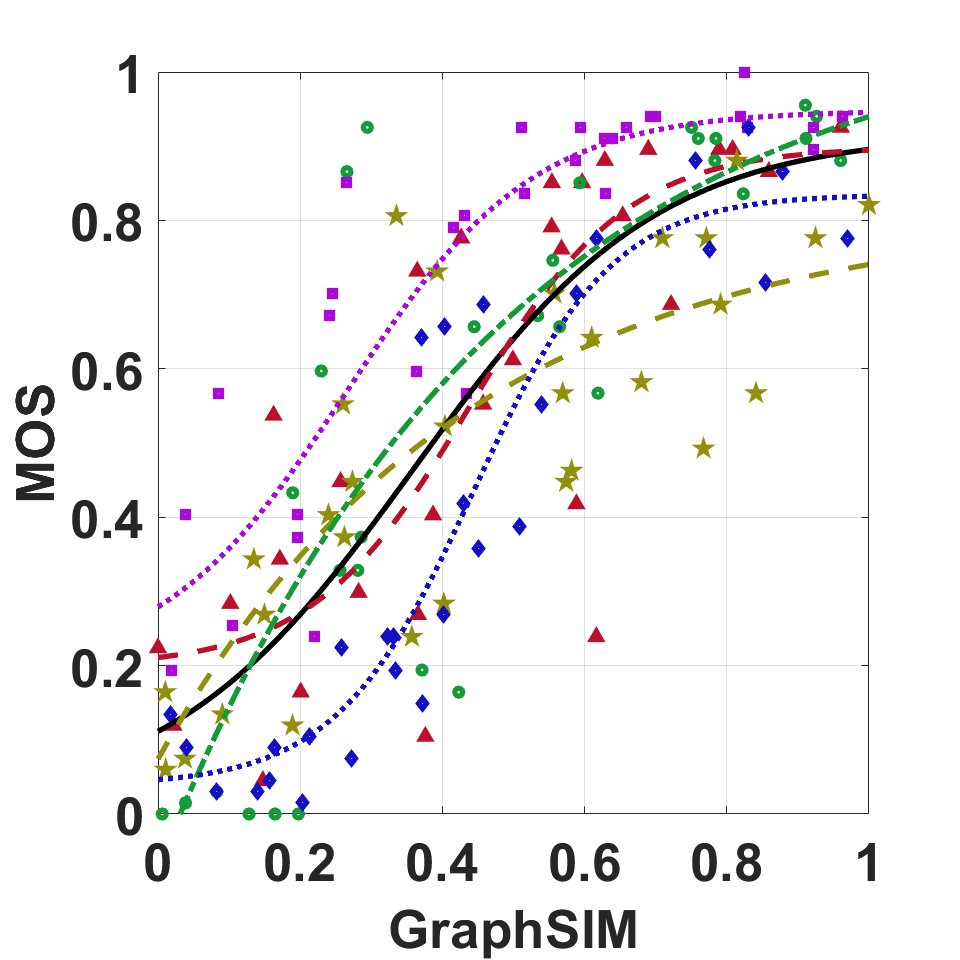}}\\
    \includegraphics[width=0.5\linewidth]{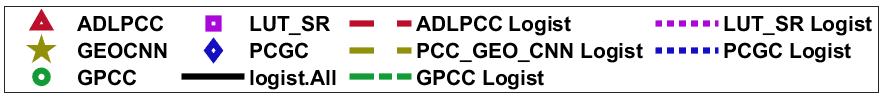}
    \caption{Objective metric vs. Evaluation 1 MOS plots, with logistic regression curves (global and for each codec).} 
    \label{fig:metrics}
\end{figure}
\begin{figure}[t!]
    \centering
    \subfloat[1 - PCQM\label{fig:PCQMgEO}]{%
        \includegraphics[width=0.14\linewidth]{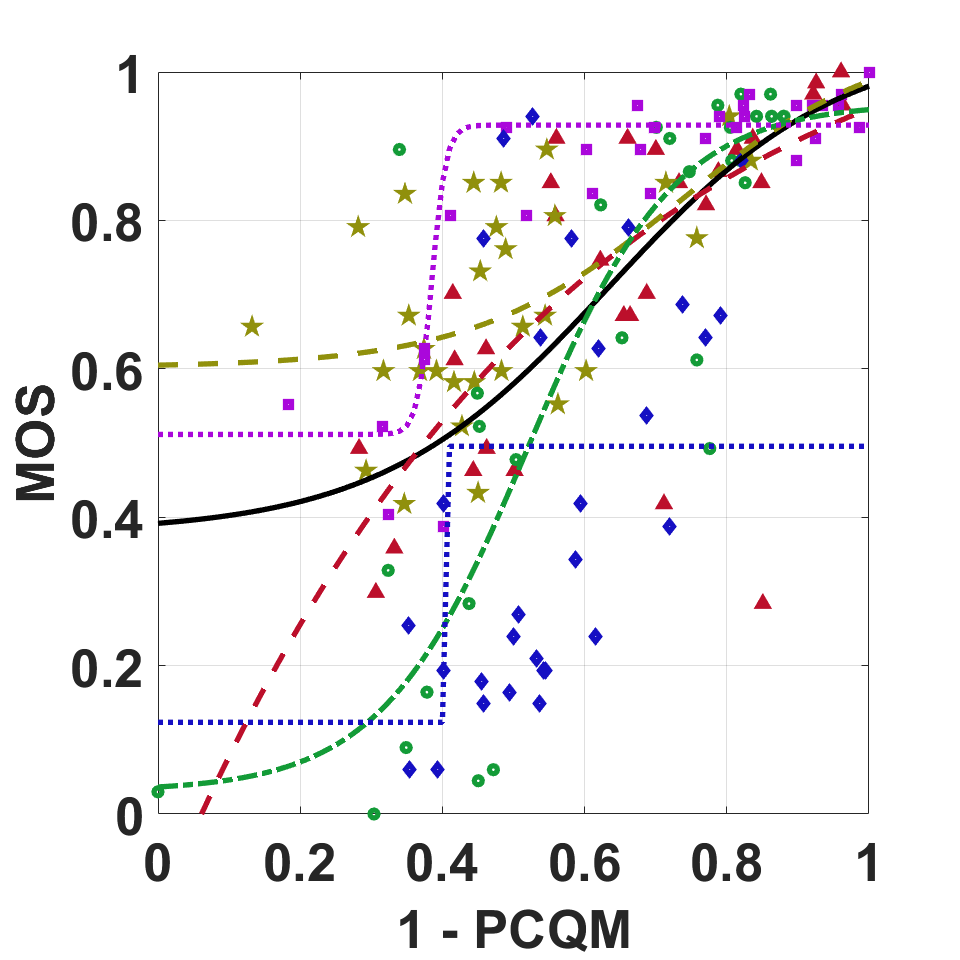}}
    \subfloat[MSE PSNR D1 \label{fig:D1MSEPSNRGeo}]{%
        \includegraphics[width=0.14\linewidth]{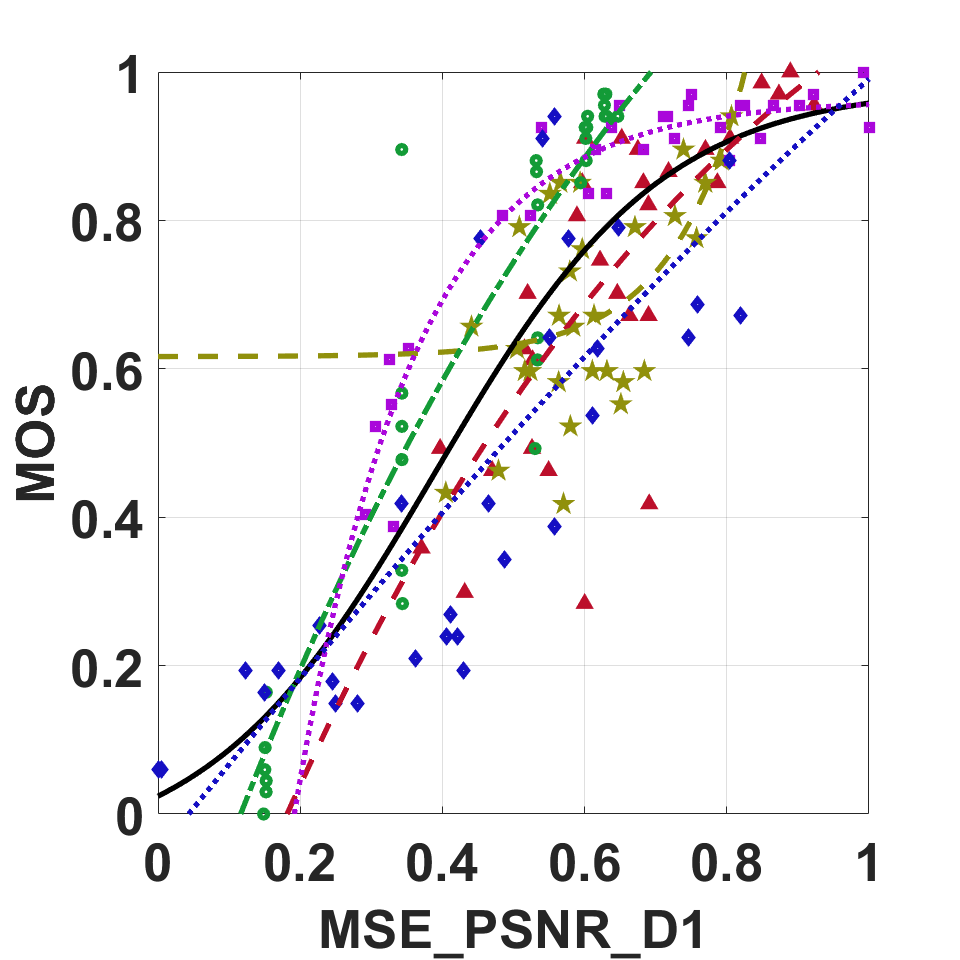}}
    \subfloat[MSE PSNR D2\label{fig:D2MSEPSNRGeo}]{%
        \includegraphics[width=0.14\linewidth]{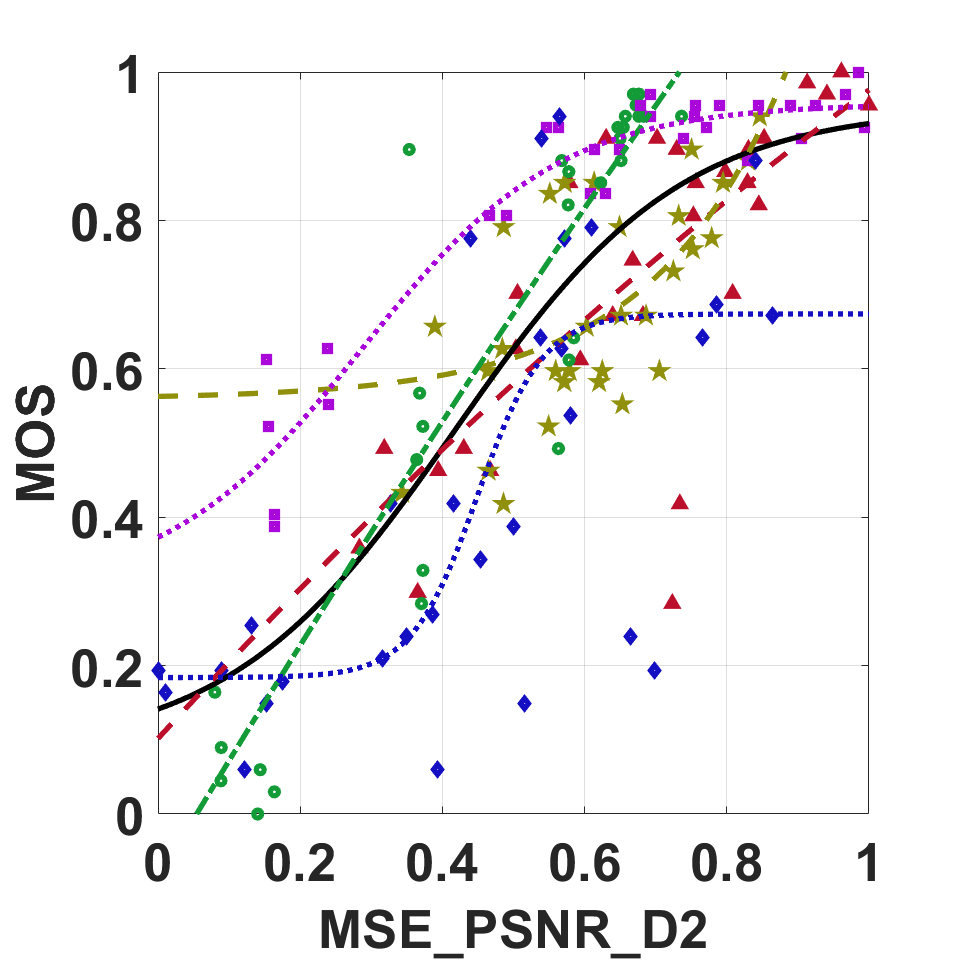}}
    \subfloat[\centering  Point 2 \mbox{Distribution}\label{fig:P2DistGeo}]{%
        \includegraphics[width=0.14\linewidth]{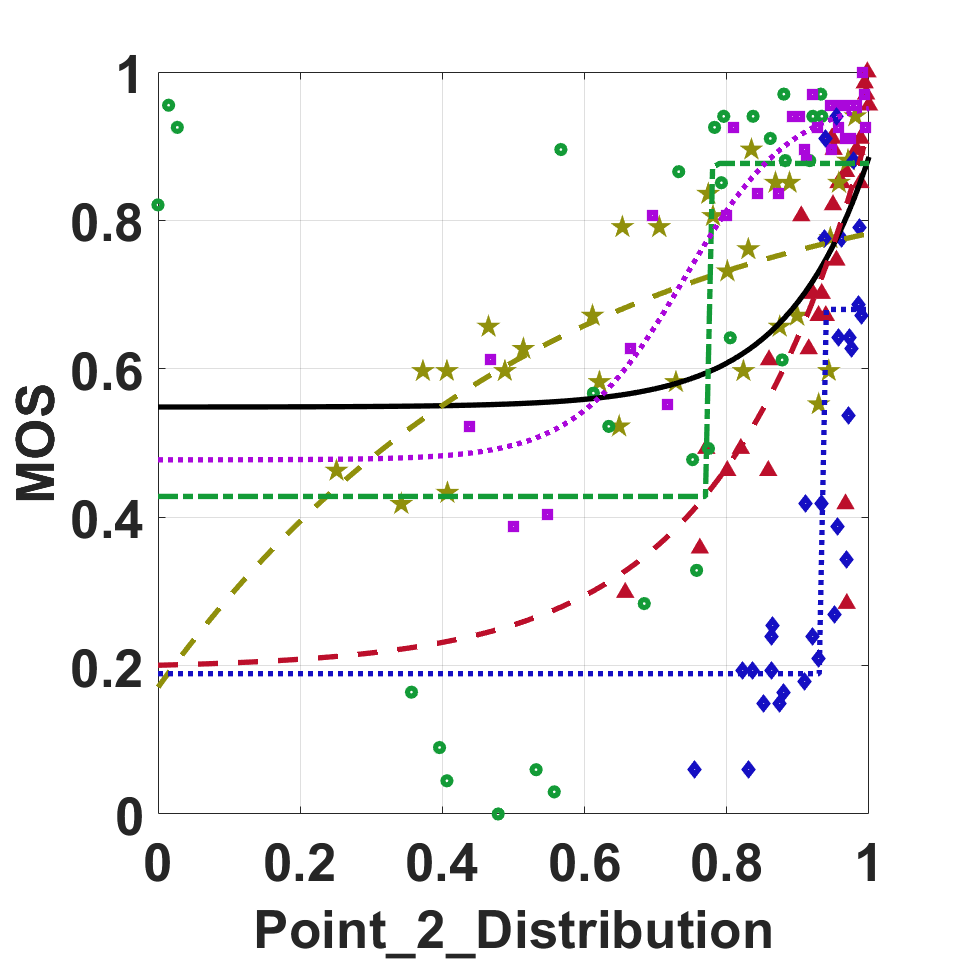}}
    \subfloat[PSSIM\label{fig:PSSIMGeo}]{%
        \includegraphics[width=0.14\linewidth]{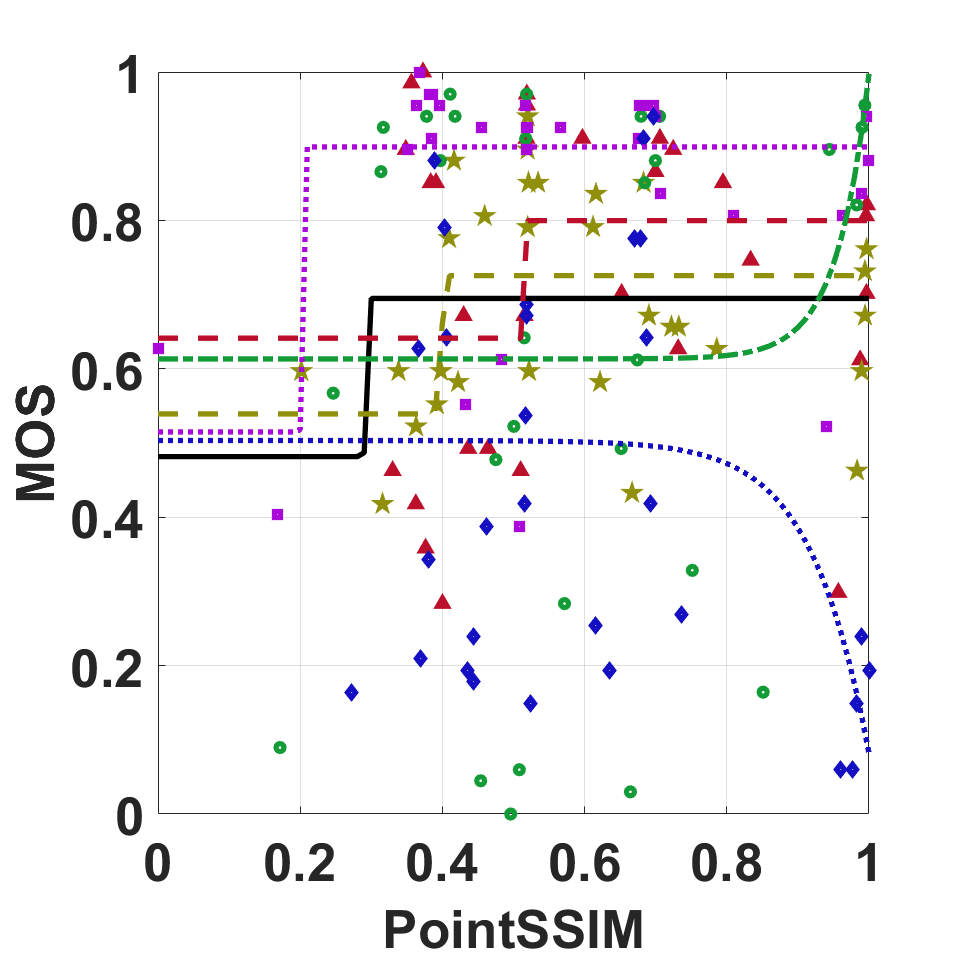}}
    \subfloat[1 - PCM-RR\label{fig:PCMRRGeo}]{%
        \includegraphics[width=0.14\linewidth]{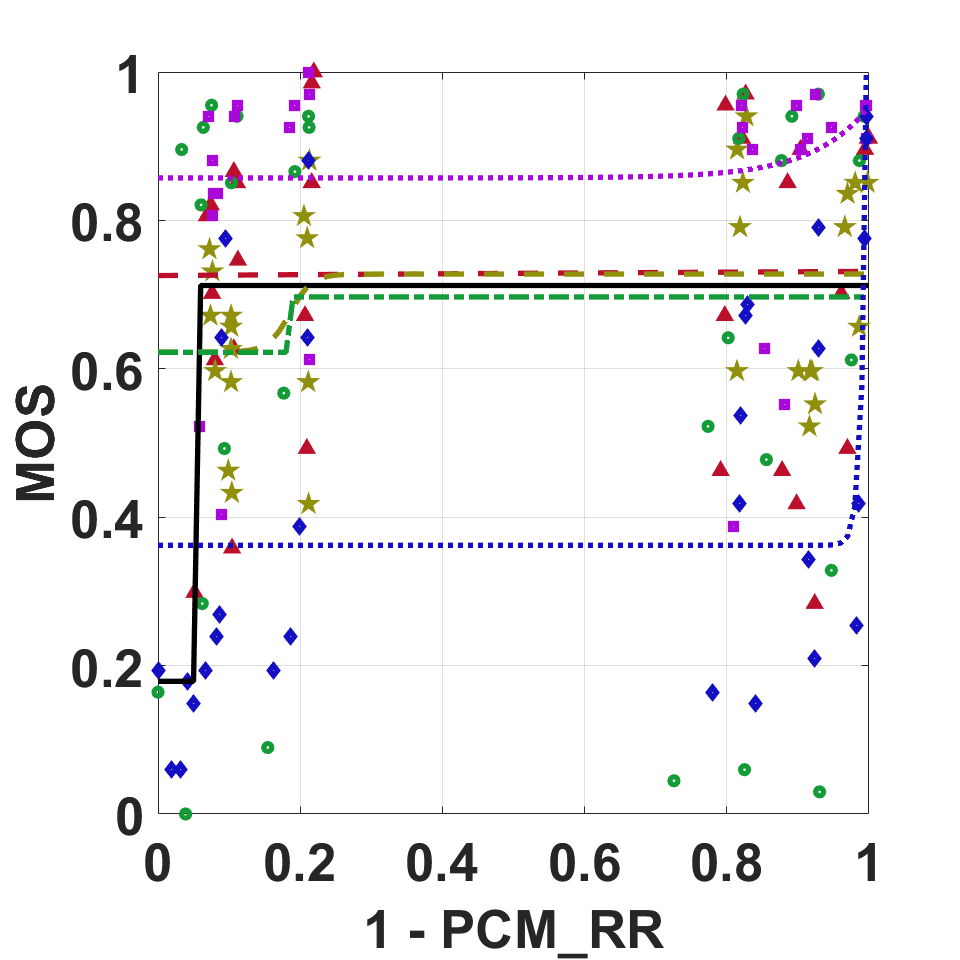}}
    \subfloat[GraphSIM\label{fig:GRAPHSIMGeo}]{%
        \includegraphics[width=0.14\linewidth]{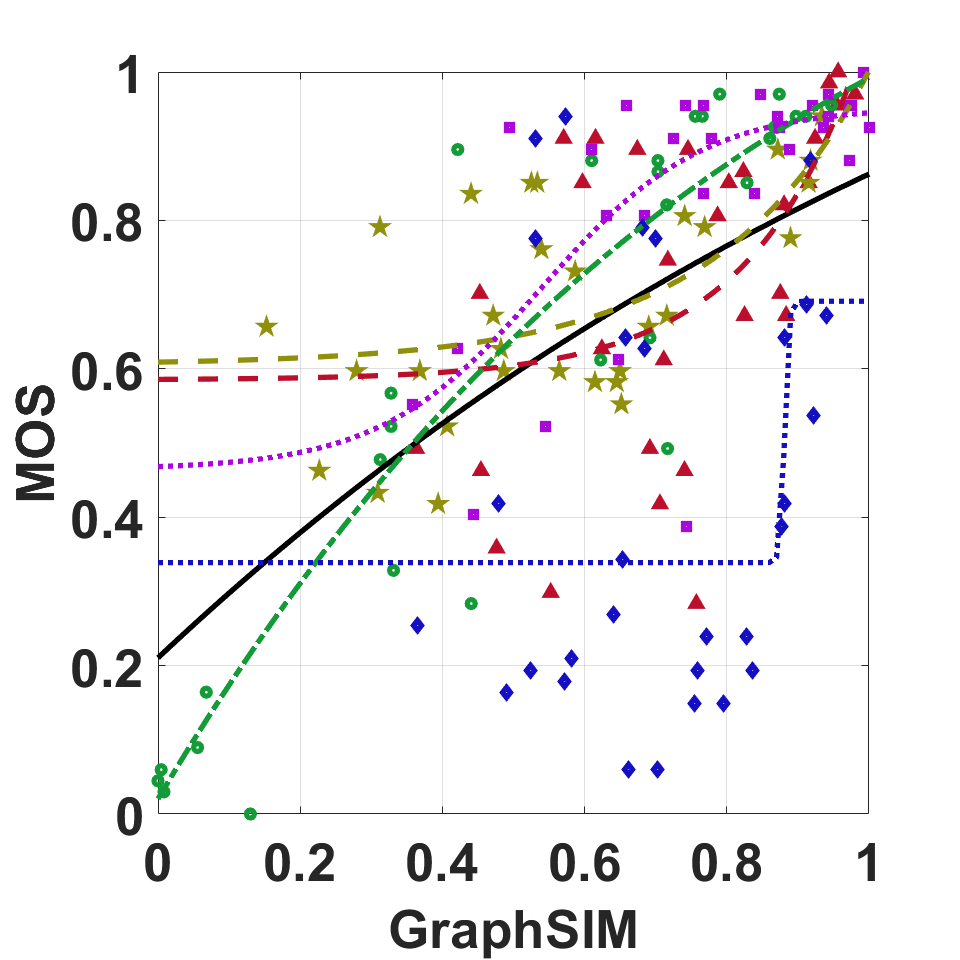}}   
    \caption{Objective metric vs. Evaluation 2 MOS plots, with logistic regression curves (global and for each codec). The symbols are the same as Fig. \ref{fig:metrics}.} 
    \label{fig:metricsGeo}
\end{figure}

To evaluate the performance of each metric, the usual benchmarking procedure~\cite{ITU-T, MarcoColor2016}. MOS predictions for each metric were obtained by logistic regression of the objective scores.
Then, PCC, SROCC, RMSE, and OR were computed to measure the correlation between these and the subjective MOS results, as specified in~\cite{ITU-T}.
For the metrics that depend on surface normals, the Cloud Compare Quadric Fitting with a radius of 5~\cite{QuadricFitting} was used, as this value usually provides the best metric performances~\cite{wg1m89044JPEGPCC}.

Figs. \ref{fig:metrics} and \ref{fig:metricsGeo} show the normalized objective metric vs. normalized MOS plots \textit{Evaluation 1} and \textit{2}, respectively, and Table \ref{tab:metricCorrel} summarizes the corresponding correlation measures for both experiments. In the table, the best values are shown in bold, and the second-best values are shown in italic.

PCQM shows the best correlation with the subjective MOS for \textit{Evaluation 1} (PCC/SROCC of 0.899/0.903), while MSE PSNR D1 shows the best correlation for \textit{Evaluation 2} (PCC/SROCC of 0.834/0.774).
The geometry-only metrics MSE PSNR D1 and MSE PSNR D2 are the only objective metrics with similar performance in the two evaluations, although both PCC and SROCC indicate a poor representation of the subjective quality. Joint objective metrics fail to predict the compression quality when the reference texture information is mapped onto the distorted geometry (\textit{Evaluation 2}), supporting the preference for the encoded texture model used in \textit{Evaluation 1}.

Quite unexpected was the very low performance of PCQM. Even though it was the best performing metric for \textit{Evaluation 1}, it did not achieve the same performance of a previous study \cite{EI2022}. The reference implementation assigns fixed weights to each feature based on a linear-optimization algorithm~\cite{Meynet2020PCQM}, and a point cloud database was used to compute them. The reasons for this behavior might be the fact that those values are not suitable for learning-based codecs.

\section{Performance Stability of Deep Learning-based Codecs}\label{sec:stabExp}

In the following section, the performance stability of the tested DL-based codecs is studied for three different training sessions with similar conditions. The evolution of the codecs throughout the learning process is first analyzed using the \textit{Guanyin}, \textit{Romanoillamp}, and \textit{Citiusp} point clouds and by computing the MSE PSNR D1 at each training epoch.
Finally, the coding performance of the resulting operating points, as well as the publicly available implementations, is assessed using PCQM and MSE PSNR D1, considering the six point clouds referred to in Section \ref{sec:subjExp}.

The three DL-based codecs studied here are trained three times, keeping all the training conditions.
The global loss function depends on the distortion of the encoded point clouds and the encoding bitrate. The encoding bitrate is estimated differently for each codec. PCGCv2 estimates the distortion from the Binary Cross-entropy loss function (BCE), $BCE = -\frac{1}{N} \sum_i (x_i \log{(p_i)} + (1 - x_i \log{(1 - p_i)}))$, where $x_i$ is the true binary occupancy value of voxel $i$, and $p_i$ is its occupancy probability output given by the model. PCC GEO CNNv2 and ADLPCC use a focal variation of the BCE to address imbalances between empty and occupied voxels in more sparse point clouds, defined as:

\begin{equation}\label{eq:focalLossADLPCC}
       \begin{cases}
     -\alpha(1-x)^\gamma \log p, & x=1 \\
     -(1-\alpha)x^\gamma \log(1-p), & x=0
    \end{cases}
\end{equation}

\subsection{PCGCv2 Model Training}

The PCGCv2 codec implementation and the training datasets are available online\footnote{available at https://github.com/NJUVISION/PCGCv2}. The model was trained with densely sampled data from the ShapeNet database~\cite{shapenet}. The final training set was obtained by random rotation and quantization with 7-bit precision and a randomized number of points.

In the original paper, different coding bitrates are targeted by varying the rate-distortion tradeoff parameter $\lambda$ between 0.75 and 16. The code made available defines the global loss function as, $J = \alpha D + \beta R$
where D is the distortion and $R$ is the coding bitrate. In this experiment, three different $\lambda$ values were tested, namely $\lambda=\{16, 4, 0.75\}$. The $\beta$ parameter was kept with the value 1, so that we have $J = \lambda D + R$.

For faster convergence, the learned weights with $\lambda=16$ were used to initialize the training for both $\lambda=4$ and $\lambda=0.75$, as recommended in~\cite{Jianqiang-PCGCv2}. For each $\lambda$, the model was trained for 50 epochs with a constant learning rate of $10^{-5}$. This training process was repeated three times under similar conditions.

\begin{figure}[t!]
 \centering
    \begin{subfigure}[t]{0.32\textwidth}
        \centering
        \includegraphics[width=\textwidth]{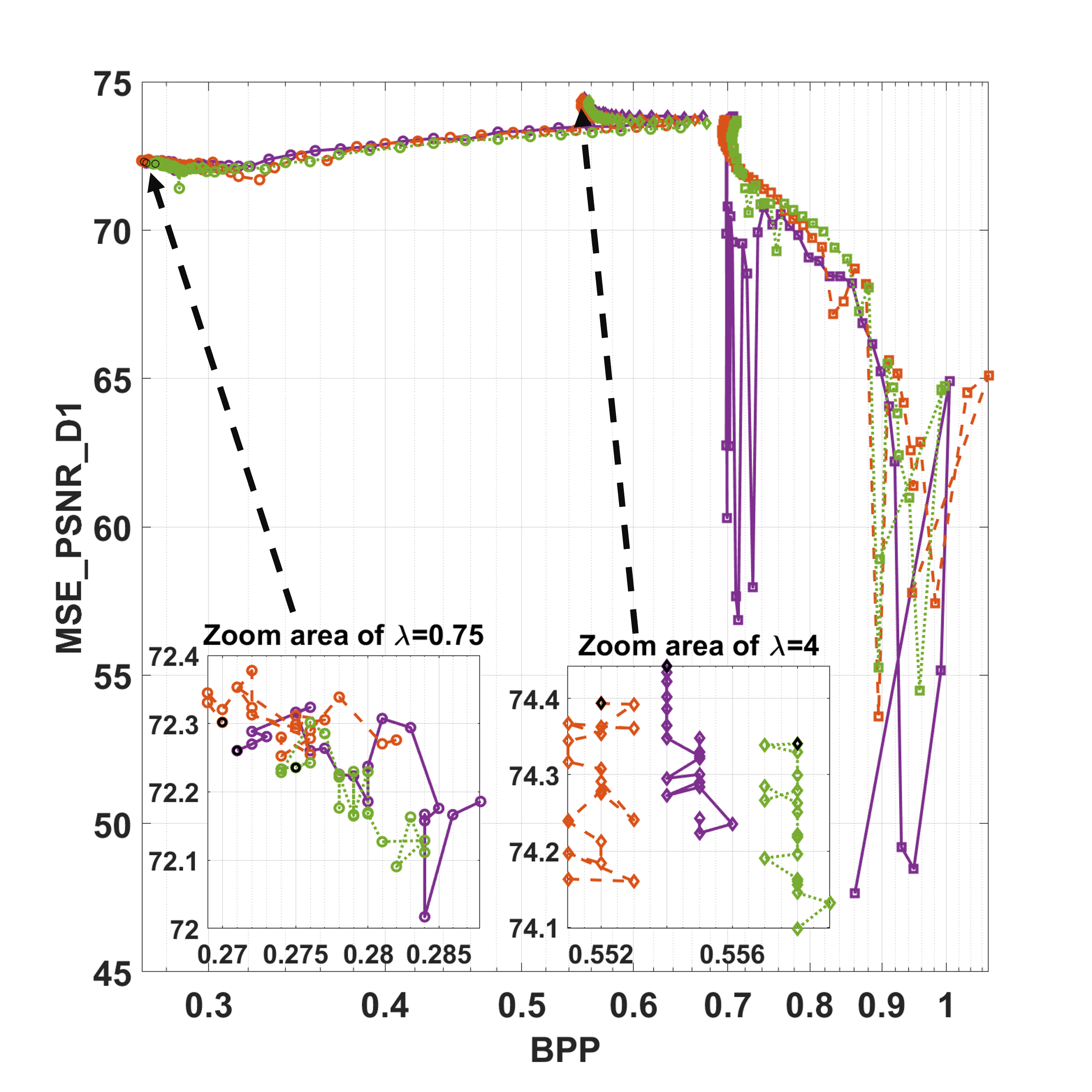}
        \caption{\textit{Guanyin}}\label{fig:trainingPlots_ABX_guanyin}
    \end{subfigure}
    \begin{subfigure}[t]{0.32\textwidth}
        \centering
        \includegraphics[width=\textwidth]{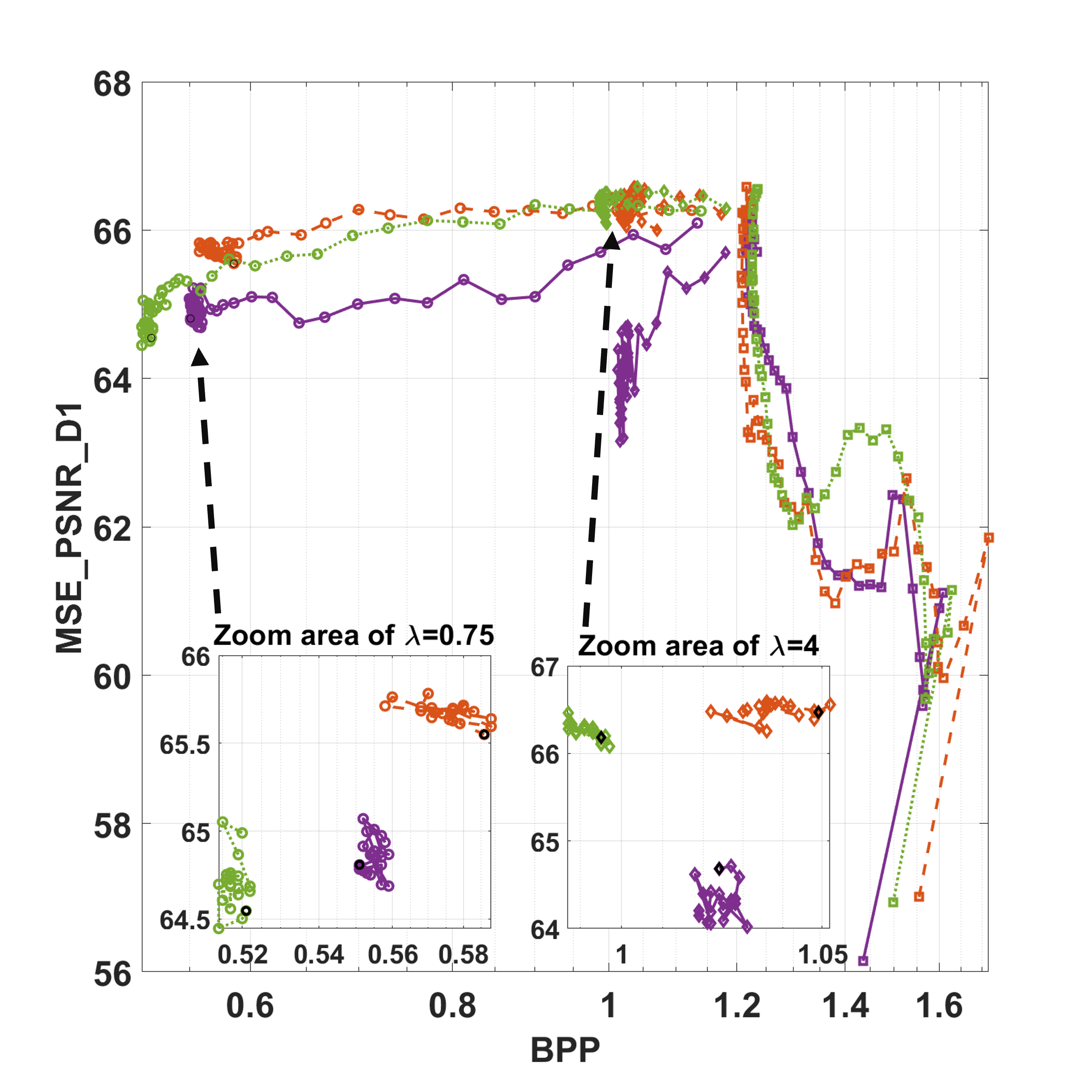}
        \caption{\textit{Romanoillamp}}
    \end{subfigure}
   \begin{subfigure}[t]{0.32\textwidth}
        \centering
        \includegraphics[width=\textwidth]{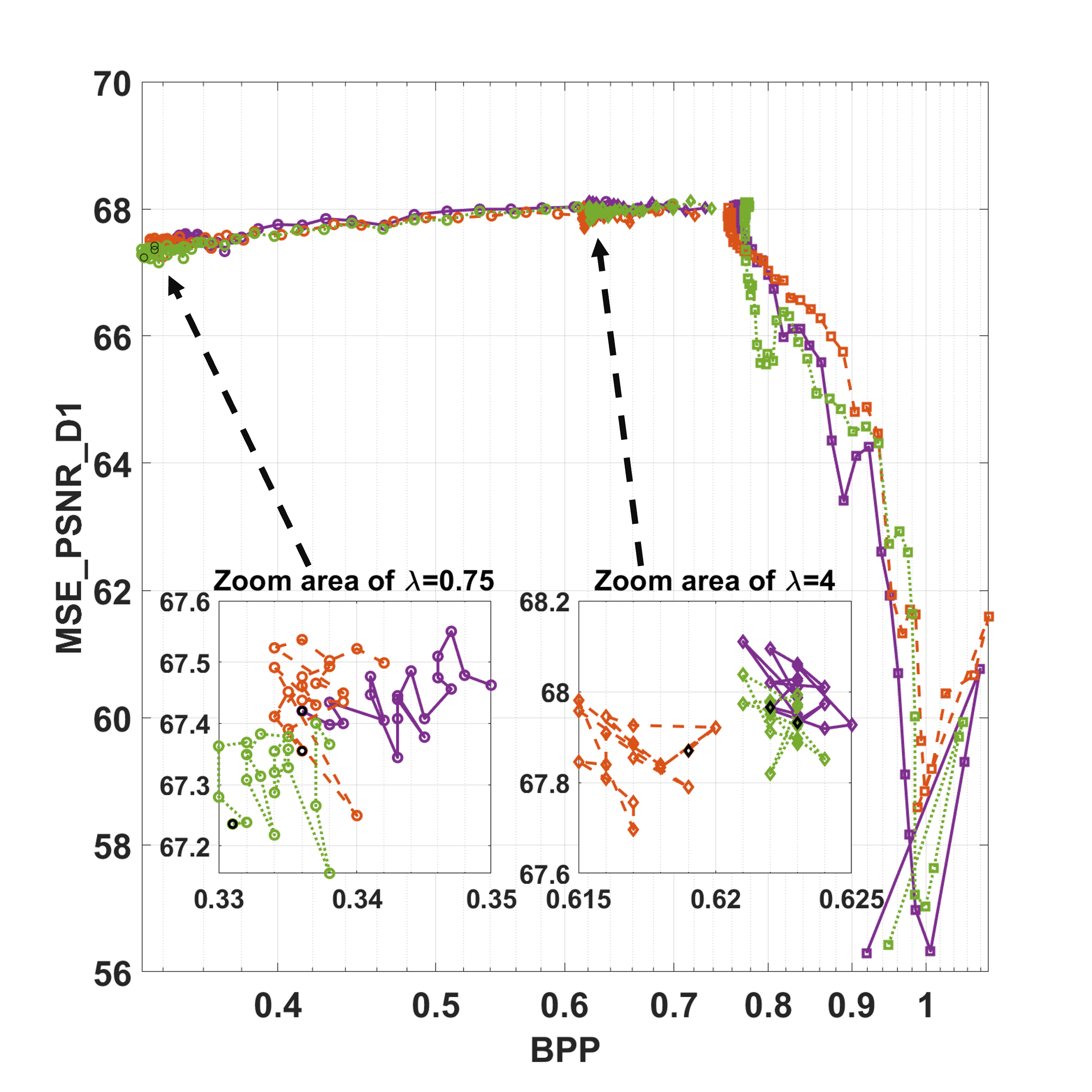}
        \caption{\textit{Citiusp}}\label{fig:trainingPlots_ABX_citiusp}
    \end{subfigure}
    \includegraphics[width=\textwidth]{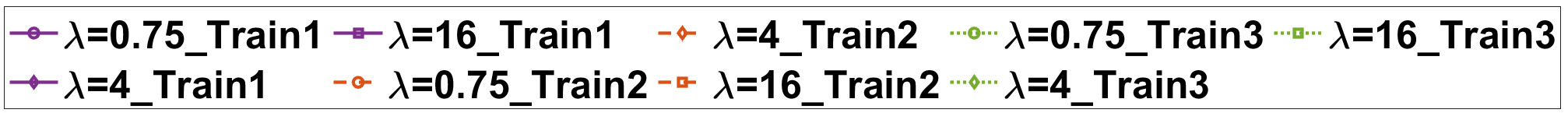}
    \caption{MSE PSNR D1 vs. bpp plots for PCGCv2, trained with $\lambda = \{16, 4, 0.75\}$.}
    \label{fig:trainingPlots_ABX}
 \end{figure}

The evolution of MSE PSNR D1 vs. the coding bitrate throughout the model training is shown in Fig. \ref{fig:trainingPlots_ABX}. The zoomed areas show the final epochs of each Rate-Distortion tradeoff. PCGCv2 reveals a good level of stability for most point clouds, notably \textit{Citiusp}.
The MSE PSNR D1 metric shows similar behavior across the three training sessions. However, even in these cases, some instability may be observed, for example, in intermediate epochs with $\lambda=16$, particularly at higher encoding bitrates for \textit{Guanyin} \textit{Citiusp} (Figs. \ref{fig:trainingPlots_ABX_guanyin} and \ref{fig:trainingPlots_ABX_citiusp}).
The encoding process of \textit{Romanoillamp} point cloud shows an extremely high level of instability.
It can also be observed that in this particular case, the training results do not converge to a stable operating point, except for $\lambda=16$.
The codec reveals a good level of stability regarding the encoding performance of the other tested point clouds.
The bitrates converge to similar operating points across all training sessions.
In conclusion, the codec usually reveals a good level of stability, but there is a possibility that other training data will produce situations like \textit{Romanoillamp}, where the performance of the codec will depend on the training.
 
\subsection{PCC GEO CNNv2 Model Training}\label{sec:GEOCNNTRAIN}

The authors of PCC GEO CNNv2 train four individual models for each Rate-Distortion tradeoff given by $ J = \lambda D + R$~\cite{quach2020improved}. They chose four values for $\lambda$, notably $3\times10^{-4}$, $10^{-4}$, $5\times10^{-5}$, $2\times10^{-5}$. In the provided implementation\footnote{available at https://github.com/mauriceqch/pcc\_geo\_cnn\_v2}, an additional value is considered, $\lambda$ = $10^{-5}$, which was also included here. This experiment followed the sequential training approach in~\cite{quach2020improved}, with successively decreasing values of $\lambda$. The trained weights for $\lambda_{i-1}$ were used to initialize the training for $\lambda_{i}$).

The models were trained on a subset of the ModelNet40~\cite{modelnet} dataset. First, the mesh data is voxelized with a resolution of $512\times512\times512$, and the 200 largest point clouds are selected. Then, the point clouds are divided into blocks with a resolution of $64\times64\times64$, and the 4000 largest blocks are selected. For each value of $\lambda$, the model was trained for 500 steps, with early stopping if the loss did not improve for more than 4 validation steps.

\begin{figure}[t!]
 \centering
    \begin{subfigure}[t]{0.32\textwidth}
        \centering
        \includegraphics[width=\textwidth]{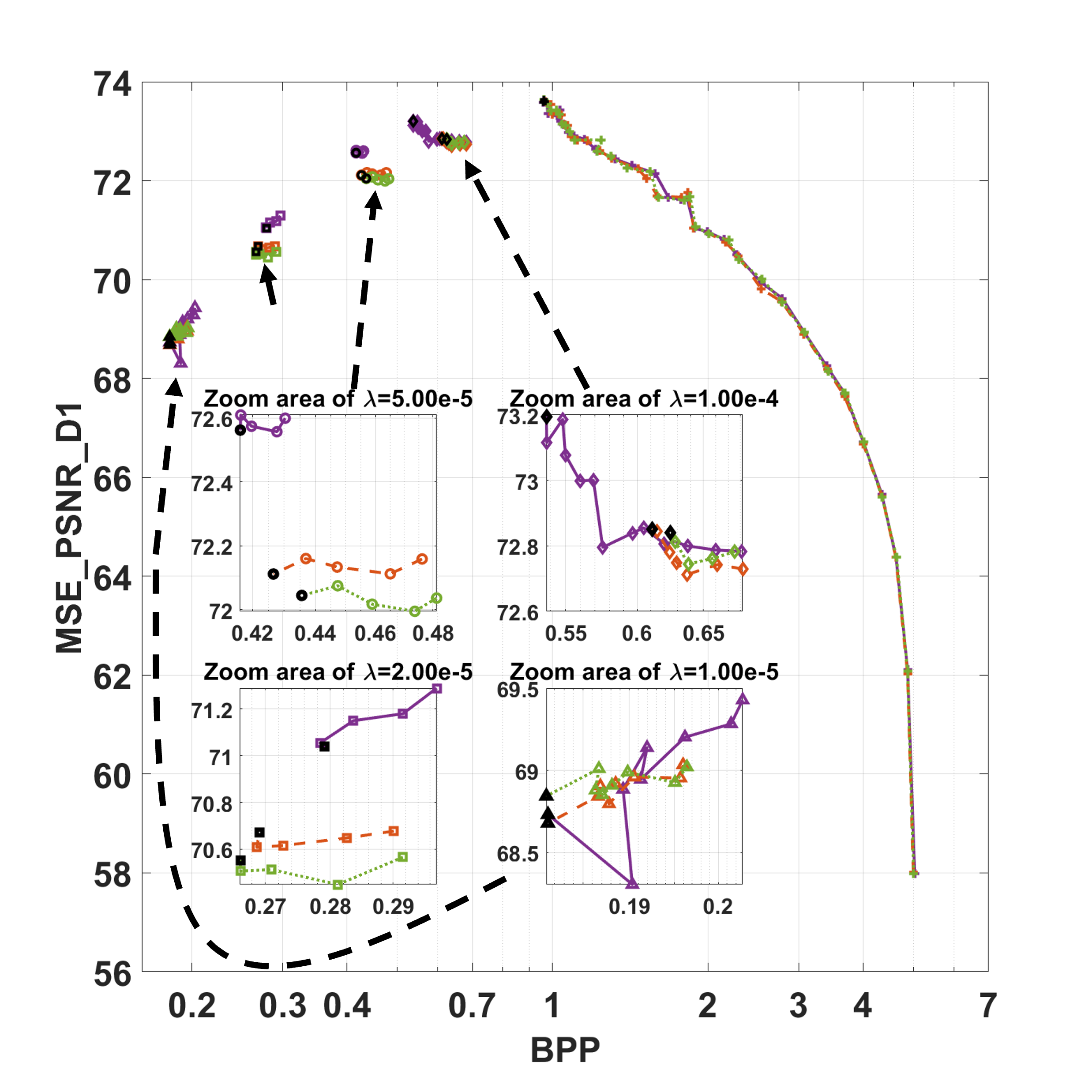}
        \caption{\textit{Guanyin}}
    \end{subfigure}
    \begin{subfigure}[t]{0.32\textwidth}
        \centering
        \includegraphics[width=\textwidth]{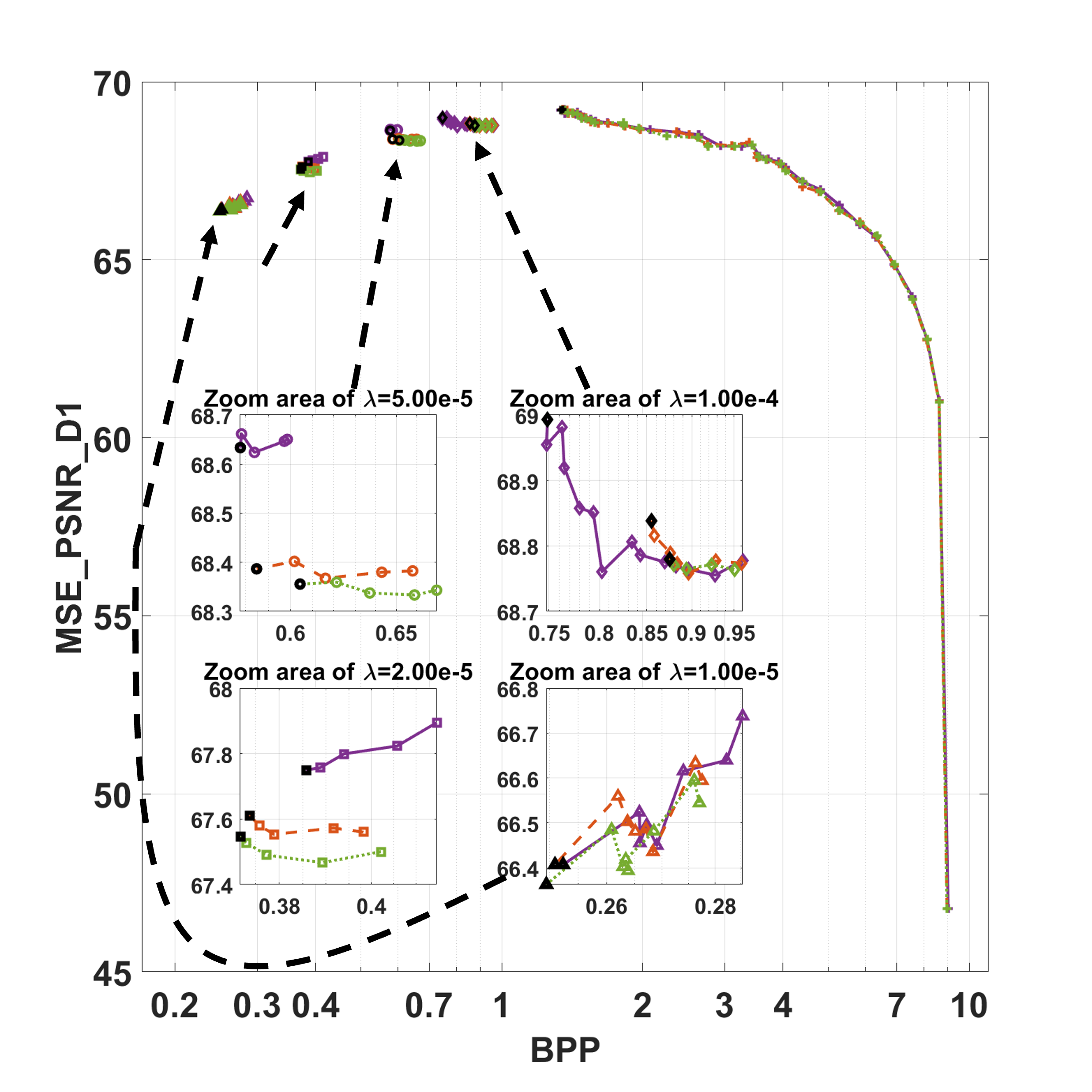}
        \caption{\textit{Romanoillamp}}
    \end{subfigure}
   \begin{subfigure}[t]{0.32\textwidth}
        \centering
        \includegraphics[width=\textwidth]{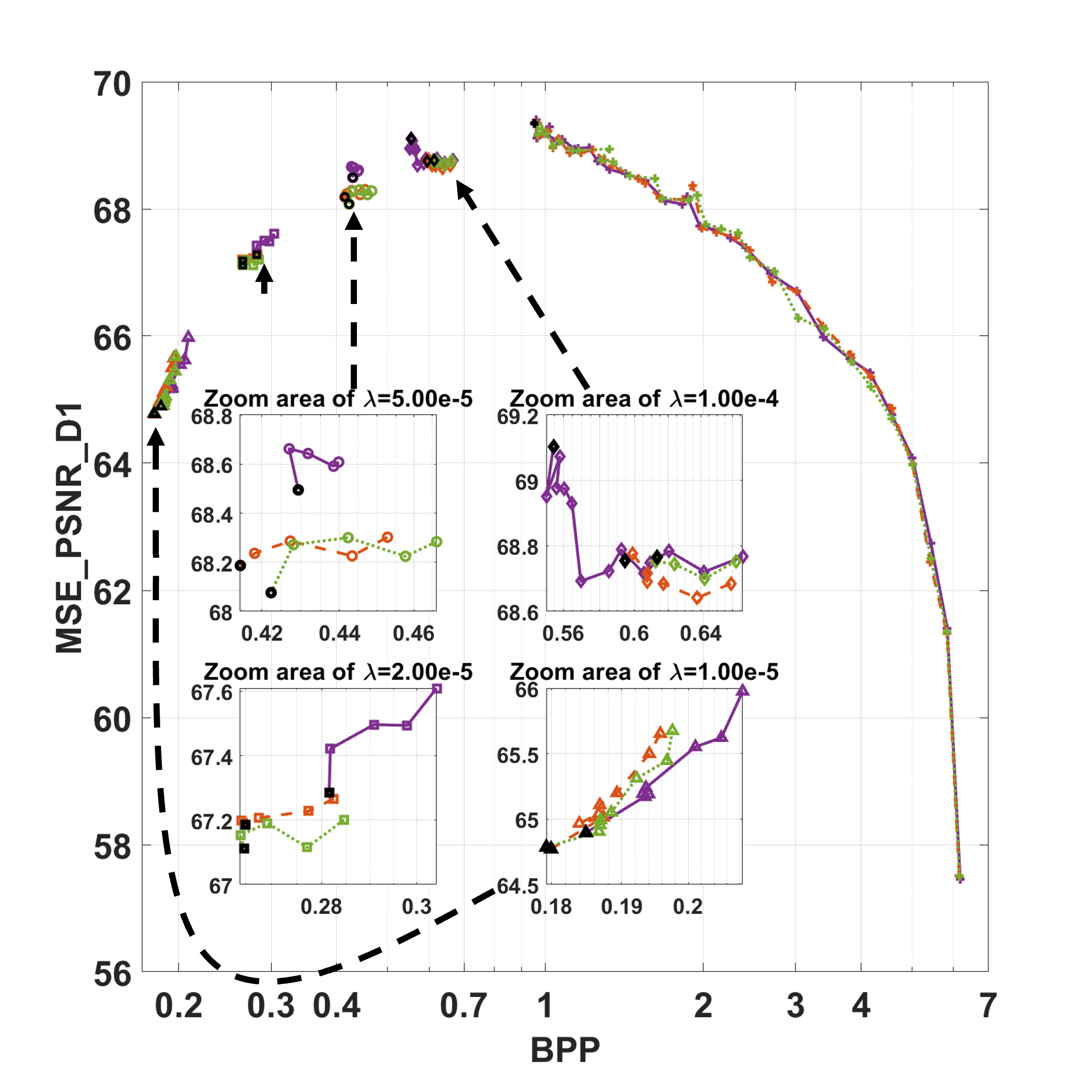}
        \caption{\textit{Citiusp}}
    \end{subfigure}
    \includegraphics[width=\textwidth]{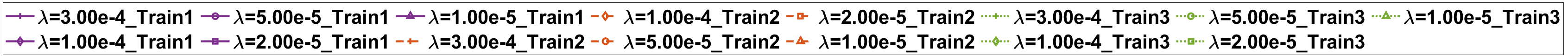}
    \caption{MSE PSNR D1 vs. bpp plots for PCC GEO CNNv2, trained with $\lambda = \{3\times10^{-4}, 10^{-4}, 5\times10^{-5}, 2\times10^{-5}, 10^{-5} \}$.}
    \label{fig:trainingPlotsGEO}
 \end{figure}

Fig. \ref{fig:trainingPlotsGEO} shows the evolution of MSE PSNR D1 for three training sessions of PCC GEO CNNv2. The zoomed areas show the final epochs of each Rate-Distortion tradeoff. This codec shows a very high level of stability. However, for intermediate rates, i.e., with $\lambda = 1\times10^{-4}$, $5\times10^{-5}$, and $2\times10^{-5}$, the MSE PSNR D1 values of the final resulting models are slightly different across training sessions. In practice, no case in which the final coding result is highly dependent on the training session was identified.

\subsection{ADLPCC Model Training}

The global loss function of ADLPCC\footnote{https://github.com/aguarda/ADLPCC} is given by
$ J =  D + \lambda R$, where the coding rate $R$ is estimated during training as the summed entropy of its autoencoder and variational autoencoder latent representations.

In order to obtain several Rate-Distortion tradeoff points, different $\lambda$ values are considered, thus varying the weight of the rate. The model was trained with a dataset consisting of JPEG and MPEG point clouds~\cite{GuardaDL}, with $\lambda=\{500, 900, 1500, 5000, 20000\}$. For each value of $\lambda$, the codec was trained with $\alpha=\{0.5, 0.6, 0.7, 0.8, 0.9\}$, which is a parameter of the BCE focal loss function (Eq. \ref{eq:focalLossADLPCC}), which allows choosing the best performing model considering the characteristics of the point cloud, such as its sparsity.

\begin{figure}[t!]
 \centering
    \begin{subfigure}[t]{0.32\textwidth}
        \centering
        \includegraphics[width=\textwidth]{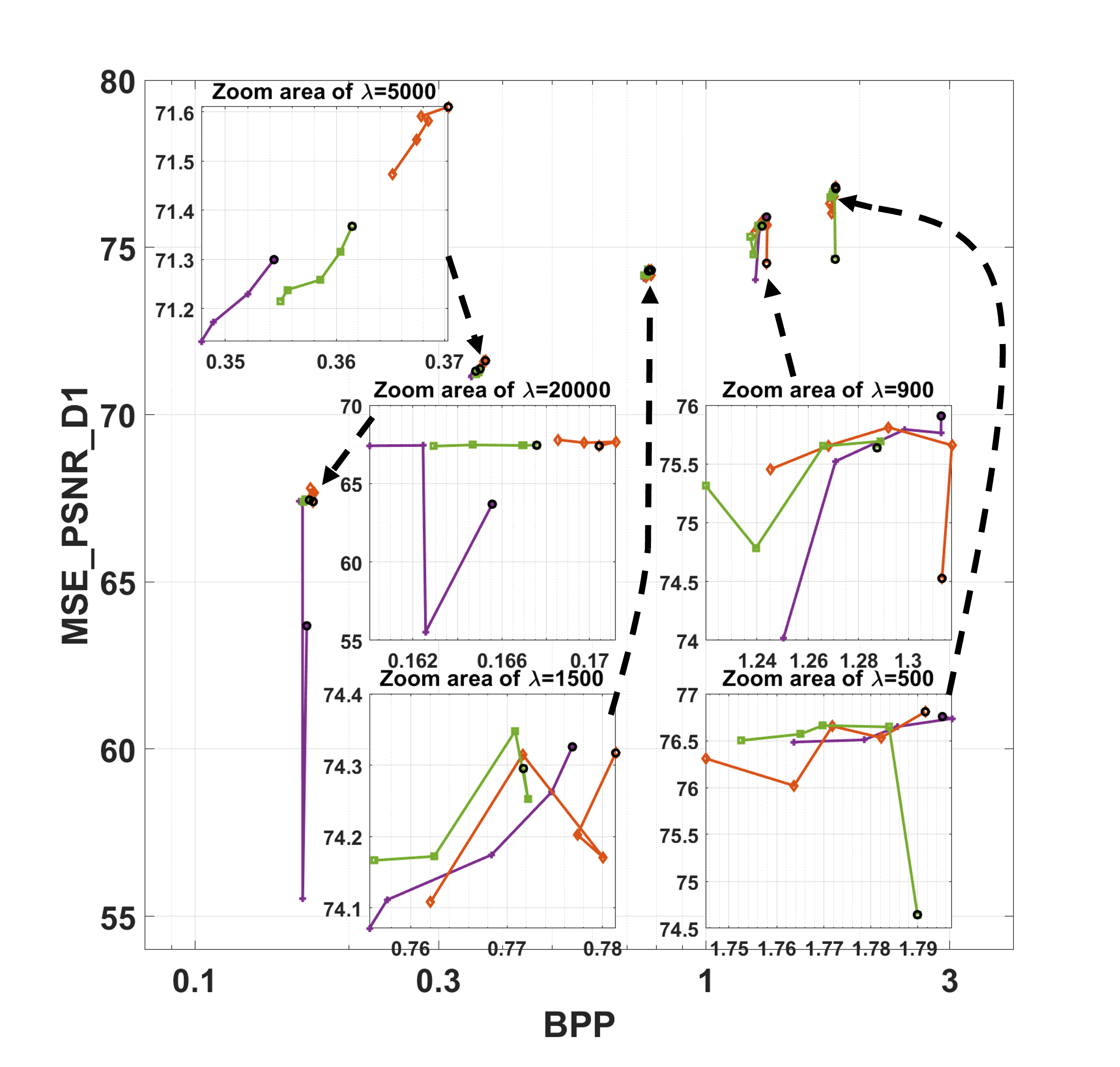}
        \caption{\textit{Guanyin}}
    \end{subfigure}
    \begin{subfigure}[t]{0.32\textwidth}
        \centering
        \includegraphics[width=\textwidth]{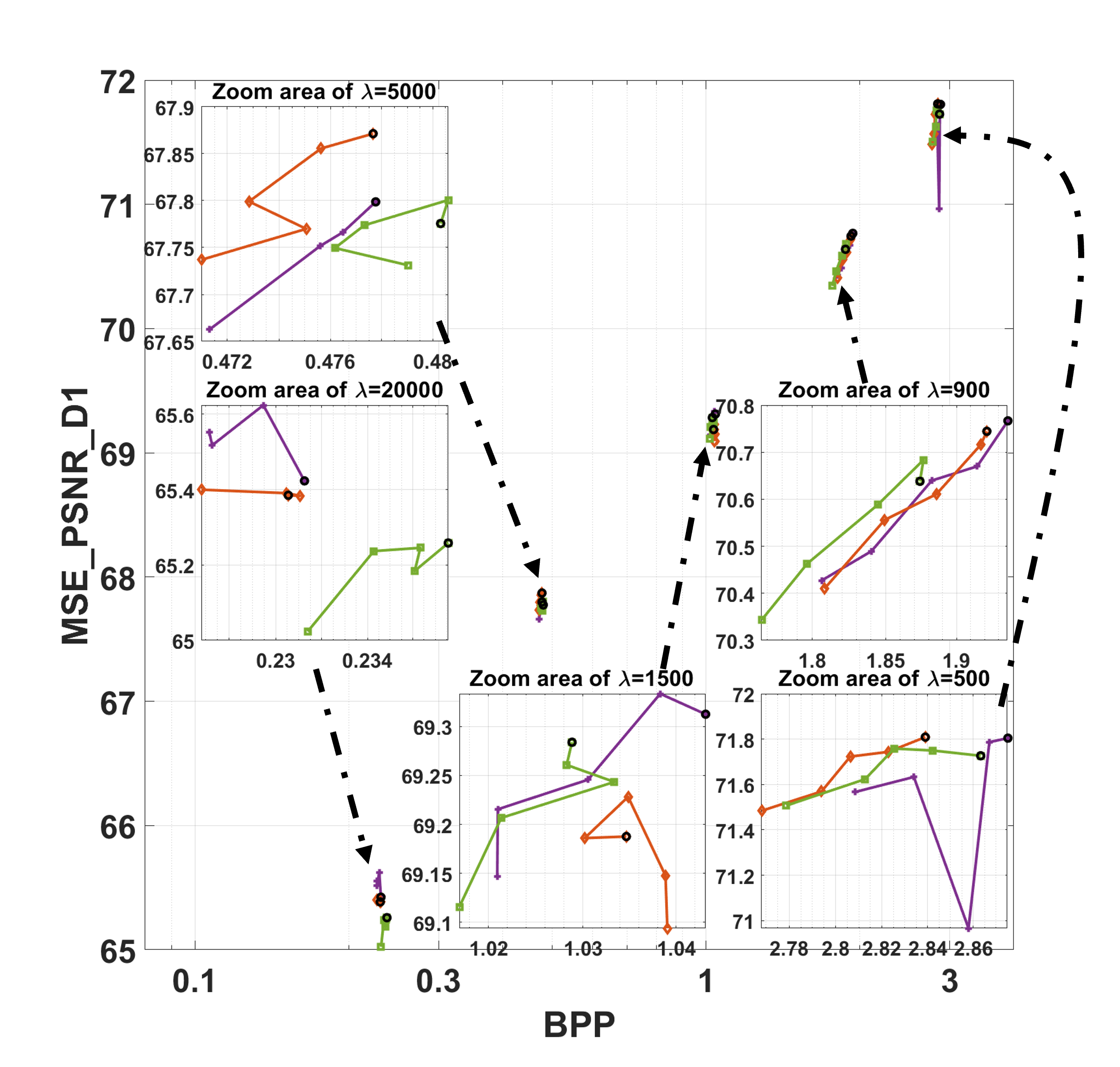}
        \caption{\textit{Romanoillamp}}
    \end{subfigure}
   \begin{subfigure}[t]{0.32\textwidth}
        \centering
        \includegraphics[width=\textwidth]{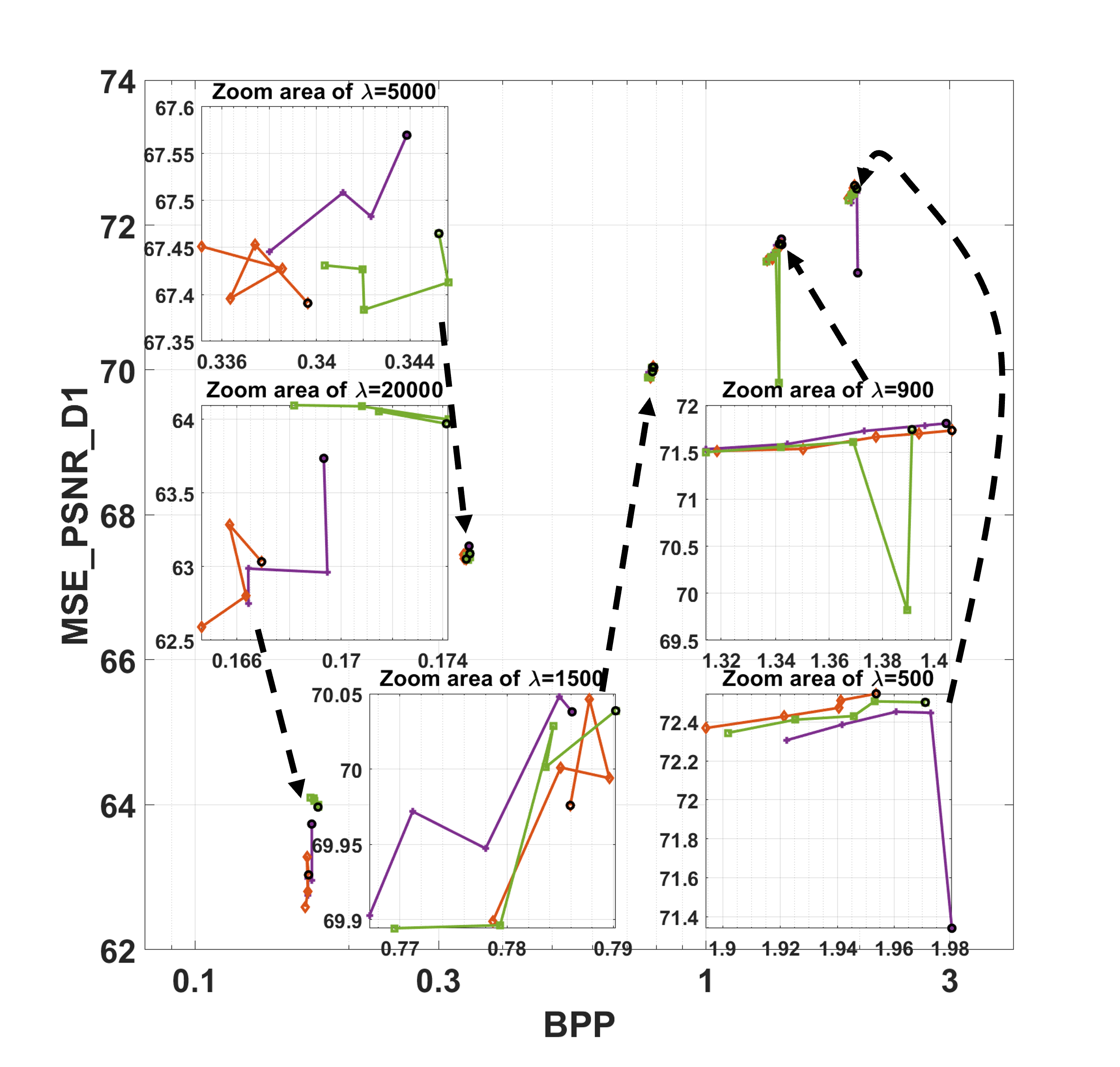}
        \caption{\textit{Citiusp}}
    \end{subfigure}
    \includegraphics[width=\textwidth]{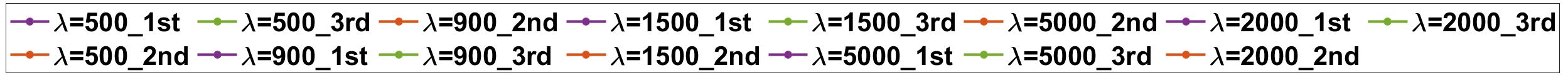}
    \caption{MSE PSNR D1 vs. bpp plots for ADLPCC, trained with $\lambda = \{500, 900, 1500, 5000, 20000\}$.}
    \label{fig:trainingPlotsADLPCC}
 \end{figure}

Fig. \ref{fig:trainingPlotsADLPCC} shows the plots for each epoch of the ADLPCC codec across training. The zoomed areas show the final epochs of each Rate-Distortion tradeoff. The codec shows a high degree of stability, as most encoding steps show little variation across the epochs. One notable exception should be noted, namely for the \textit{Guanyin} point cloud when trained with $\lambda = 20000$. The first train revealed a sudden drop in one of the intermediate epochs. The rest of the training process for that specific $\lambda$ showed little variation in quality across bitrates.

\begin{figure}[t!]
\begin{center}      
    \subfloat[\textit{Longdress}]{%
        \includegraphics[width=0.17\linewidth]{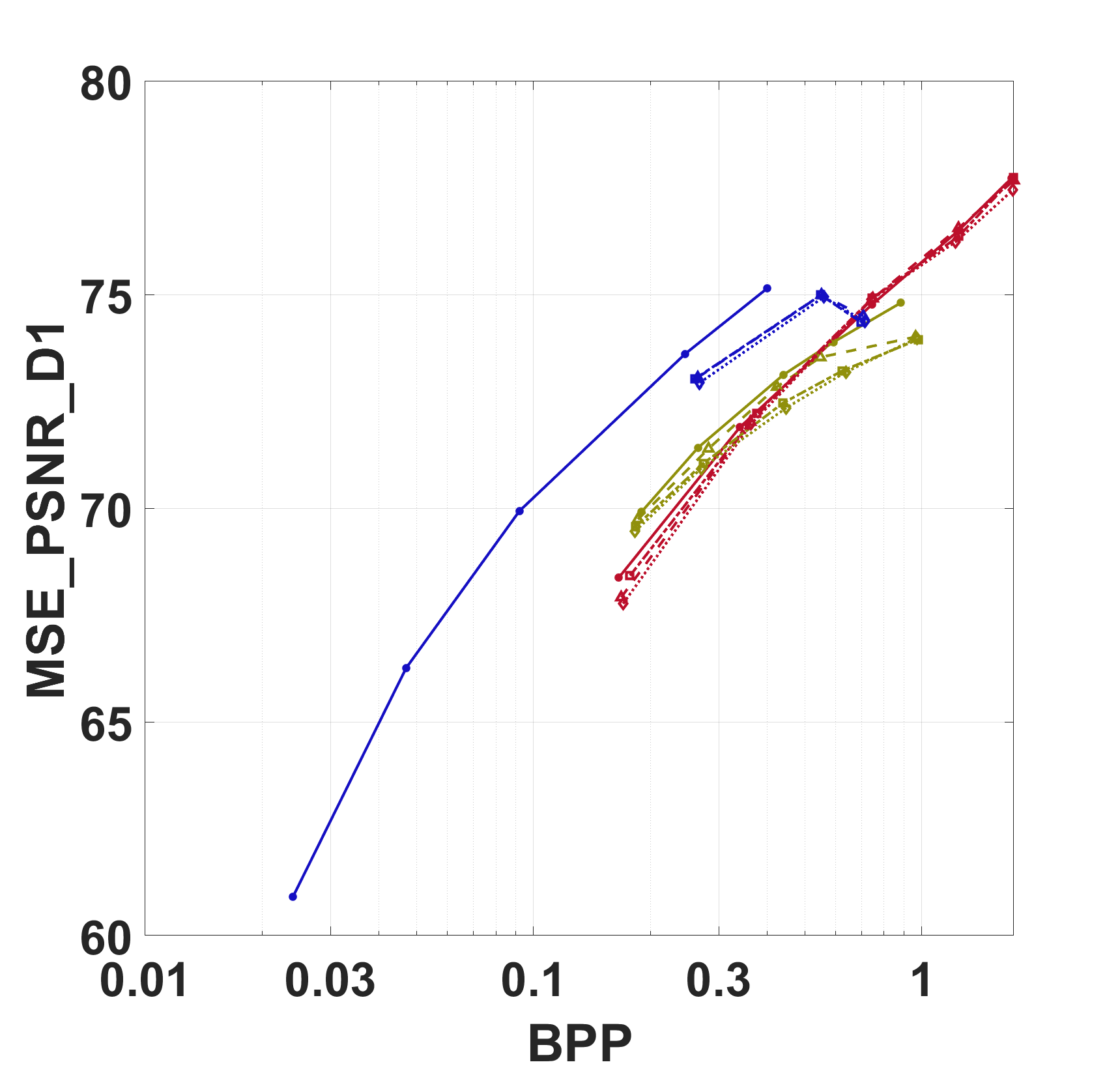}}
    \subfloat[\textit{Guanyin}]{%
        \includegraphics[width=0.17\linewidth]{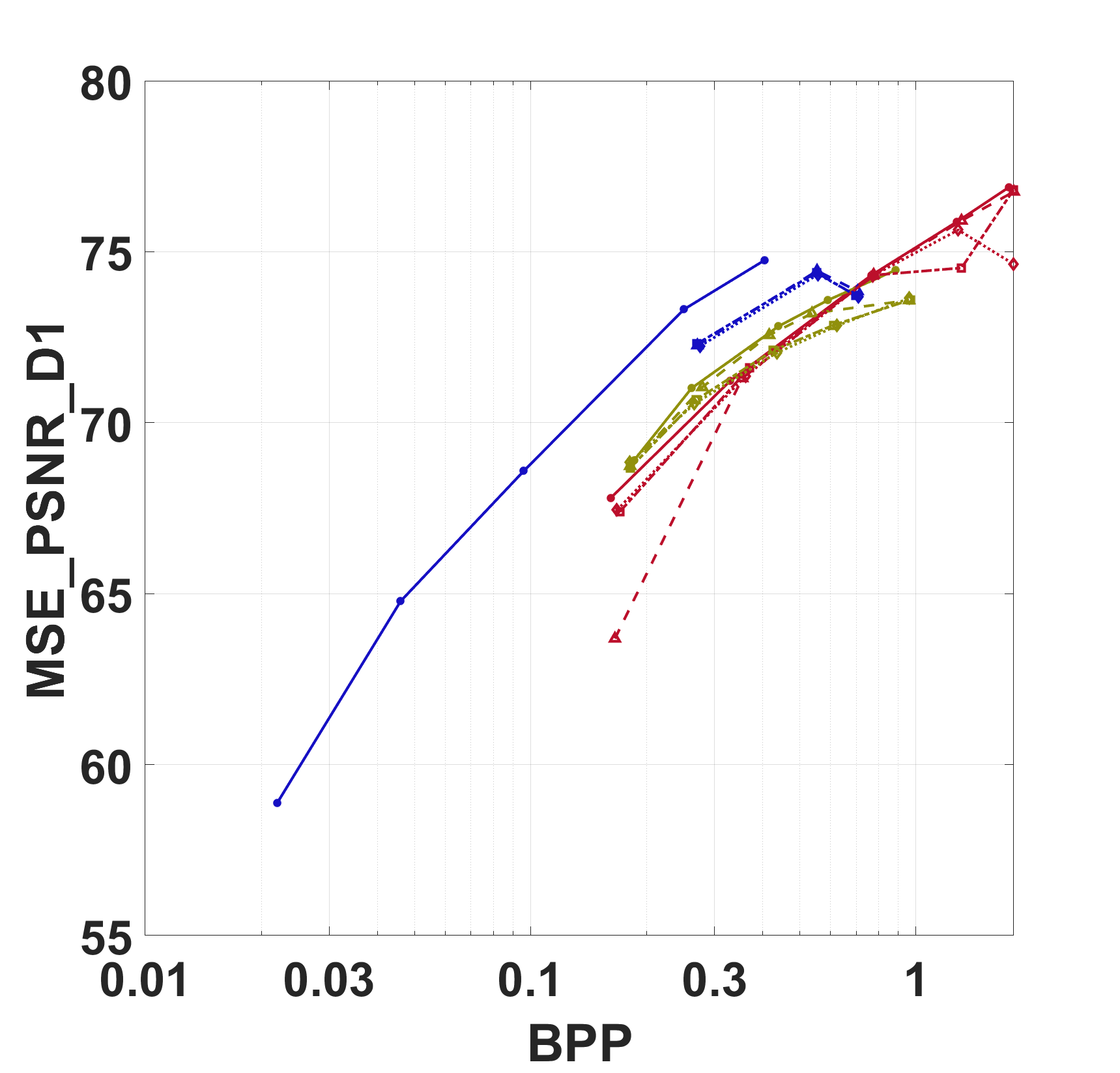}}
    \subfloat[\textit{Romanoillamp}]{%
        \includegraphics[width=0.17\linewidth]{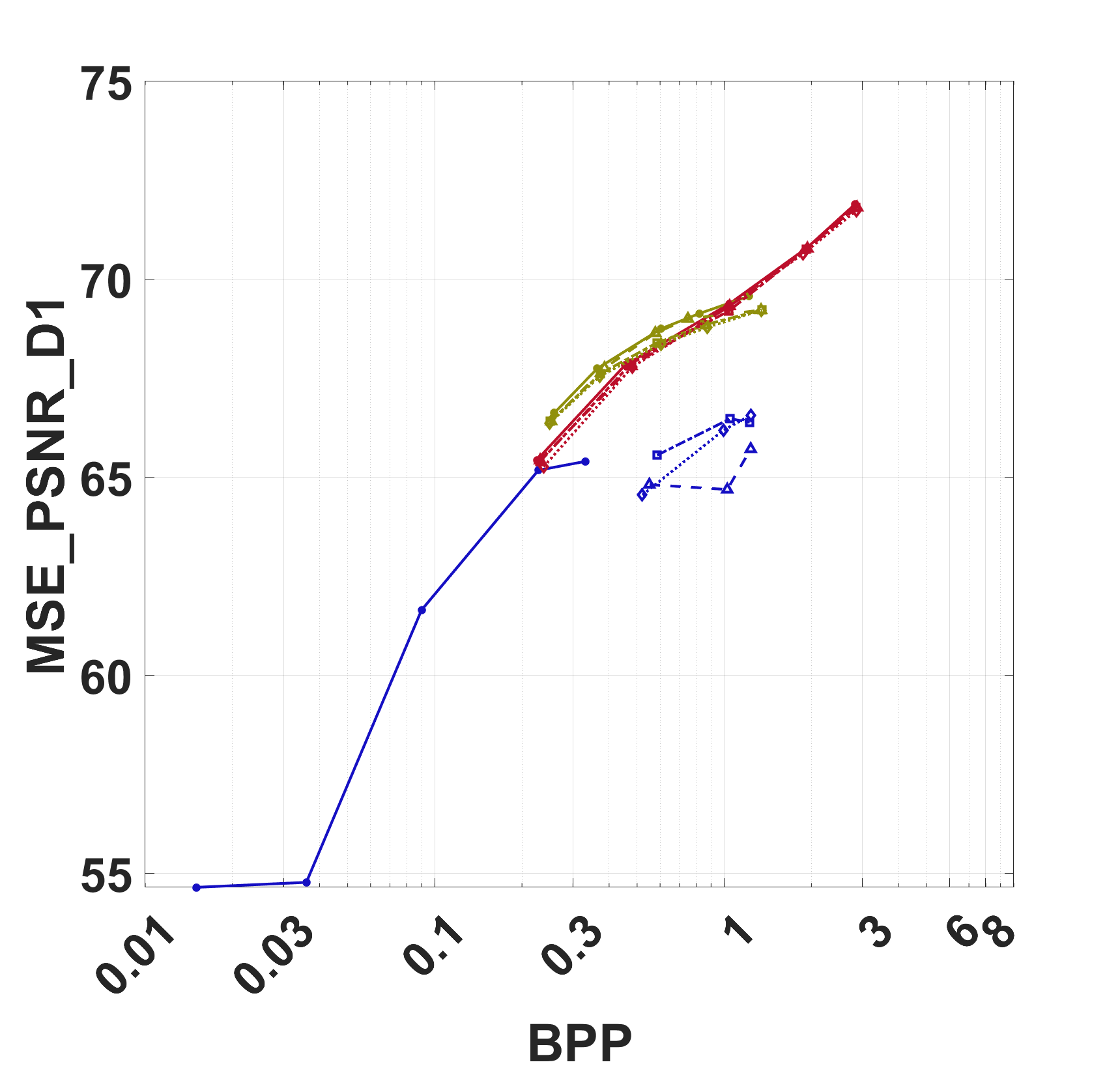}}
    \subfloat[\textit{Citiusp}]{%
        \includegraphics[width=0.17\linewidth]{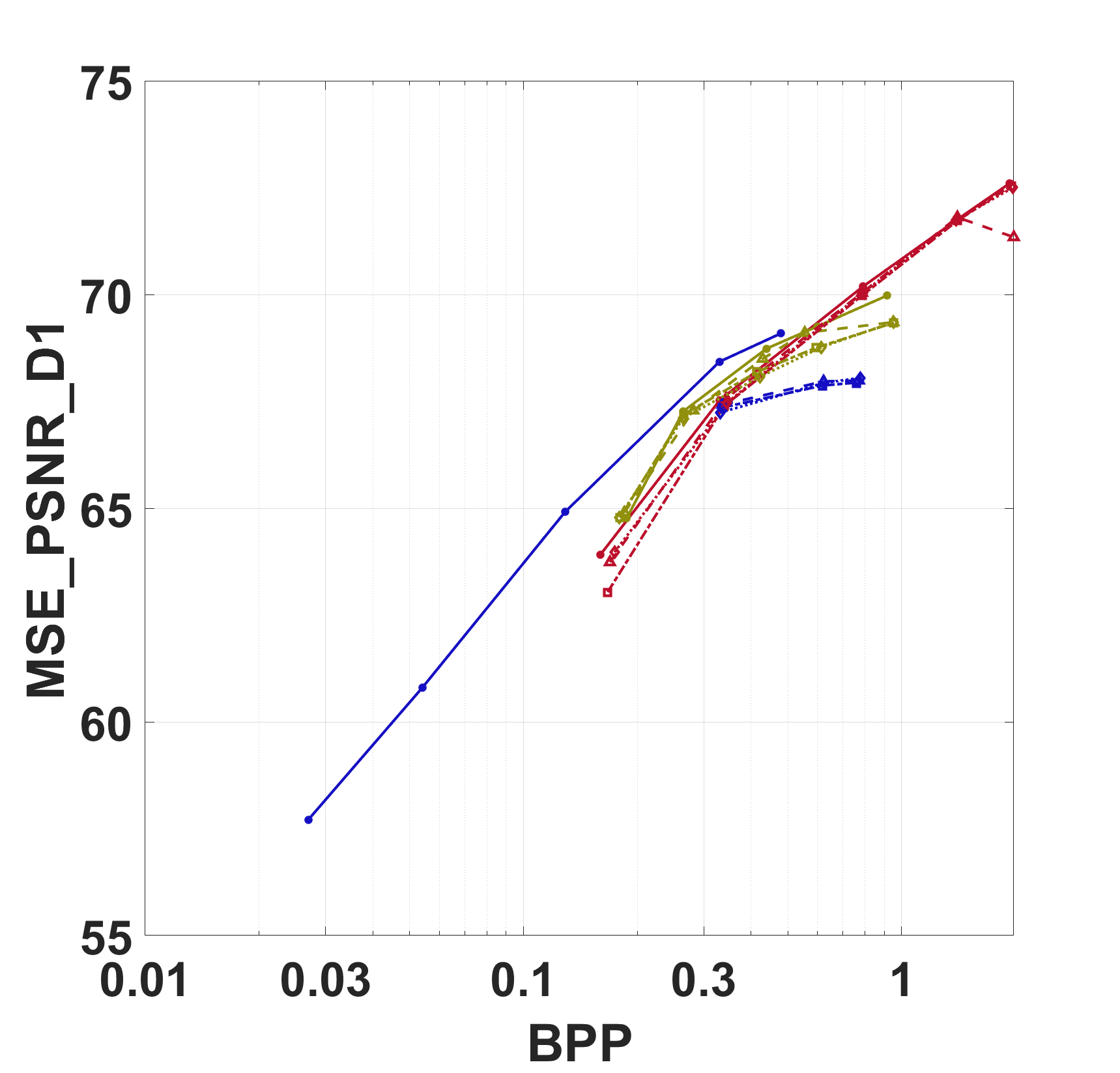}}
    \subfloat[\textit{IpanemaCut}]{%
        \includegraphics[width=0.17\linewidth]{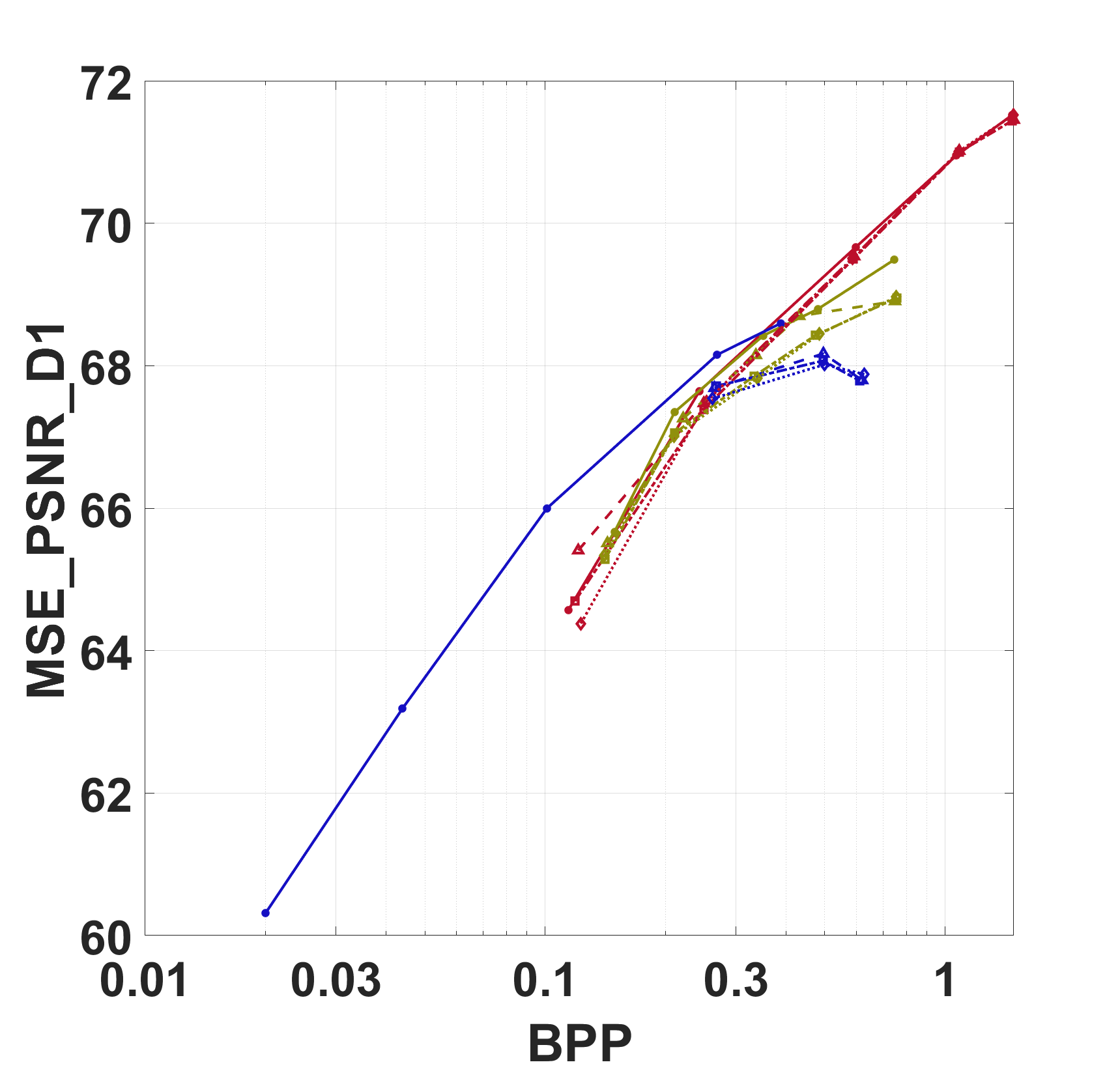}}
    \subfloat[\textit{Ramos}]{%
        \includegraphics[width=0.17\linewidth]{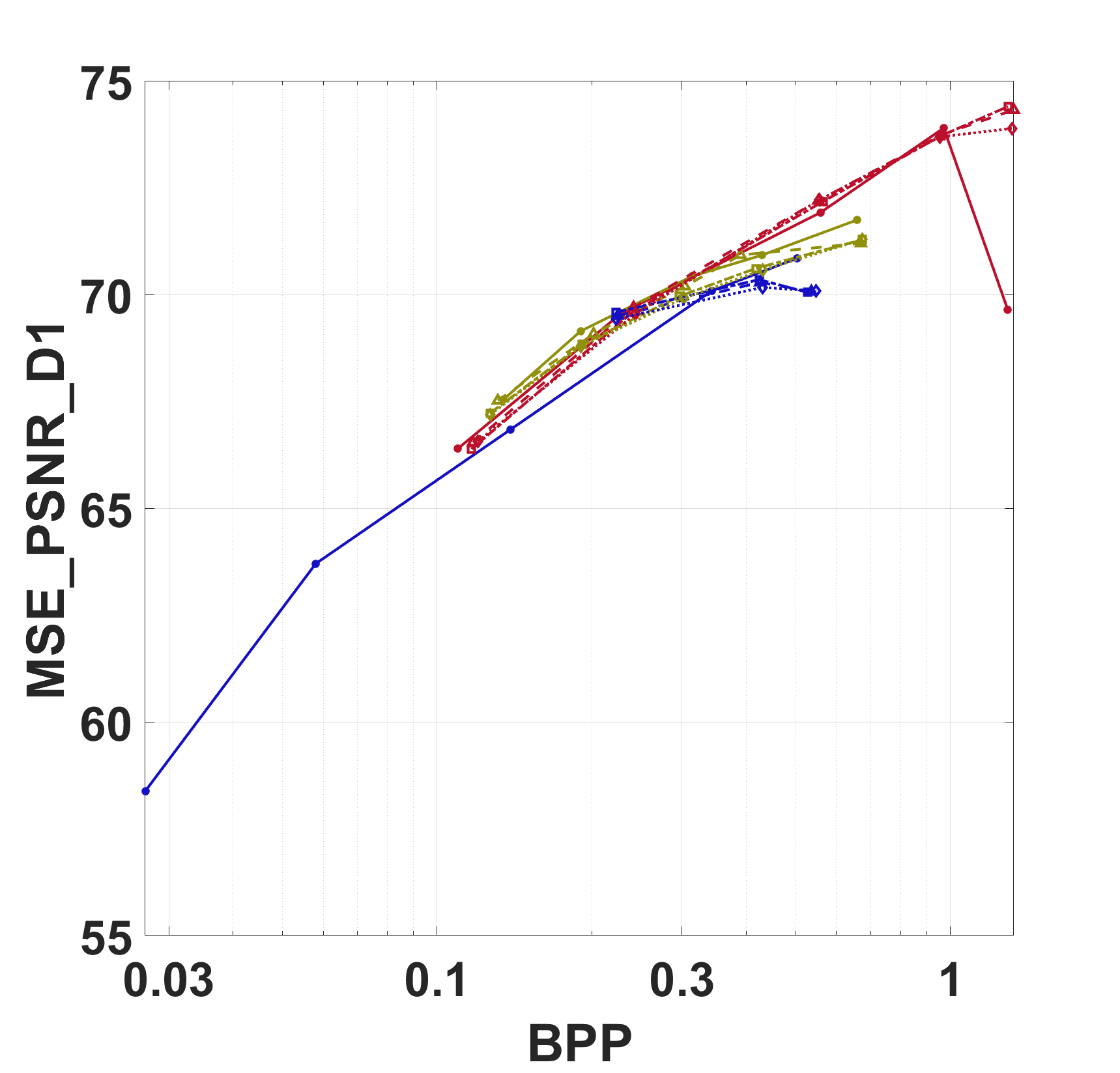}}
        \\
        \caption{MSE PSNR D1 plots for each of the defined operating points for each codec.}
        \label{finalEpochs}
        \end{center}
\end{figure}
\begin{figure}
\begin{center}      

    \subfloat[\textit{Longdress}]{%
         \includegraphics[width=0.17\linewidth]{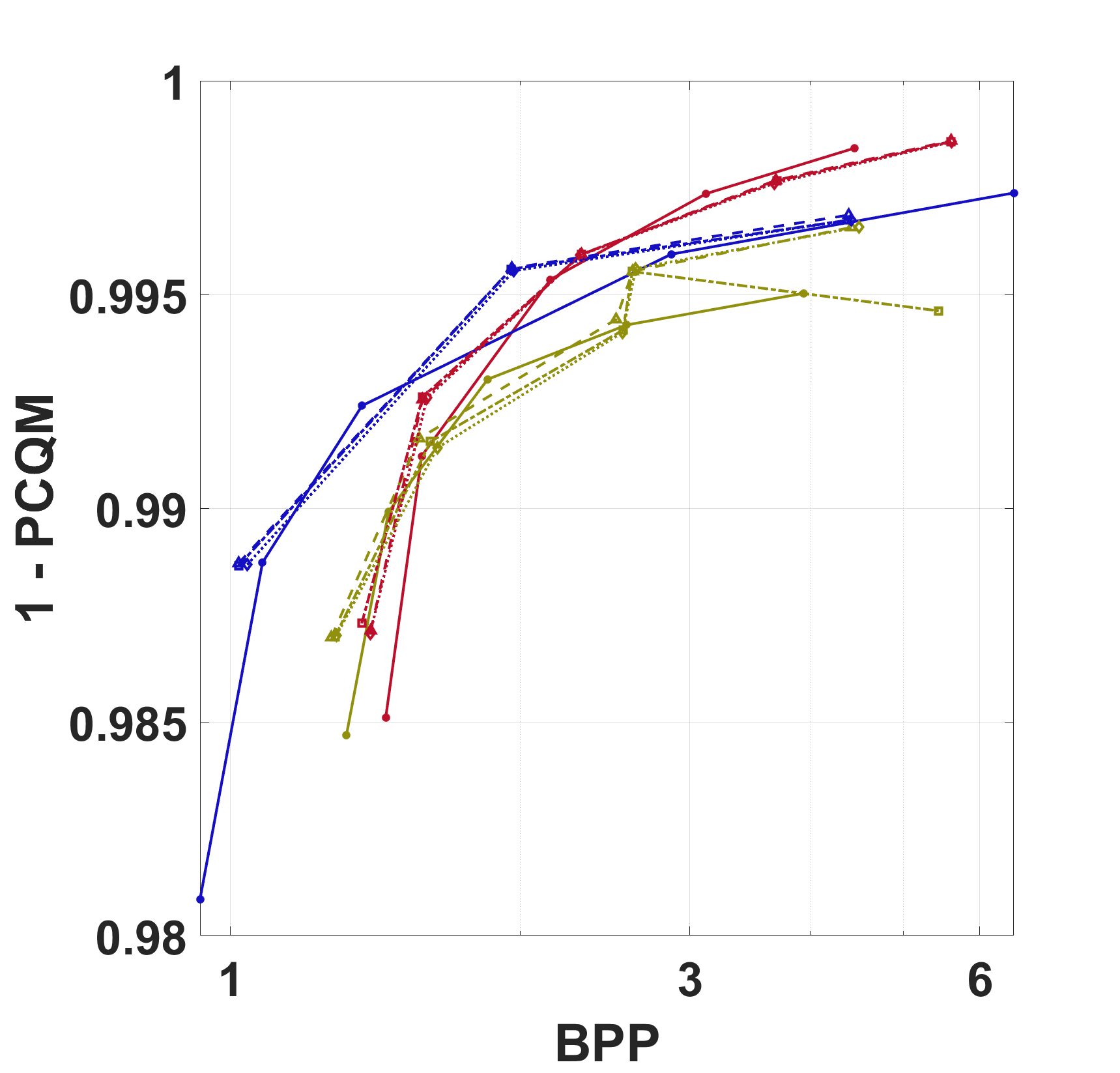}}
    \subfloat[\textit{Guanyin}]{%
        \includegraphics[width=0.17\linewidth]{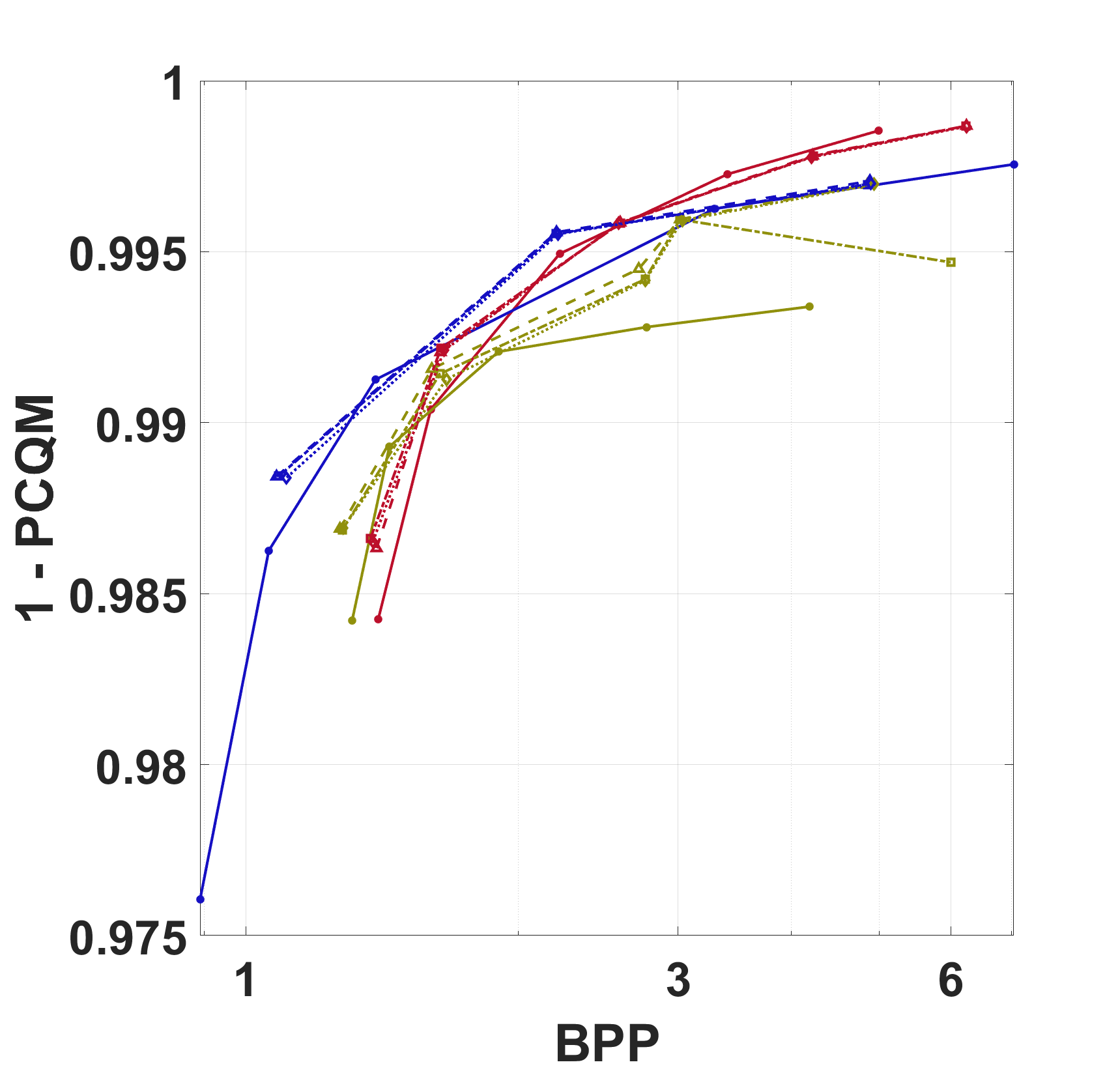}}
    \subfloat[\textit{Romanoillamp}]{%
        \includegraphics[width=0.17\linewidth]{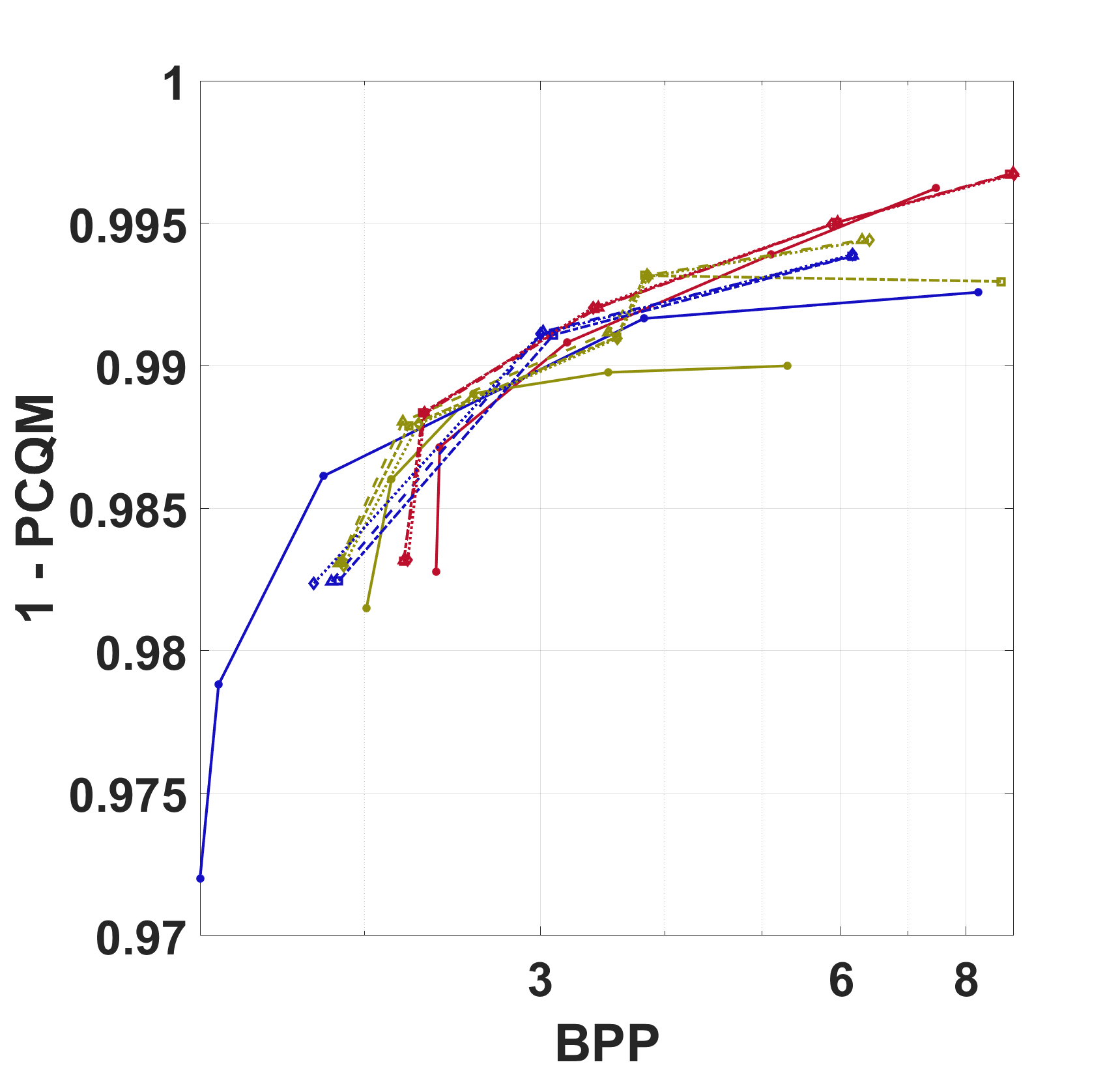}}
    \subfloat[\textit{Citiusp}]{%
        \includegraphics[width=0.17\linewidth]{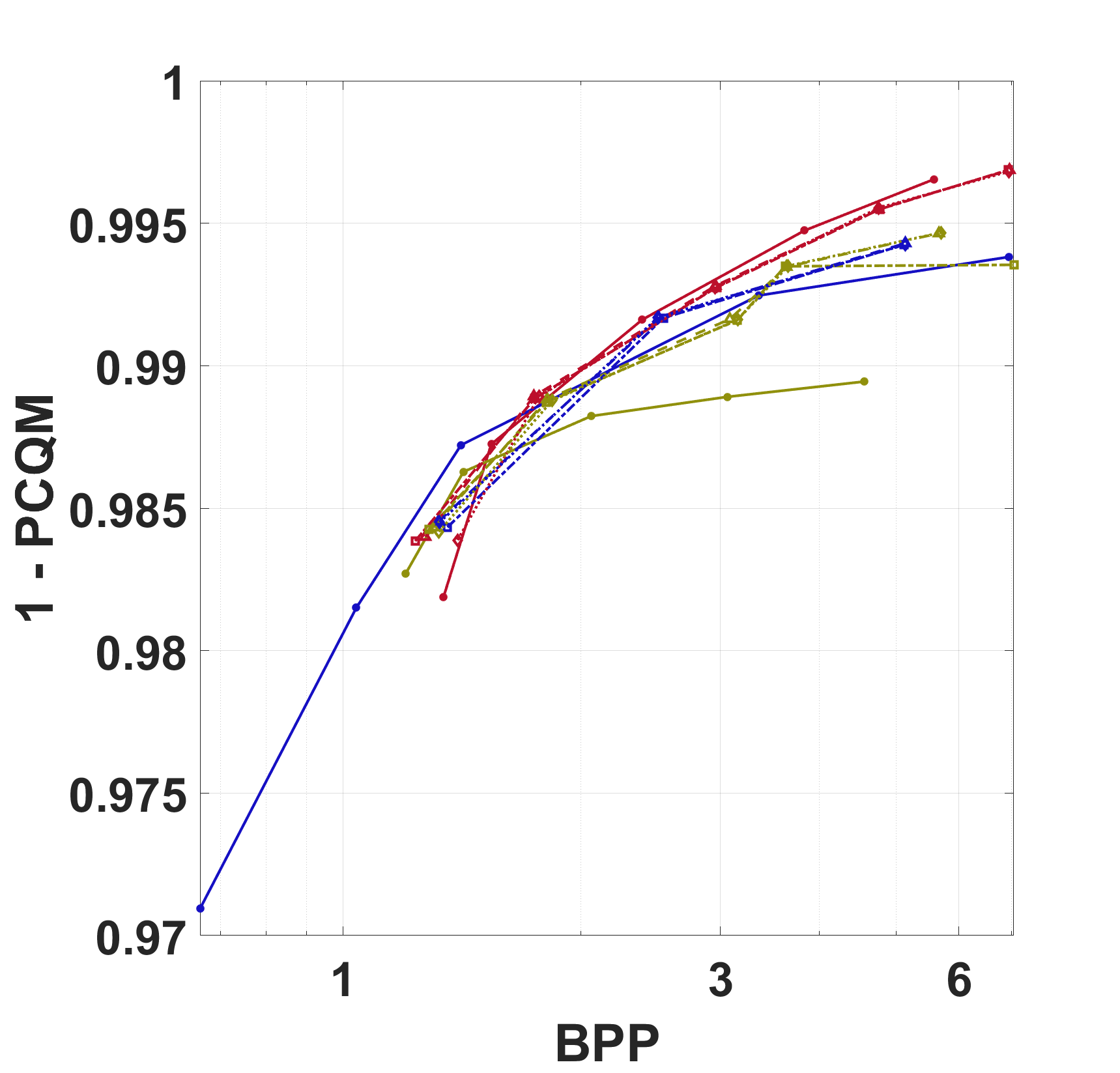}}
    \subfloat[\textit{IpanemaCut}]{%
        \includegraphics[width=0.17\linewidth]{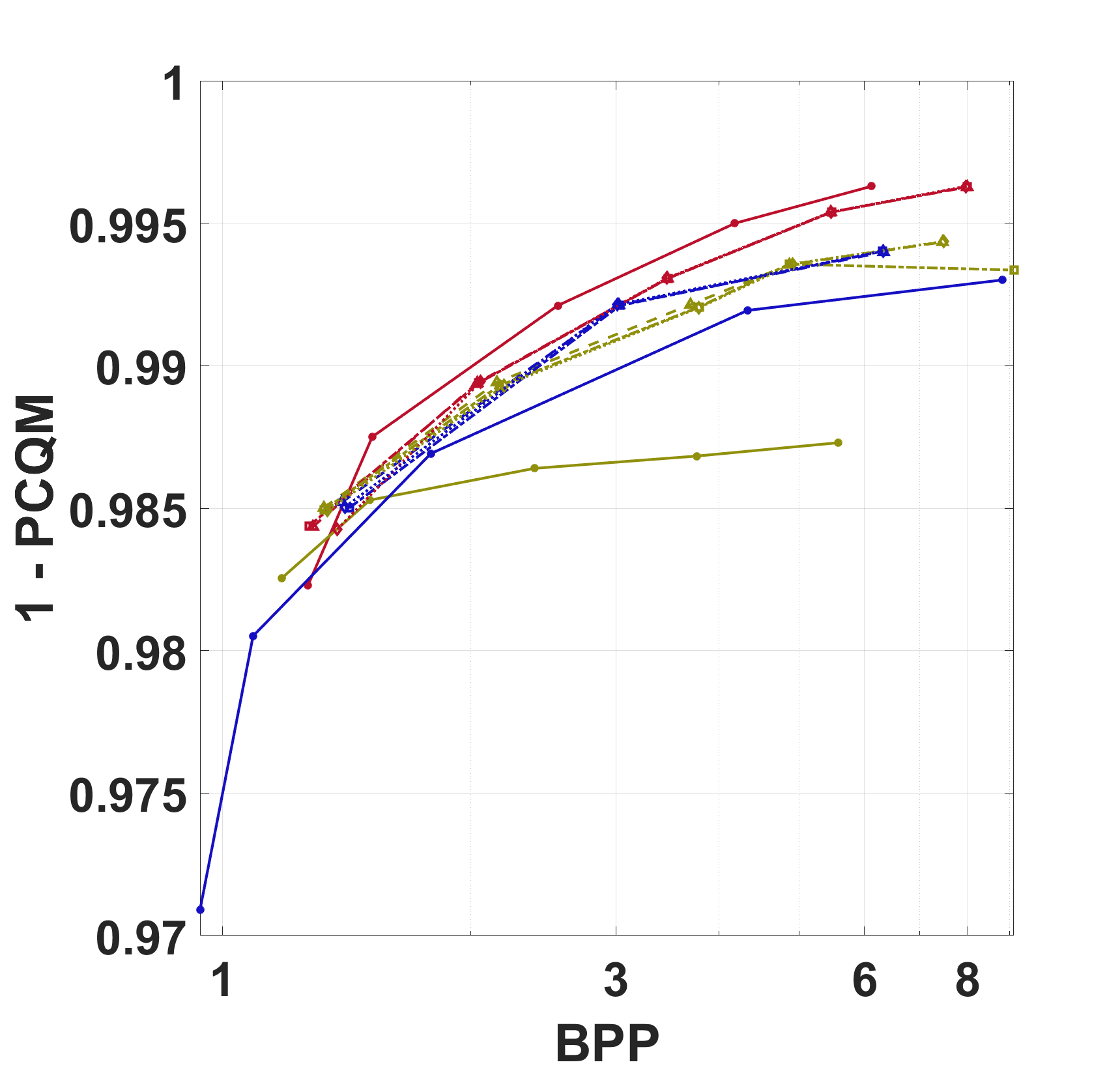}}
    \subfloat[\textit{Ramos}]{%
        \includegraphics[width=0.17\linewidth]{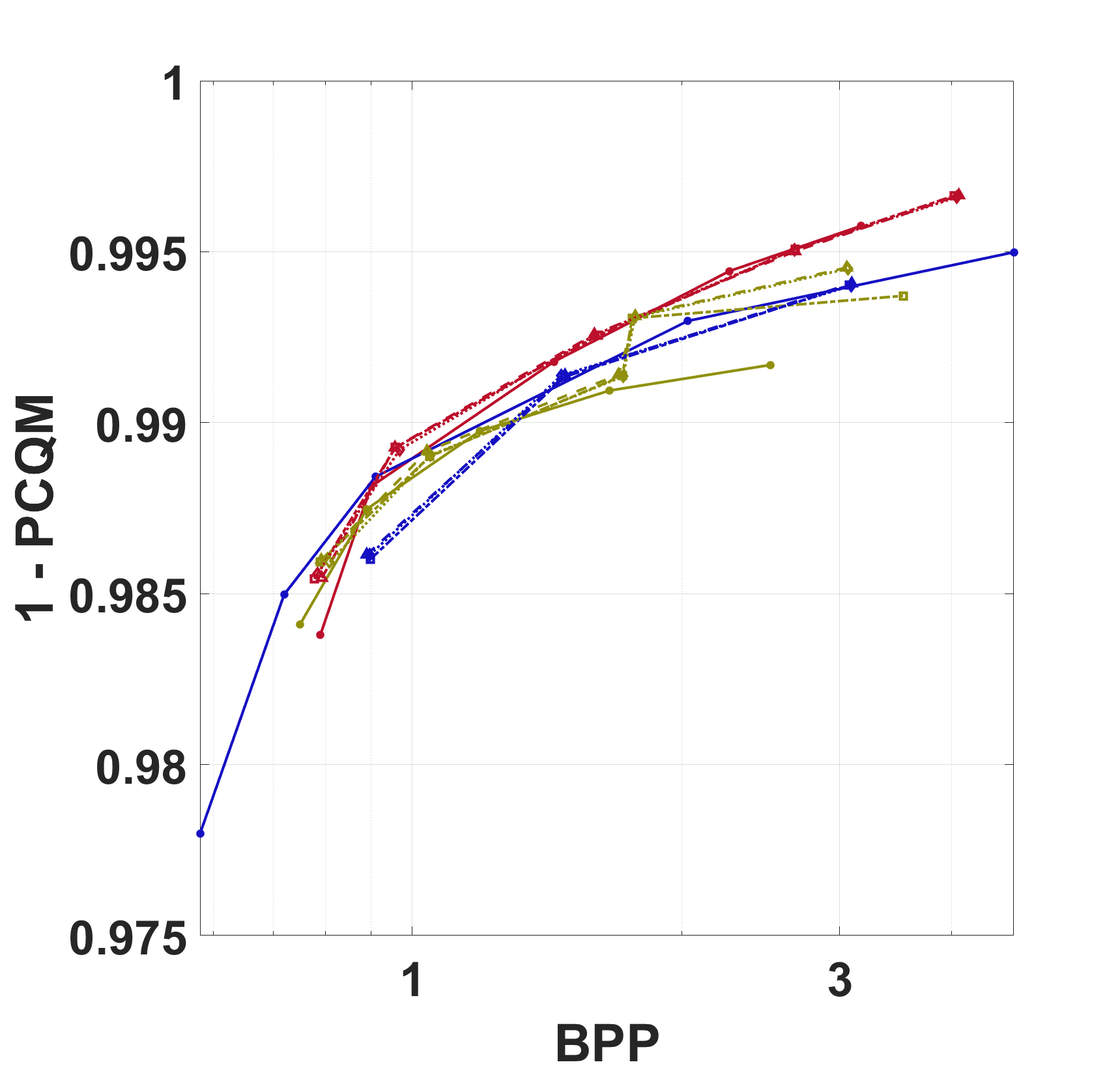}}
        \\
        \includegraphics[width=0.9\linewidth]{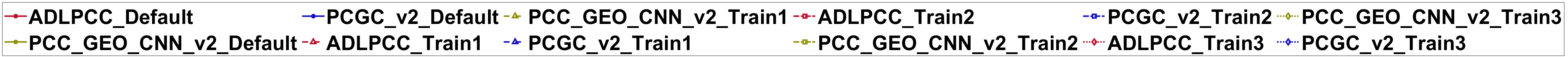}
        \caption{1 - PCQM plots for each of the defined operating points for each codec.}
        \label{finalEpochs_PCQM}
\end{center}        
\end{figure}

\subsection{Final comments on the codecs performance stability}\label{sec:resultsStab}

Figs. \ref{finalEpochs} and \ref{finalEpochs_PCQM} show the performance of the three codecs that resulted from the three training sessions. It is important to understand that the three codecs have different performances for different training sessions. The figures also depict the performance of the default implementations of ADLPCC, PCGCv2 and PCC GEO CNNv2.

The ADLPCC is the one that depends less on the training session, but has a performance decrease in the higher bit rate for three point clouds and the second higher bit rate for one point cloud.
It is important to emphasize that the higher bit rates are the ones that provide an acceptable quality, and they might be the ones most commonly used in practical applications, which makes this inconsistency a problem.
It should be noted that the three different training sessions did not produce the quality drop at the higher rate.

PCC GEO CNNv2 has one training situation that consistently leads to much better performance in the middle bit rates than the other training situations. This level of variation establishes some unreliability in the codecs performance.

The PGCCv2 is the most stable codec and presents the best performance on the middle bit rates for the \textit{Longdress} and \textit{Guanyin} point clouds. The three training sessions also produced poor results for the \textit{Romanoillamp} point cloud.
The larger bit rate can result in lower performance and is not reliable. This is caused by the training model that starts to obtain this bit rate and then adjusts the cost function to the remaining operating points. It has the problem of not reaching near-perceptual lossless qualities.
It should be noted that the authors do not specify the lambda trade-off that they use in the current implementation. As such, some results will vary depending on the default codec and the training sessions conducted.

\section{Conclusions}\label{sec:conclusions}

A study on the performance of machine learning-based codecs for static point clouds was reported, notably, PCGCv2, PCC GEO CNNv2, and ADLPCC.

Because these codecs are geometry-only codecs, two different models that included texture were tested.
Based on previous studies and observations, texture is essential for allowing reliable subjective evaluation.
From the results, it was concluded that the two tests are statistically different. Furthermore, it was concluded that encoding the texture provides a more reliable subjective evaluation than just mapping the original one to the resulting geometry. 
Several objective metrics were computed and correlated with the results of both evaluations.
The PCQM reveals the best representation of the subjective results for \textit{Evaluation 1}, and MSE PSNR D1 showed the best results for \textit{Evaluation 2}. The point cloud objective metrics did not provide a good representation for the subjective evaluation, where texture was just mapped on the decoded point clouds without any compression.

Finally, the stability of different training sessions was analyzed for the three codecs. Although in most cases the performance slightly changes for different training sessions, there were cases where a significant quality variation resulted.
This is highly undesirable, as it results in a reduction in the point cloud coding performance, depending on the training session. Moreover, the performance can depend too much on the content, which is also undesirable (like in the case of \textit{Romanoillamp} for PCGCv2).

It is important to emphasize the difficulty of DL-based codecs in reaching near-perceptual lossless coding, which potentially makes their use difficult to be adopted by the community.
In some cases, it is observed an undesired decrease in quality with the increase in bit rate, mainly for the high bit rates.
This might be caused by a lack of training data or overfitting.
DL-based codecs define different decoders for each bit rate. One possible solution is to define multiple encoders and decoders for high bit rates and use the one that provides the best bit-rate distortion ratio. Eventually, the decoder and encoder settings could be defined in the metadata, and non-default decoders could even be included in the metadata. Furthermore, these non-default codecs could also be compressed. MPEG recently created a standard for neural network compression that could be used. However, the bit-rate distortion consequences of using it in this scenario still need to be researched and evaluated.

\section*{Data availability}\vspace{-0.3cm}
All the data generated or analyzed during this study are available at \\https://github.com/JoeyPrazeres/MMTA2023-PAoDLbPCCS.
\section*{Conflict of interests}\vspace{-0.3cm}
The authors declare that they have no conflict of interest.
\bibliography{sn-bibliography}


\begin{thebibliography}{58}
\ifx \bisbn   \undefined \def \bisbn  #1{ISBN #1}\fi
\ifx \binits  \undefined \def \binits#1{#1}\fi
\ifx \bauthor  \undefined \def \bauthor#1{#1}\fi
\ifx \batitle  \undefined \def \batitle#1{#1}\fi
\ifx \bjtitle  \undefined \def \bjtitle#1{#1}\fi
\ifx \bvolume  \undefined \def \bvolume#1{\textbf{#1}}\fi
\ifx \byear  \undefined \def \byear#1{#1}\fi
\ifx \bissue  \undefined \def \bissue#1{#1}\fi
\ifx \bfpage  \undefined \def \bfpage#1{#1}\fi
\ifx \blpage  \undefined \def \blpage #1{#1}\fi
\ifx \burl  \undefined \def \burl#1{\textsf{#1}}\fi
\ifx \doiurl  \undefined \def \doiurl#1{\url{https://doi.org/#1}}\fi
\ifx \betal  \undefined \def \betal{\textit{et al.}}\fi
\ifx \binstitute  \undefined \def \binstitute#1{#1}\fi
\ifx \binstitutionaled  \undefined \def \binstitutionaled#1{#1}\fi
\ifx \bctitle  \undefined \def \bctitle#1{#1}\fi
\ifx \beditor  \undefined \def \beditor#1{#1}\fi
\ifx \bpublisher  \undefined \def \bpublisher#1{#1}\fi
\ifx \bbtitle  \undefined \def \bbtitle#1{#1}\fi
\ifx \bedition  \undefined \def \bedition#1{#1}\fi
\ifx \bseriesno  \undefined \def \bseriesno#1{#1}\fi
\ifx \blocation  \undefined \def \blocation#1{#1}\fi
\ifx \bsertitle  \undefined \def \bsertitle#1{#1}\fi
\ifx \bsnm \undefined \def \bsnm#1{#1}\fi
\ifx \bsuffix \undefined \def \bsuffix#1{#1}\fi
\ifx \bparticle \undefined \def \bparticle#1{#1}\fi
\ifx \barticle \undefined \def \barticle#1{#1}\fi
\bibcommenthead
\ifx \bconfdate \undefined \def \bconfdate #1{#1}\fi
\ifx \botherref \undefined \def \botherref #1{#1}\fi
\ifx \url \undefined \def \url#1{\textsf{#1}}\fi
\ifx \bchapter \undefined \def \bchapter#1{#1}\fi
\ifx \bbook \undefined \def \bbook#1{#1}\fi
\ifx \bcomment \undefined \def \bcomment#1{#1}\fi
\ifx \oauthor \undefined \def \oauthor#1{#1}\fi
\ifx \citeauthoryear \undefined \def \citeauthoryear#1{#1}\fi
\ifx \endbibitem  \undefined \def \endbibitem {}\fi
\ifx \bconflocation  \undefined \def \bconflocation#1{#1}\fi
\ifx \arxivurl  \undefined \def \arxivurl#1{\textsf{#1}}\fi
\csname PreBibitemsHook\endcsname

\bibitem[\protect\citeauthoryear{Prazeres et~al.}{2022}]{euvip22}
\begin{bchapter}
\bauthor{\bsnm{Prazeres}, \binits{J.}},
\bauthor{\bsnm{Rodrigues}, \binits{R.}},
\bauthor{\bsnm{Pereira}, \binits{M.}},
\bauthor{\bsnm{Pinheiro}, \binits{A.M.G.}}:
\bctitle{On the stability of point cloud machine learning based coding}.
In: \bbtitle{2022 10th European Workshop on Visual Information Processing (EUVIP)},
pp. \bfpage{1}--\blpage{6}
(\byear{2022}).
\doiurl{10.1109/EUVIP53989.2022.9922676}
\end{bchapter}
\endbibitem

\bibitem[\protect\citeauthoryear{Mammou et~al.}{2019}]{G-PCC}
\begin{botherref}
\oauthor{\bsnm{Mammou}, \binits{K.}},
\oauthor{\bsnm{Chou}, \binits{P.A.}},
\oauthor{\bsnm{Flynn}, \binits{D.}},
\oauthor{\bsnm{Krivoku\'ca}, \binits{M.}}:
G-pcc codec description v2.
ISO/IEC JTC1/SC29/WG11 N18189
(2019)
\end{botherref}
\endbibitem

\bibitem[\protect\citeauthoryear{Borges et~al.}{2022}]{QueirozTechRxiv2021}
\begin{barticle}
\bauthor{\bsnm{Borges}, \binits{T.M.}},
\bauthor{\bsnm{Garcia}, \binits{D.C.}},
\bauthor{\bsnm{De~Queiroz}, \binits{R.L.}}:
\batitle{Fractional super-resolution of voxelized point clouds}.
\bjtitle{IEEE Transactions on Image Processing}
\bvolume{31},
\bfpage{1380}--\blpage{1390}
(\byear{2022})
\end{barticle}
\endbibitem

\bibitem[\protect\citeauthoryear{da~Silva~Cruz et~al.}{2019}]{Silva2019a}
\begin{bchapter}
\bauthor{\bsnm{Silva~Cruz}, \binits{L.A.}},
\bauthor{\bsnm{Dumi{\'c}}, \binits{E.}},
\bauthor{\bsnm{Alexiou}, \binits{E.}},
\bauthor{\bsnm{Prazeres}, \binits{J.}},
\bauthor{\bsnm{Duarte}, \binits{R.}},
\bauthor{\bsnm{Pereira}, \binits{M.}},
\bauthor{\bsnm{Pinheiro}, \binits{A.}},
\bauthor{\bsnm{Ebrahimi}, \binits{T.}}:
\bctitle{Point cloud quality evaluation: {Towards} a definition for test conditions}.
In: \bbtitle{2019 Eleventh International Conference on Quality of Multimedia Experience (QoMEX)},
pp. \bfpage{1}--\blpage{6}
(\byear{2019})
\end{bchapter}
\endbibitem

\bibitem[\protect\citeauthoryear{Perry et~al.}{2021}]{MMSP2021}
\begin{bchapter}
\bauthor{\bsnm{Perry}, \binits{S.}},
\bauthor{\bsnm{Cruz}, \binits{L.A.D.S.}},
\bauthor{\bsnm{Dumic}, \binits{E.}},
\bauthor{\bsnm{Nguyen}, \binits{N.H.T.}},
\bauthor{\bsnm{Pinheiro}, \binits{A.}},
\bauthor{\bsnm{Alexiou}, \binits{E.}}:
\bctitle{Comparison of remote subjective assessment strategies in the context of the \mbox{JPEG Pleno} point cloud activity}.
In: \bbtitle{2021 IEEE 23rd International Workshop on Multimedia Signal Processing (MMSP)},
pp. \bfpage{1}--\blpage{6}
(\byear{2021})
\end{bchapter}
\endbibitem

\bibitem[\protect\citeauthoryear{Alexiou et~al.}{2018}]{Alexiou2018a}
\begin{bchapter}
\bauthor{\bsnm{Alexiou}, \binits{E.}},
\bauthor{\bsnm{Ebrahimi}, \binits{T.}},
\bauthor{\bsnm{Bernardo}, \binits{M.V.}},
\bauthor{\bsnm{Pereira}, \binits{M.}},
\bauthor{\bsnm{Pinheiro}, \binits{A.}},
\bauthor{\bsnm{Cruz}, \binits{L.A.D.S.}},
\bauthor{\bsnm{Duarte}, \binits{C.}},
\bauthor{\bsnm{Dmitrovic}, \binits{L.G.}},
\bauthor{\bsnm{Dumic}, \binits{E.}},
\bauthor{\bsnm{Matkovics}, \binits{D.}}, \betal:
\bctitle{Point cloud subjective evaluation methodology based on {2D} rendering}.
In: \bbtitle{2018 Tenth International Conference on Quality of Multimedia Experience (QoMEX)},
pp. \bfpage{1}--\blpage{6}
(\byear{2018}).
\bcomment{IEEE}
\end{bchapter}
\endbibitem

\bibitem[\protect\citeauthoryear{Astola et~al.}{2020}]{astola2020jpeg}
\begin{botherref}
\oauthor{\bsnm{Astola}, \binits{P.}},
\oauthor{\bsnm{Silva~Cruz}, \binits{L.A.}},
\oauthor{\bsnm{Silva}, \binits{E.A.B.}},
\oauthor{\bsnm{Ebrahimi}, \binits{T.}},
\oauthor{\bsnm{Freitas}, \binits{P.G.}},
\oauthor{\bsnm{Gilles}, \binits{A.}},
\oauthor{\bsnm{Oh}, \binits{K.-J.}},
\oauthor{\bsnm{Pagliari}, \binits{C.}},
\oauthor{\bsnm{Pereira}, \binits{F.}},
\oauthor{\bsnm{Perra}, \binits{C.}},
\oauthor{\bsnm{Perry}, \binits{S.}},
\oauthor{\bsnm{Pinheiro}, \binits{A.M.G.}},
\oauthor{\bsnm{Schelkens}, \binits{P.}},
\oauthor{\bsnm{Seidel}, \binits{I.}},
\oauthor{\bsnm{Tabus}, \binits{I.}}:
\mbox{JPEG Pleno}: Standardizing a coding framework and tools for plenoptic imaging modalities.
ITU Journal: ICT Discoveries
\textbf{3}
(2020)
\end{botherref}
\endbibitem

\bibitem[\protect\citeauthoryear{Rusu and Cousins}{2011}]{rusu2011a}
\begin{bchapter}
\bauthor{\bsnm{Rusu}, \binits{R.B.}},
\bauthor{\bsnm{Cousins}, \binits{S.}}:
\bctitle{{3D} is here: {P}oint {C}loud {L}ibrary ({PCL})}.
In: \bbtitle{2011 {IEEE} International Conference on Robotics and Automation},
pp. \bfpage{1}--\blpage{4}
(\byear{2011})
\end{bchapter}
\endbibitem

\bibitem[\protect\citeauthoryear{Alexiou et~al.}{2019}]{alexiou2019a}
\begin{barticle}
\bauthor{\bsnm{Alexiou}, \binits{E.}},
\bauthor{\bsnm{Viola}, \binits{I.}},
\bauthor{\bsnm{Borges}, \binits{T.M.}},
\bauthor{\bsnm{Fonseca}, \binits{T.A.}},
\bauthor{\bsnm{De~Queiroz}, \binits{R.L.}},
\bauthor{\bsnm{Ebrahimi}, \binits{T.}}:
\batitle{A comprehensive study of the rate-distortion performance in {MPEG} point cloud compression}.
\bjtitle{APSIPA Transactions on Signal and Information Processing}
\bvolume{8},
\bfpage{27}
(\byear{2019})
\end{barticle}
\endbibitem

\bibitem[\protect\citeauthoryear{Perry et~al.}{2020}]{icip2020}
\begin{bchapter}
\bauthor{\bsnm{Perry}, \binits{S.}},
\bauthor{\bsnm{Cong}, \binits{H.P.}},
\bauthor{\bsnm{Silva~Cruz}, \binits{L.A.}},
\bauthor{\bsnm{Prazeres}, \binits{J.}},
\bauthor{\bsnm{Pereira}, \binits{M.}},
\bauthor{\bsnm{Pinheiro}, \binits{A.}},
\bauthor{\bsnm{Dumic}, \binits{E.}},
\bauthor{\bsnm{Alexiou}, \binits{E.}},
\bauthor{\bsnm{Ebrahimi}, \binits{T.}}:
\bctitle{Quality evaluation of static point clouds encoded using mpeg codecs}.
In: \bbtitle{2020 IEEE International Conference on Image Processing (ICIP)},
pp. \bfpage{3428}--\blpage{3432}
(\byear{2020}).
\doiurl{10.1109/ICIP40778.2020.9191308}
\end{bchapter}
\endbibitem

\bibitem[\protect\citeauthoryear{Zakharchenko}{2018}]{vpcc}
\begin{botherref}
\oauthor{\bsnm{Zakharchenko}, \binits{V.}}:
“algorithm description of \mbox{mpeg-pcc-tmc2}”.
ISO/IEC JTC1/SC29/WG11 MPEG2018/N17767
(2018)
\end{botherref}
\endbibitem

\bibitem[\protect\citeauthoryear{Prazeres et~al.}{2022}]{EI2022}
\begin{bchapter}
\bauthor{\bsnm{Prazeres}, \binits{J.}},
\bauthor{\bsnm{Pereira}, \binits{M.}},
\bauthor{\bsnm{Pinheiro}, \binits{A.M.}}:
\bctitle{Quality analysis of point cloud coding solutions}.
In: \bbtitle{2022 Electronic Imaging Symposium}
(\byear{2022})
\end{bchapter}
\endbibitem

\bibitem[\protect\citeauthoryear{Guarda et~al.}{2019}]{8954537}
\begin{bchapter}
\bauthor{\bsnm{Guarda}, \binits{A.F.}},
\bauthor{\bsnm{Rodrigues}, \binits{N.M.}},
\bauthor{\bsnm{Pereira}, \binits{F.}}:
\bctitle{Point cloud coding: {Adopting} a deep learning-based approach}.
In: \bbtitle{2019 Picture Coding Symposium (PCS)},
pp. \bfpage{1}--\blpage{5}
(\byear{2019})
\end{bchapter}
\endbibitem

\bibitem[\protect\citeauthoryear{Guarda et~al.}{2020}]{9287060}
\begin{bchapter}
\bauthor{\bsnm{Guarda}, \binits{A.F.}},
\bauthor{\bsnm{Rodrigues}, \binits{N.M.}},
\bauthor{\bsnm{Pereira}, \binits{F.}}:
\bctitle{Deep learning-based point cloud geometry coding with resolution scalability}.
In: \bbtitle{2020 IEEE 22nd International Workshop on Multimedia Signal Processing (MMSP)},
pp. \bfpage{1}--\blpage{6}
(\byear{2020})
\end{bchapter}
\endbibitem

\bibitem[\protect\citeauthoryear{Wang et~al.}{2019}]{Jianqiang-PCGC}
\begin{botherref}
\oauthor{\bsnm{Wang}, \binits{J.}},
\oauthor{\bsnm{Zhu}, \binits{H.}},
\oauthor{\bsnm{Ma}, \binits{Z.}},
\oauthor{\bsnm{Chen}, \binits{T.}},
\oauthor{\bsnm{Liu}, \binits{H.}},
\oauthor{\bsnm{Shen}, \binits{Q.}}:
Learned point cloud geometry compression.
arXiv preprint arXiv:1909.12037
(2019)
\end{botherref}
\endbibitem

\bibitem[\protect\citeauthoryear{Wang et~al.}{2021}]{Jianqiang-PCGCv2}
\begin{bchapter}
\bauthor{\bsnm{Wang}, \binits{J.}},
\bauthor{\bsnm{Ding}, \binits{D.}},
\bauthor{\bsnm{Li}, \binits{Z.}},
\bauthor{\bsnm{Ma}, \binits{Z.}}:
\bctitle{Multiscale point cloud geometry compression}.
In: \bbtitle{2021 Data Compression Conference (DCC)},
pp. \bfpage{73}--\blpage{82}
(\byear{2021}).
\bcomment{IEEE}
\end{bchapter}
\endbibitem

\bibitem[\protect\citeauthoryear{Quach et~al.}{2019}]{quach2019-geocnn}
\begin{bchapter}
\bauthor{\bsnm{Quach}, \binits{M.}},
\bauthor{\bsnm{Valenzise}, \binits{G.}},
\bauthor{\bsnm{Dufaux}, \binits{F.}}:
\bctitle{Learning convolutional transforms for lossy point cloud geometry compression}.
In: \bbtitle{2019 IEEE International Conference on Image Processing (ICIP)},
pp. \bfpage{4320}--\blpage{4324}
(\byear{2019})
\end{bchapter}
\endbibitem

\bibitem[\protect\citeauthoryear{Quach et~al.}{2020}]{quach2020improved}
\begin{bchapter}
\bauthor{\bsnm{Quach}, \binits{M.}},
\bauthor{\bsnm{Valenzise}, \binits{G.}},
\bauthor{\bsnm{Dufaux}, \binits{F.}}:
\bctitle{Improved deep point cloud geometry compression}.
In: \bbtitle{2020 IEEE 22nd International Workshop on Multimedia Signal Processing (MMSP)},
pp. \bfpage{1}--\blpage{6}
(\byear{2020})
\end{bchapter}
\endbibitem

\bibitem[\protect\citeauthoryear{Guarda et~al.}{2020}]{GuardaDL}
\begin{barticle}
\bauthor{\bsnm{Guarda}, \binits{A.F.}},
\bauthor{\bsnm{Rodrigues}, \binits{N.M.}},
\bauthor{\bsnm{Pereira}, \binits{F.}}:
\batitle{Adaptive deep learning-based point cloud geometry coding}.
\bjtitle{IEEE Journal of Selected Topics in Signal Processing}
\bvolume{15}(\bissue{2}),
\bfpage{415}--\blpage{430}
(\byear{2020})
\end{barticle}
\endbibitem

\bibitem[\protect\citeauthoryear{Pang et~al.}{2022}]{pang2022grasp}
\begin{botherref}
\oauthor{\bsnm{Pang}, \binits{J.}},
\oauthor{\bsnm{Lodhi}, \binits{M.A.}},
\oauthor{\bsnm{Tian}, \binits{D.}}:
Grasp-net: Geometric residual analysis and synthesis for point cloud compression
(2022)
\end{botherref}
\endbibitem

\bibitem[\protect\citeauthoryear{Wang et~al.}{2022}]{WangPAMI2022}
\begin{botherref}
\oauthor{\bsnm{Wang}, \binits{J.}},
\oauthor{\bsnm{Ding}, \binits{D.}},
\oauthor{\bsnm{Li}, \binits{Z.}},
\oauthor{\bsnm{Feng}, \binits{X.}},
\oauthor{\bsnm{Cao}, \binits{C.}},
\oauthor{\bsnm{Ma}, \binits{Z.}}:
Sparse tensor-based multiscale representation for point cloud geometry compression.
IEEE Transactions on Pattern Analysis and Machine Intelligence,
1--18
(2022)
\doiurl{10.1109/TPAMI.2022.3225816}
\end{botherref}
\endbibitem

\bibitem[\protect\citeauthoryear{{ISO/IEC JTC1/SC29 WG7 (MPEG 3D Graphics Coding) }}{2022}]{MPEGDOC}
\begin{botherref}
\oauthor{\bsnm{{ISO/IEC JTC1/SC29 WG7 (MPEG 3D Graphics Coding) }}}:
Performance analysis of currently AI-based available solutions for PCC.
ISO/IEC
(2022)
\end{botherref}
\endbibitem

\bibitem[\protect\citeauthoryear{Szegedy et~al.}{2017}]{SzegedyInceptionv4}
\begin{bchapter}
\bauthor{\bsnm{Szegedy}, \binits{C.}},
\bauthor{\bsnm{Ioffe}, \binits{S.}},
\bauthor{\bsnm{Vanhoucke}, \binits{V.}},
\bauthor{\bsnm{Alemi}, \binits{A.}}:
\bctitle{{Inception-v4}, {Inception-ResNet} and the impact of residual connections on learning}.
In: \bbtitle{Proceedings of the AAAI Conference on Artificial Intelligence},
vol. \bseriesno{31}
(\byear{2017})
\end{bchapter}
\endbibitem

\bibitem[\protect\citeauthoryear{Javaheri et~al.}{2017}]{Javaheri2017b}
\begin{bchapter}
\bauthor{\bsnm{Javaheri}, \binits{A.}},
\bauthor{\bsnm{Brites}, \binits{C.}},
\bauthor{\bsnm{Pereira}, \binits{F.}},
\bauthor{\bsnm{Ascenso}, \binits{J.}}:
\bctitle{Subjective and objective quality evaluation of compressed point clouds}.
In: \bbtitle{2017 IEEE 19th International Workshop on Multimedia Signal Processing (MMSP)},
pp. \bfpage{1}--\blpage{6}
(\byear{2017}).
\doiurl{10.1109/MMSP.2017.8122239}
\end{bchapter}
\endbibitem

\bibitem[\protect\citeauthoryear{Alexious et~al.}{2018}]{alexious2018point}
\begin{bchapter}
\bauthor{\bsnm{Alexious}, \binits{E.}},
\bauthor{\bsnm{Pinheiro}, \binits{A.M.}},
\bauthor{\bsnm{Duarte}, \binits{C.}},
\bauthor{\bsnm{Matkovi{\'c}}, \binits{D.}},
\bauthor{\bsnm{Dumi{\'c}}, \binits{E.}},
\bauthor{\bsnm{Silva~Cruz}, \binits{L.A.}},
\bauthor{\bsnm{Dmitrovi{\'c}}, \binits{L.G.}},
\bauthor{\bsnm{Bernardo}, \binits{M.V.}},
\bauthor{\bsnm{Pereira}, \binits{M.}},
\bauthor{\bsnm{Ebrahimi}, \binits{T.}}:
\bctitle{Point cloud subjective evaluation methodology based on reconstructed surfaces}.
In: \bbtitle{Applications of Digital Image Processing XLI},
vol. \bseriesno{10752},
pp. \bfpage{160}--\blpage{173}
(\byear{2018}).
\bcomment{SPIE}
\end{bchapter}
\endbibitem

\bibitem[\protect\citeauthoryear{Su et~al.}{2019}]{Su2019a}
\begin{bchapter}
\bauthor{\bsnm{Su}, \binits{H.}},
\bauthor{\bsnm{Duanmu}, \binits{Z.}},
\bauthor{\bsnm{Liu}, \binits{W.}},
\bauthor{\bsnm{Liu}, \binits{Q.}},
\bauthor{\bsnm{Wang}, \binits{Z.}}:
\bctitle{Perceptual quality assessment of {3D} point clouds}.
In: \bbtitle{2019 IEEE International Conference on Image Processing (ICIP)},
pp. \bfpage{3182}--\blpage{3186}
(\byear{2019})
\end{bchapter}
\endbibitem

\bibitem[\protect\citeauthoryear{Prazeres et~al.}{2022}]{acm22}
\begin{bchapter}
\bauthor{\bsnm{Prazeres}, \binits{J.}},
\bauthor{\bsnm{Rodrigues}, \binits{R.}},
\bauthor{\bsnm{Pereira}, \binits{M.}},
\bauthor{\bsnm{Pinheiro}, \binits{A.M.}}:
\bctitle{Quality evaluation of machine learning-based point cloud coding solutions}.
In: \bbtitle{Proceedings of the 1st International Workshop on Advances in Point Cloud Compression, Processing and Analysis},
pp. \bfpage{57}--\blpage{65}
(\byear{2022})
\end{bchapter}
\endbibitem

\bibitem[\protect\citeauthoryear{Alexiou et~al.}{2017}]{AlexiouMMSP2017}
\begin{bchapter}
\bauthor{\bsnm{Alexiou}, \binits{E.}},
\bauthor{\bsnm{Upenik}, \binits{E.}},
\bauthor{\bsnm{Ebrahimi}, \binits{T.}}:
\bctitle{Towards subjective quality assessment of point cloud imaging in augmented reality}.
In: \bbtitle{2017 IEEE 19th International Workshop on Multimedia Signal Processing (MMSP)},
pp. \bfpage{1}--\blpage{6}
(\byear{2017}).
\bcomment{IEEE}
\end{bchapter}
\endbibitem

\bibitem[\protect\citeauthoryear{Subramanyam et~al.}{2020}]{ViolaVr}
\begin{bchapter}
\bauthor{\bsnm{Subramanyam}, \binits{S.}},
\bauthor{\bsnm{Li}, \binits{J.}},
\bauthor{\bsnm{Viola}, \binits{I.}},
\bauthor{\bsnm{Cesar}, \binits{P.}}:
\bctitle{Comparing the quality of highly realistic digital humans in {3DoF} and {6DoF}: A volumetric video case study}.
In: \bbtitle{2020 IEEE Conference on Virtual Reality and 3D User Interfaces (VR)},
pp. \bfpage{127}--\blpage{136}
(\byear{2020}).
\bcomment{IEEE}
\end{bchapter}
\endbibitem

\bibitem[\protect\citeauthoryear{Mekuria et~al.}{2016}]{MekuriaVR}
\begin{barticle}
\bauthor{\bsnm{Mekuria}, \binits{R.}},
\bauthor{\bsnm{Blom}, \binits{K.}},
\bauthor{\bsnm{Cesar}, \binits{P.}}:
\batitle{Design, implementation, and evaluation of a point cloud codec for tele-immersive video}.
\bjtitle{IEEE Transactions on Circuits and Systems for Video Technology}
\bvolume{27}(\bissue{4}),
\bfpage{828}--\blpage{842}
(\byear{2016})
\end{barticle}
\endbibitem

\bibitem[\protect\citeauthoryear{Subramanyam et~al.}{2022}]{SubramanyamQoMEX2022}
\begin{bchapter}
\bauthor{\bsnm{Subramanyam}, \binits{S.}},
\bauthor{\bsnm{Viola}, \binits{I.}},
\bauthor{\bsnm{Jansen}, \binits{J.}},
\bauthor{\bsnm{Alexiou}, \binits{E.}},
\bauthor{\bsnm{Hanjalic}, \binits{A.}},
\bauthor{\bsnm{Cesar}, \binits{P.}}:
\bctitle{Subjective qoe evaluation of user-centered adaptive streaming of dynamic point clouds}.
In: \bbtitle{2022 14th International Conference on Quality of Multimedia Experience (QoMEX)},
pp. \bfpage{1}--\blpage{6}
(\byear{2022}).
\doiurl{10.1109/QoMEX55416.2022.9900879}
\end{bchapter}
\endbibitem

\bibitem[\protect\citeauthoryear{Alexiou et~al.}{2020}]{Alexiou:277378}
\begin{bchapter}
\bauthor{\bsnm{Alexiou}, \binits{E.}},
\bauthor{\bsnm{Yang}, \binits{N.}},
\bauthor{\bsnm{Ebrahimi}, \binits{T.}}:
\bctitle{{PointXR}: A toolbox for visualization and subjective evaluation of point clouds in virtual reality}.
In: \bbtitle{2020 Twelfth International Conference on Quality of Multimedia Experience (QoMEX)},
pp. \bfpage{1}--\blpage{6}
(\byear{2020}).
\bcomment{IEEE}
\end{bchapter}
\endbibitem

\bibitem[\protect\citeauthoryear{Prazeres et~al.}{2022}]{ICIP2022}
\begin{bchapter}
\bauthor{\bsnm{Prazeres}, \binits{J.}},
\bauthor{\bsnm{Pereira}, \binits{M.}},
\bauthor{\bsnm{Pinheiro}, \binits{A.M.G.}}:
\bctitle{Subjective quality evaluation of point clouds with \mbox{3D} stereoscopic visualization}.
In: \bbtitle{IEEE International Conference on Image Processing (ICIP)}
(\byear{2022})
\end{bchapter}
\endbibitem

\bibitem[\protect\citeauthoryear{Ak et~al.}{2024}]{ak2023basics}
\begin{botherref}
\oauthor{\bsnm{Ak}, \binits{A.}},
\oauthor{\bsnm{Zerman}, \binits{E.}},
\oauthor{\bsnm{Quach}, \binits{M.}},
\oauthor{\bsnm{Chetouani}, \binits{A.}},
\oauthor{\bsnm{Smolic}, \binits{A.}},
\oauthor{\bsnm{Valenzise}, \binits{G.}},
\oauthor{\bsnm{Le~Callet}, \binits{P.}}:
Basics: Broad quality assessment of static point clouds in a compression scenario.
IEEE Transactions on Multimedia,
1--13
(2024)
\doiurl{10.1109/TMM.2024.3355642}
\end{botherref}
\endbibitem

\bibitem[\protect\citeauthoryear{Lazzarotto et~al.}{2022}]{LazzarottoQoMEX2022}
\begin{bchapter}
\bauthor{\bsnm{Lazzarotto}, \binits{D.}},
\bauthor{\bsnm{Testolina}, \binits{M.}},
\bauthor{\bsnm{Ebrahimi}, \binits{T.}}:
\bctitle{On the impact of spatial rendering on point cloud subjective visual quality assessment}.
In: \bbtitle{2022 14th International Conference on Quality of Multimedia Experience (QoMEX)},
pp. \bfpage{1}--\blpage{6}
(\byear{2022}).
\doiurl{10.1109/QoMEX55416.2022.9900898}
\end{bchapter}
\endbibitem

\bibitem[\protect\citeauthoryear{Liu et~al.}{2022}]{Liu2022TVCG}
\begin{botherref}
\oauthor{\bsnm{Liu}, \binits{Q.}},
\oauthor{\bsnm{Su}, \binits{H.}},
\oauthor{\bsnm{Duanmu}, \binits{Z.}},
\oauthor{\bsnm{Liu}, \binits{W.}},
\oauthor{\bsnm{Wang}, \binits{Z.}}:
Perceptual quality assessment of colored 3d point clouds.
IEEE Transactions on Visualization and Computer Graphics,
1--1
(2022)
\doiurl{10.1109/TVCG.2022.3167151}
\end{botherref}
\endbibitem

\bibitem[\protect\citeauthoryear{Yang et~al.}{2021}]{projectionYANG}
\begin{barticle}
\bauthor{\bsnm{Yang}, \binits{Q.}},
\bauthor{\bsnm{Chen}, \binits{H.}},
\bauthor{\bsnm{Ma}, \binits{Z.}},
\bauthor{\bsnm{Xu}, \binits{Y.}},
\bauthor{\bsnm{Tang}, \binits{R.}},
\bauthor{\bsnm{Sun}, \binits{J.}}:
\batitle{Predicting the perceptual quality of point cloud: A 3d-to-2d projection-based exploration}.
\bjtitle{IEEE Transactions on Multimedia}
\bvolume{23},
\bfpage{3877}--\blpage{3891}
(\byear{2021})
\doiurl{10.1109/TMM.2020.3033117}
\end{barticle}
\endbibitem

\bibitem[\protect\citeauthoryear{Liu et~al.}{2023}]{LiuYipengACM2023}
\begin{botherref}
\oauthor{\bsnm{Liu}, \binits{Y.}},
\oauthor{\bsnm{Yang}, \binits{Q.}},
\oauthor{\bsnm{Xu}, \binits{Y.}},
\oauthor{\bsnm{Yang}, \binits{L.}}:
Point cloud quality assessment: Dataset construction and learning-based no-reference metric
\textbf{19}(2s)
(2023)
\doiurl{10.1145/3550274}
\end{botherref}
\endbibitem

\bibitem[\protect\citeauthoryear{Lavou{\'e} et~al.}{2015}]{7272102}
\begin{barticle}
\bauthor{\bsnm{Lavou{\'e}}, \binits{G.}},
\bauthor{\bsnm{Larabi}, \binits{M.C.}},
\bauthor{\bsnm{V{\'a}{\v{s}}a}, \binits{L.}}:
\batitle{On the efficiency of image metrics for evaluating the visual quality of {3D} models}.
\bjtitle{IEEE Transactions on Visualization and Computer Graphics}
\bvolume{22}(\bissue{8}),
\bfpage{1987}--\blpage{1999}
(\byear{2015})
\end{barticle}
\endbibitem

\bibitem[\protect\citeauthoryear{Tian et~al.}{2017}]{Dtian}
\begin{bchapter}
\bauthor{\bsnm{Tian}, \binits{D.}},
\bauthor{\bsnm{Ochimizu}, \binits{H.}},
\bauthor{\bsnm{Feng}, \binits{C.}},
\bauthor{\bsnm{Cohen}, \binits{R.}},
\bauthor{\bsnm{Vetro}, \binits{A.}}:
\bctitle{Geometric distortion metrics for point cloud compression}.
In: \bbtitle{2017 IEEE International Conference on Image Processing (ICIP)},
pp. \bfpage{3460}--\blpage{3464}
(\byear{2017}).
\bcomment{IEEE}
\end{bchapter}
\endbibitem

\bibitem[\protect\citeauthoryear{Lazzarotto et~al.}{2021}]{LazzarottoMMSP2021}
\begin{bchapter}
\bauthor{\bsnm{Lazzarotto}, \binits{D.}},
\bauthor{\bsnm{Alexiou}, \binits{E.}},
\bauthor{\bsnm{Ebrahimi}, \binits{T.}}:
\bctitle{Benchmarking of objective quality metrics for point cloud compression}.
In: \bbtitle{2021 IEEE 23rd International Workshop on Multimedia Signal Processing (MMSP)},
pp. \bfpage{1}--\blpage{6}
(\byear{2021})
\end{bchapter}
\endbibitem

\bibitem[\protect\citeauthoryear{Yang et~al.}{2022}]{Yang_2022_CVPR}
\begin{bchapter}
\bauthor{\bsnm{Yang}, \binits{Q.}},
\bauthor{\bsnm{Liu}, \binits{Y.}},
\bauthor{\bsnm{Chen}, \binits{S.}},
\bauthor{\bsnm{Xu}, \binits{Y.}},
\bauthor{\bsnm{Sun}, \binits{J.}}:
\bctitle{No-reference point cloud quality assessment via domain adaptation}.
In: \bbtitle{Proceedings of the IEEE/CVF Conference on Computer Vision and Pattern Recognition (CVPR)},
pp. \bfpage{21179}--\blpage{21188}
(\byear{2022})
\end{bchapter}
\endbibitem

\bibitem[\protect\citeauthoryear{Zhang et~al.}{2022}]{ZhangCSVT2022}
\begin{barticle}
\bauthor{\bsnm{Zhang}, \binits{Z.}},
\bauthor{\bsnm{Sun}, \binits{W.}},
\bauthor{\bsnm{Min}, \binits{X.}},
\bauthor{\bsnm{Wang}, \binits{T.}},
\bauthor{\bsnm{Lu}, \binits{W.}},
\bauthor{\bsnm{Zhai}, \binits{G.}}:
\batitle{No-reference quality assessment for 3d colored point cloud and mesh models}.
\bjtitle{IEEE Transactions on Circuits and Systems for Video Technology}
\bvolume{32}(\bissue{11}),
\bfpage{7618}--\blpage{7631}
(\byear{2022})
\doiurl{10.1109/TCSVT.2022.3186894}
\end{barticle}
\endbibitem

\bibitem[\protect\citeauthoryear{Tian et~al.}{2017}]{DongICIP2017}
\begin{bchapter}
\bauthor{\bsnm{Tian}, \binits{D.}},
\bauthor{\bsnm{Ochimizu}, \binits{H.}},
\bauthor{\bsnm{Feng}, \binits{C.}},
\bauthor{\bsnm{Cohen}, \binits{R.}},
\bauthor{\bsnm{Vetro}, \binits{A.}}:
\bctitle{Geometric distortion metrics for point cloud compression}.
In: \bbtitle{2017 IEEE International Conference on Image Processing (ICIP)},
pp. \bfpage{3460}--\blpage{3464}
(\byear{2017}).
\bcomment{IEEE}
\end{bchapter}
\endbibitem

\bibitem[\protect\citeauthoryear{Alexiou and Ebrahimi}{2020}]{AlexiouPSSIM}
\begin{bchapter}
\bauthor{\bsnm{Alexiou}, \binits{E.}},
\bauthor{\bsnm{Ebrahimi}, \binits{T.}}:
\bctitle{Towards a point cloud structural similarity metric}.
In: \bbtitle{2020 IEEE International Conference on Multimedia \& Expo Workshops (ICMEW)},
pp. \bfpage{1}--\blpage{6}
(\byear{2020}).
\bcomment{IEEE}
\end{bchapter}
\endbibitem

\bibitem[\protect\citeauthoryear{Meynet et~al.}{2020}]{Meynet2020PCQM}
\begin{bchapter}
\bauthor{\bsnm{Meynet}, \binits{G.}},
\bauthor{\bsnm{Nehm{\'e}}, \binits{Y.}},
\bauthor{\bsnm{Digne}, \binits{J.}},
\bauthor{\bsnm{Lavou{\'e}}, \binits{G.}}:
\bctitle{{PCQM}: A full-reference quality metric for colored 3d point clouds}.
In: \bbtitle{2020 Twelfth International Conference on Quality of Multimedia Experience (QoMEX)},
pp. \bfpage{1}--\blpage{6}
(\byear{2020}).
\bcomment{IEEE}
\end{bchapter}
\endbibitem

\bibitem[\protect\citeauthoryear{Javaheri et~al.}{2021}]{javaheri2021pointtodistribution}
\begin{bchapter}
\bauthor{\bsnm{Javaheri}, \binits{A.}},
\bauthor{\bsnm{Brites}, \binits{C.}},
\bauthor{\bsnm{Pereira}, \binits{F.}},
\bauthor{\bsnm{Ascenso}, \binits{J.}}:
\bctitle{A point-to-distribution joint geometry and color metric for point cloud quality assessment}.
In: \bbtitle{2021 IEEE 23rd International Workshop on Multimedia Signal Processing (MMSP)},
pp. \bfpage{1}--\blpage{6}
(\byear{2021})
\end{bchapter}
\endbibitem

\bibitem[\protect\citeauthoryear{Viola and Cesar}{2020}]{Viola2020PCMRR}
\begin{barticle}
\bauthor{\bsnm{Viola}, \binits{I.}},
\bauthor{\bsnm{Cesar}, \binits{P.}}:
\batitle{A reduced reference metric for visual quality evaluation of point cloud contents}.
\bjtitle{IEEE Signal Processing Letters}
\bvolume{27},
\bfpage{1660}--\blpage{1664}
(\byear{2020})
\end{barticle}
\endbibitem

\bibitem[\protect\citeauthoryear{Yang et~al.}{2020}]{QiYangGraphSIM2022}
\begin{barticle}
\bauthor{\bsnm{Yang}, \binits{Q.}},
\bauthor{\bsnm{Ma}, \binits{Z.}},
\bauthor{\bsnm{Xu}, \binits{Y.}},
\bauthor{\bsnm{Li}, \binits{Z.}},
\bauthor{\bsnm{Sun}, \binits{J.}}:
\batitle{Inferring point cloud quality via graph similarity}.
\bjtitle{IEEE Transactions on Pattern Analysis and Machine Intelligence}
\bvolume{44}(\bissue{6}),
\bfpage{3015}--\blpage{3029}
(\byear{2020})
\end{barticle}
\endbibitem

\bibitem[\protect\citeauthoryear{H.264}{2021}]{H264Text}
\begin{bchapter}
\bauthor{\bsnm{H.264}, \binits{I.}}:
\bctitle{Recommendation h.264}.
(\byear{2021})
\end{bchapter}
\endbibitem

\bibitem[\protect\citeauthoryear{BT.500-13}{2012}]{conditions}
\begin{bchapter}
\bauthor{\bsnm{BT.500-13}, \binits{I.-R.}}:
\bctitle{Methodology for the subjective assessment of the quality of television pictures}.
(\byear{2012})
\end{bchapter}
\endbibitem

\bibitem[\protect\citeauthoryear{P.1401}{2012}]{ITU-T}
\begin{bchapter}
\bauthor{\bsnm{P.1401}, \binits{I.-T.}}:
\bctitle{International telecommunication union,}.
In: \bbtitle{Methods, Metrics and Procedures for Statistical Evaluation, Qualification and Comparison of Objective Quality Prediction Models}
(\byear{2012})
\end{bchapter}
\endbibitem

\bibitem[\protect\citeauthoryear{Kruskal and Wallis}{1952}]{KruskalWallis}
\begin{barticle}
\bauthor{\bsnm{Kruskal}, \binits{W.H.}},
\bauthor{\bsnm{Wallis}, \binits{W.A.}}:
\batitle{Use of ranks in one-criterion variance analysis}.
\bjtitle{Journal of the American Statistical Association}
\bvolume{47}(\bissue{260}),
\bfpage{583}--\blpage{621}
(\byear{1952})
\end{barticle}
\endbibitem

\bibitem[\protect\citeauthoryear{Bernardo et~al.}{2016}]{MarcoColor2016}
\begin{botherref}
\oauthor{\bsnm{Bernardo}, \binits{M.V.}},
\oauthor{\bsnm{Pinheiro}, \binits{A.M.G.}},
\oauthor{\bsnm{Fiadeiro}, \binits{P.T.}},
\oauthor{\bsnm{Pereira}, \binits{M.}}:
Image quality under chromatic impairments.
ACM Trans. Appl. Percept.
\textbf{14}(1)
(2016)
\end{botherref}
\endbibitem

\bibitem[\protect\citeauthoryear{Bo et~al.}{2012}]{QuadricFitting}
\begin{barticle}
\bauthor{\bsnm{Bo}, \binits{P.}},
\bauthor{\bsnm{Ling}, \binits{R.}},
\bauthor{\bsnm{Wang}, \binits{W.}}:
\batitle{A revisit to fitting parametric surfaces to point clouds}.
\bjtitle{Computers \& Graphics}
\bvolume{36}(\bissue{5}),
\bfpage{534}--\blpage{540}
(\byear{2012})
\end{barticle}
\endbibitem

\bibitem[\protect\citeauthoryear{{ISO/IEC JTC1/SC29/WG1}}{2020}]{wg1m89044JPEGPCC}
\begin{botherref}
\oauthor{\bsnm{{ISO/IEC JTC1/SC29/WG1}}}:
\mbox{JPEG Pleno PC} Exploration Study 4 Results.
ISO/IEC.
\mbox{WG1M89044}, 89th Meeting
(2020)
\end{botherref}
\endbibitem

\bibitem[\protect\citeauthoryear{Chang et~al.}{2015}]{shapenet}
\begin{botherref}
\oauthor{\bsnm{Chang}, \binits{A.X.}},
\oauthor{\bsnm{Funkhouser}, \binits{T.}},
\oauthor{\bsnm{Guibas}, \binits{L.}},
\oauthor{\bsnm{Hanrahan}, \binits{P.}},
\oauthor{\bsnm{Huang}, \binits{Q.}},
\oauthor{\bsnm{Li}, \binits{Z.}},
\oauthor{\bsnm{Savarese}, \binits{S.}},
\oauthor{\bsnm{Savva}, \binits{M.}},
\oauthor{\bsnm{Song}, \binits{S.}},
\oauthor{\bsnm{Su}, \binits{H.}}, et al.:
Shapenet: An information-rich {3D} model repository.
arXiv preprint arXiv:1512.03012
(2015)
\end{botherref}
\endbibitem

\bibitem[\protect\citeauthoryear{Wu et~al.}{2015}]{modelnet}
\begin{bchapter}
\bauthor{\bsnm{Wu}, \binits{Z.}},
\bauthor{\bsnm{Song}, \binits{S.}},
\bauthor{\bsnm{Khosla}, \binits{A.}},
\bauthor{\bsnm{Yu}, \binits{F.}},
\bauthor{\bsnm{Zhang}, \binits{L.}},
\bauthor{\bsnm{Tang}, \binits{X.}},
\bauthor{\bsnm{Xiao}, \binits{J.}}:
\bctitle{{3D} shapenets: A deep representation for volumetric shapes}.
In: \bbtitle{Proceedings of the IEEE Conference on Computer Vision and Pattern Recognition},
pp. \bfpage{1912}--\blpage{1920}
(\byear{2015})
\end{bchapter}
\endbibitem

\end{thebibliography}

\end{document}